\documentstyle[aps,prd,epsf]{revtex}
\def\lsim{\raise0.3ex\hbox{$<$\kern-0.75em\raise-1.1ex\hbox{$\sim$}}}
\def\gsim{\raise0.3ex\hbox{$>$\kern-0.75em\raise-1.1ex\hbox{$\sim$}}}
\newcommand{\be} {\begin{equation}}
\newcommand{\ee} {\end{equation}}
\newcommand{\bea} {\begin{eqnarray}}
\newcommand{\eea} {\end{eqnarray}}
\newcommand{\dM} {\Delta M}

\begin{document}

\draft
\tightenlines

\title{
\begin{flushright}
\normalsize
UTHEP-454 \\
UTCCP-P-120 \\
December  2001\\
\end{flushright}
Charmonium Spectrum from Quenched Anisotropic Lattice QCD
}

\author{
  M.~Okamoto\rlap,$^{\rm a}$
  S.~Aoki\rlap,$^{\rm a}$
  R.~Burkhalter\rlap,$^{\rm a,b}$ 
  S.~Ejiri\rlap,$^{\rm b}$\thanks{present address :
            Department of Physics, University of Wales, 
            Swansea SA2 8PP, U.K.} 
  M.~Fukugita\rlap,$^{\rm c}$
  S.~Hashimoto\rlap,$^{\rm d}$
  K-I.~Ishikawa\rlap,$^{\rm b}$
  N.~Ishizuka\rlap,$^{\rm a,b}$
  Y.~Iwasaki\rlap,$^{\rm a,b}$
  K.~Kanaya\rlap,$^{\rm a}$ 
  T.~Kaneko\rlap,$^{\rm d}$ 
  Y.~Kuramashi\rlap,$^{\rm d}$
  V.~Lesk\rlap,$^{\rm b}$
  K.~Nagai\rlap,$^{\rm b}$\thanks{present address :
            CERN, Theory Division, CH--1211 Geneva 23, Switzerland}
  M.~Okawa\rlap,$^{\rm d}$ 
  Y.~Taniguchi\rlap,$^{\rm a}$
  A.~Ukawa$^{\rm a,b}$ and
  T.~Yoshi\'e$^{\rm a,b}$\\
  (CP-PACS Collaboration)
 }

\address{
$^{\rm a}$Institute of Physics,
    University of Tsukuba, Tsukuba, Ibaraki 305-8571, Japan \\
$^{\rm b}$Center for Computational Physics,
    University of Tsukuba, Tsukuba, Ibaraki 305-8577, Japan \\
$^{\rm c}$Institute for Cosmic Ray Research,
    University of Tokyo, Kashiwa 277-8582, Japan \\
$^{\rm d}$High Energy Accelerator Research Organization
    (KEK), Tsukuba, Ibaraki 305-0801, Japan
}

\date{\today}

\maketitle

\begin{abstract}

We present a detailed study of the charmonium spectrum using 
anisotropic lattice QCD.
We first derive a tree-level improved clover quark action on the 
anisotropic lattice for arbitrary quark mass by matching the Hamiltonian 
on the lattice and in the continuum.
The heavy quark mass dependence of the improvement coefficients, {\it i.e.}
the ratio of the hopping parameters $\zeta=K_t/K_s$ and the clover
coefficients $c_{s,t}$, are examined at the tree level, and effects of 
the choice of the spatial Wilson parameter $r_s$ are discussed. 
We then compute the charmonium 
spectrum in the quenched approximation 
employing $\xi = a_s/a_t = 3$ anisotropic lattices.
Simulations are made with the standard 
anisotropic gauge action and the anisotropic clover quark action with 
$r_s=1$ at four lattice spacings in the range
$a_s$=0.07-0.2 fm. The clover coefficients $c_{s,t}$
are estimated from tree-level tadpole improvement.
On the other hand, for the ratio of the hopping parameters $\zeta$,
we adopt both the tree-level tadpole-improved value and a non-perturbative one.  
The latter employs the condition that the speed of light calculated from 
the meson energy-momentum relation be unity.  
We calculate the spectrum of S- and P-states and their excitations 
using both the pole and kinetic masses.  

We find that the combination of the pole mass and the tadpole-improved 
value of $\zeta$ to yield the smoothest approach to the continuum limit, 
which we then adopt for the continuum extrapolation of the spectrum. 
The results largely depend on the scale input even in the continuum limit,  
showing a quenching effect.   
When the lattice spacing is determined from the $1P-1S$ splitting, 
the deviation from the experimental value
is estimated to be $\sim$30\% for the S-state hyperfine splitting
and $\sim$20\% for the P-state fine structure.
Our results are consistent with previous results at $\xi = 2$
obtained by Chen when the lattice spacing 
is determined from the Sommer scale $r_0$.

We also address the problem with the hyperfine splitting
that different choices of the clover coefficients lead to disagreeing
results in the continuum limit.
Making a leading order analysis based on potential models we show that 
a large hyperfine splitting $\sim$95 MeV obtained by Klassen
with a different choice of the clover coefficients 
is likely an overestimate.

\end{abstract}

\section{Introduction}

Lattice study of heavy quark physics is indispensable
for determining the standard model parameters such as the quark masses
and CKM matrix elements, and for finding signals of new physics beyond it.
Obtaining accurate results for heavy quark observables, however, is 
a non-trivial task.  Since lattice spacings of order 
$a\approx (2 \mbox{GeV})^{-1}$ currently accessible is comparable or even
larger than the Compton wavelength of heavy quark given by $1/m_q$ for 
charm and bottom, a naive lattice calculation with conventional fermion 
actions suffers from large uncontrolled systematic errors. 
For this reason, effective theory approaches for heavy quark have been 
pursued.

One of the approaches is the lattice version of the
non-relativistic QCD (NRQCD), which is applicable for 
$a > 1/m_q$ \cite{Thacker:1991bm,Lepage:1992tx}. 
Since the expansion parameter of NRQCD is the quark velocity squared
$v^2$, lattice NRQCD works well for sufficiently heavy quark such as
the the bottom ($v^2\sim 0.1$), and the bottomonium spectrum
\cite{Davies:1994mp,Davies:1998im,Trottier:1997ce,Shakespeare:1998dt}
and the $b\bar{b}g$ hybrid spectrum 
\cite{Manke:1998yg,Manke:1999qc,Manke:2000dg,Manke:2001ft} have been
studied successfully using lattice NRQCD.
An serious constraint with the approach, however, is that the 
continuum limit cannot be taken due to the condition $a > 1/m_q$. 
Thus scaling violation from the gauge and light quark sectors should 
be sufficiently small.  In practice it is often difficult to quantify 
the magnitude of systematic errors arising from this origin. 
Another difficulty is that there are a number of 
parameters in the NRQCD action which have to be determined.
Since in the present calculations the tuning of 
parameters is made at
the tree level (or tadpole improved tree level) of perturbation
theory, the accuracy achieved is rather limited.

Another approach for heavy quark uses a space-time asymmetric quark action,
aiming to implement the $O(a)$ improvement for arbitrary quark
mass \cite{El-Khadra:1997mp}.   
With appropriate parameter tunings, this action is unitarily
equivalent to the NRQCD action up to higher order corrections 
for $a > 1/m_q$, and goes over into the light quark
Sheikholeslami-Wohlert (SW) action \cite{Sheikholeslami:1985ij}
for $am_q \ll 1$. 
This approach has been originally proposed by the Fermilab group and
the action is hence called the ``Fermilab action'', whose first
application is found in \cite{Sroczynski:2000he}.
Since the necessary tuning of mass-dependent parameters is in general
difficult, in practice one uses the usual SW quark action even for 
$a > 1/m_q$, where the SW action is unitarily equivalent to 
NRQCD.
This simplified approach, called the ``non-relativistic
interpretation'' for the SW quark, has been widely used in current
lattice simulations of heavy quark, such as the calculation of the
$B$ meson decay constant
\cite{Aoki:1998ji,El-Khadra:1998hq,Bernard:1998xi,AliKhan:2001eg}.
Toward the continuum limit $a\to 0$ the lattice action approaches the
usual $O(a)$-improved action and the systematic error becomes smaller
as $(am_q)^2$.  However, the $am_q$ dependence at $am_q\gsim 1$ is 
quite non-linear, and it is not trivial how the systematic error 
could be controlled. 

Recently use of the anisotropic lattice for heavy quark simulations
has been proposed \cite{Klassen:1999fh,Klassen} as a possible alternative 
to solve the difficulties of the effective approach.
On an anisotropic lattice, where the temporal lattice spacing 
$a_t$ is smaller than the spatial one $a_s$, one can achieve
$a_tm_q \ll 1$ while keeping $a_sm_q \sim 1$. 
Therefore, using anisotropic lattices, one can reduce 
$O((a_tm_q)^n)$ ($n=1,2,\cdots$) discretization errors while the
computer cost is much less than that needed for the isotropic lattice
at the same $a_t$. 
Naively it is expected that the reduction of $O((a_tm_q)^n)$ errors 
entails the reduction of most of discretization errors due to large
quark mass, since the on-shell condition ensures that the large energy
scale flows only into the temporal direction as far as one considers
the static particle, {\it with zero or small spatial momentum}.
If such a naive expectation is correct, the discretization error is
controlled by a small parameter $a_tm_q$ as it is for light quarks,
and one can achieve even better accuracy by taking a continuum limit. 
However, it is not obvious that one can eliminate all $O((a_sm_q)^n)$ errors
at the quantum level, even if it is possible at the tree level.

Another advantage of the anisotropic lattice, which is more practical,
is that a finer temporal resolution allows us to determine large
masses more accurately. 
This has been already demonstrated in simulations of the glueball
\cite{Morningstar:1997ff,Morningstar:1999rf} and the hybrid meson
\cite{Manke:1999qc}.

Klassen calculated the charmonium spectrum in the quenched
approximation, employing lattices with the ratio of the temporal and
spatial lattice spacings $\xi \equiv a_s/a_t=2$ and 3, as a
feasibility study of the anisotropic approach
\cite{Klassen:1999fh,Klassen}. 
He tuned the ratio of the temporal and spatial hopping parameters 
$\zeta \equiv K_t/K_s$ non-perturbatively by demanding the
relativistic dispersion relation for mesons. 
For the spatial clover coefficient $c_s$, he adopted two choices,
the tree level tadpole improved value correct for any mass
($a_{t}m_q\ge0$) and that correct only in the massless ($a_{t}m_q=0$)
limit, in order to make a comparison. 
He mainly studied the spin splitting of the spectrum, and obtained an 
unexpected result that two different choices of the clover coefficients
lead to two different values of the S-state hyperfine splitting 
even in the continuum limit \cite{Klassen:1999fh,Klassen}. 
The continuum limit is of course unique, and clearly, at least
one of the two continuum extrapolations is misleading.
Since the hyperfine splitting is sensitive to the clover coefficients, 
it is plausible that the disagreement is due to a large discretization
error arising from the choice of the clover coefficients.
In an unpublished paper \cite{Klassen}, he pointed out the possibility 
that the $O((\xi a_tm_q)^n) = O((a_sm_q)^n)$ errors still remain with
his choice of the parameters, which we review in the next section.
A similar statement can be found in some recent studies
\cite{Harada:2001ei,Aoki:2001ra}.
In fact, he adopted rather coarse lattice spacings 
$a_s \simeq 0.17-$0.30 fm where $a_sm_q \sim 1$. 
It is then questionable whether the reliable continuum extrapolation
is performed at such coarse lattice spacings.

Using the same anisotropic approach as Klassen, Chen has recently
calculated the quenched charmonium spectrum \cite{Chen:2001ej}. 
She employed $\xi =2$ and finer ($a_s \simeq 0.10-$0.25 fm) lattices, 
and adopted the tree level tadpole improved clover coefficient $c_s$
correct for any mass, which is expected to be better than the other
choice that is correct only in the massless limit. 
She computed not only the ground state masses but also the first
excited state masses, and extrapolated them to the continuum limit. 
Her results at $\xi =2$ are consistent with Klassen's results at 
$\xi =2$ and 3 with the same choice of the clover coefficients.

Since Chen's calculation was performed only at $\xi =2$, similar
calculations at different values of $\xi$ using fine lattices are
needed to check the reliability of the continuum limit from the
anisotropic approach. 
In addition, the complete P-state fine structure has not yet obtained
in this approach so far, since the mass of $^3P_2 (\chi_{c2})$  
state has not been measured in previous studies.

In this work, we present a detailed study of the charmonium spectrum
from the anisotropic lattice QCD.
We perform simulations in the quenched approximation at $\xi =3$,
employing fine lattice spacings in the range $a_s=0.07-$0.2 fm.
We attempt to determine the ground state masses of all the S- and
P-states (including $^3P_2$) as well as their first excited state
masses. 
To estimate the systematic errors accurately, we adopt both the tree
level tadpole improved value and non-perturbative one for $\zeta$, 
and both the pole mass and kinetic mass for $M_{\rm lat}(1\bar{S})$
which is tuned to the experimental value. 
We focus on the lattice spacing dependence and continuum limit of the 
mass splittings. 
We compare our results with the previous anisotropic results by 
Klassen and Chen to check the consistency, and with experimental
values \cite{Caso:tx} to estimate the quenching effect. 

In addition, to understand the discrepancy of the hyperfine splitting
mentioned above, we make a leading order analysis using the potential
model. 
To examine the effect of clover coefficients, we estimate the
hyperfine splitting at leading order.
Comparing the leading order estimates with numerical results for the
hyperfine splitting, we attempt to find a probable solution for this
problem. 
Our preliminary results are already reported in
Refs.~\cite{AliKhan:2001bv,Aoki:2001dh}.

This paper is organized as follows. 
In Section \ref{sec:ansioQCDact}, we summarize and discuss the
theoretical aspect of the anisotropic lattice QCD.
In Section \ref{sec:simulations}, we give details of our simulation.
Our results for the charmonium spectra 
are shown in Section \ref{sec:results}, where
we attempt to take
the continuum limit and estimate the quenching effect. 
We address the problem 
of the discrepancy of the hyperfine splitting 
and study the effect of clover coefficients 
in Section \ref{sec:hfsrevisit}.
Section \ref{sec:conclusion} is devoted to our conclusions.

\section{Anisotropic lattice QCD action}\label{sec:ansioQCDact}

In this section we first define the anisotropic lattice action used 
in this work and fix notations. We then derive the tree level values
of bare parameters in our massive quark action, and discuss effects of  
the anisotropy. 
Although it was already discussed in earlier papers
\cite{Harada:2001ei,Aoki:2001ra}, we briefly describe the outline of
derivations in order to be self-contained. 
We also consider the tadpole improvement of bare parameters and see
how tree level values are modified. 

\subsection{Anisotropic gauge action }

In this work, we use the standard Wilson gauge action defined on an
anisotropic lattice:
\begin{equation}
S_g = \beta\left[\frac{1}{\xi_0}\sum_{x,s>s'}(1-P_{ss'}(x)) +
\xi_0 \sum_{x,s}(1-P_{st}(x)) \right]
\label{gaugeact}
\end{equation}
where $\beta =6/g^2$ is the gauge coupling, and $P_{ss'}(x)$
and $P_{st}(x)$ are the spatial and temporal plaquettes with $P_{\mu 
\nu}(x) = \frac{1}{3} \mbox{Re}\ \mbox{Tr}\  U_{\mu \nu}(x)$.
The anisotropy is introduced by the parameter $\xi_0$ and we
call this the ``bare anisotropy''. We denote spatial and temporal
lattice spacings as $a_s$ and $a_t$ and define the  ``renormalized
anisotropy'' $\xi \equiv a_s/a_t$. We have $\xi = \xi_0$ at
the tree level, and the $\xi = \xi (\xi_0,\beta)$ at finite
$\beta$ can be determined non-perturbatively by Wilson loop
matching \cite{Fujisaki:1997vv,Klassen:1998ua,Engels:2000tk}. 
In numerical simulations, there are two methods for anisotropy tuning,
either varying $\xi_0$ to keep $\xi$ constant or {\it vice versa}. 
Since the former is more convenient for keeping the physical size
constant, and easier for performing the continuum extrapolation, we
adopt it in this work.

\subsection{Anisotropic quark action }

For the quark action, we employ the space-time asymmetric clover 
quark action on an anisotropic lattice 
proposed in Ref. \cite{Klassen:1999fh,Klassen}:
\begin{eqnarray}
S_f &=& \sum_x  \bar{\psi}_x Q \psi_x, \\
Q    &=& m_0 + \nu_0 \hat{W}_0\gamma_0 + \frac{\nu}{\xi_0}\sum_i
\hat{W}_i\gamma_i + \frac{i}{2} \left[\omega_0\sum_{x, i} \sigma_{0i} 
\hat{F}_{0i}(x) + \frac{\omega}{\xi_0}\sum_{x, i < j} \sigma_{ij}
\hat{F}_{ij}(x)\right],
\label{massform}
\end{eqnarray}
where $\nu_0 = 1$ and $m_0 \equiv a_t m_{q0}$ is the bare quark mass,
and $\hat{W}_\mu\gamma_\mu \equiv a_\mu W_\mu\gamma_\mu$ 
and $\hat{F}_{\mu\nu} \equiv a_\mu a_\nu F_{\mu\nu}$ with 
$(a_0,a_i)=(a_t,a_s)$.
The Wilson operator $W_\mu$ is defined by 
\begin{equation}
W_\mu\gamma_\mu \equiv D_\mu\gamma_\mu -
\frac{a_\mu}{2}r_\mu D^2_\mu  \;\;\;\; (\mu = 0,1,2,3)
\end{equation}
with the Wilson coefficients
$(r_0,r_i)=(r_t,r_s)$ and 
\begin{eqnarray}
D_\mu \psi_x &\equiv& \frac{1}{2a_\mu} \left[  U_{\mu,x}
\psi_{x+\hat{\mu}} - U^{\dagger}_{\mu,x-\hat{\mu}}
\psi_{x-\hat{\mu}} \right], \\
D^2_\mu \psi_x &\equiv& \frac{1}{a_\mu^2} \left[  U_{\mu,x}
\psi_{x+\hat{\mu}} + U^{\dagger}_{\mu,x-\hat{\mu}} \psi_{x-\hat{\mu}} 
-2\psi_x \right]. 
\end{eqnarray}
For the field tensor $F_{\mu\nu}$, we adopt the standard clover leaf 
definition. 
Note that, in Eq.~(\ref{massform}),
 the factors in front of spatial Wilson and clover operators
include $\xi_0$ rather than $\xi$. This is merely a convention
and there is no deep theoretical reason. 
This action is essentially the same as the one employed by 
Klassen\cite{Klassen} and Chen\cite{Chen:2001ej}. 
In Chen's work, however, $\nu_0$ was a tuning parameter with $\nu =1$ fixed.
The two parameterizations are related to each other by a field rescaling 
$\psi_x \rightleftharpoons \psi_x/\sqrt{\nu}$.
Therefore $\{m_{0}, \nu_0, \omega, \omega_0\}$\footnote{More 
precisely, Chen used the language $\{\hat{m_{0}}, \nu_t,
C_{\rm SW}^s, C_{\rm SW}^t\}$ 
instead of $\{m_{0}, \nu_0, \omega, \omega_0\}$}
corresponds to $\{m_{0}/\nu, 1/\nu, \omega/\nu,
\omega_0/\nu\}$ in our convention. 
Among these six parameters
$\{m_{0}, \nu, r_s, r_t, \omega, \omega_0\}$, 
at least one is redundant, so that we take $r_t$ as a redundant parameter and
use it to remove the fermion doublers.
Although $r_s$ may not be taken arbitrary in the $O(a)$ improved 
anisotropic quark action \cite{Aoki:2001ra} for the matrix elements, 
it can be taken arbitrary for the hadron mass calculation.
Therefore we always set $r_t=1$ and leave $r_s$ free in this work.
The remaining parameters $\{m_{0}, \nu, \omega, \omega_0\}$
are used to tune the quark mass and reduce the lattice discretization error.

For convenience in numerical simulations, we also present the quark
action in a different form. 
Rescaling the fields $\psi_x$,
the quark action can be transformed into a form 
given by
\begin{eqnarray}
S_f^{\prime}
 = \sum_x \{ \bar{\psi}_x \psi_x &-& K_t [ \bar{\psi}_x (1- \gamma_0) 
U_{0,x} \psi_{x+\hat{0}} + \bar{\psi}_x (1+\gamma_0)
U^{\dagger}_{0,x-\hat{0}} \psi_{x-\hat{0}}]  \nonumber\\ 
&-& K_s \sum_i [ \bar{\psi}_x (r_s- \gamma_i) 
U_{i,x} \psi_{x+\hat{i}} + \bar{\psi}_x (r_s+\gamma_i)
U^{\dagger}_{i,x-\hat{i}} \psi_{x-\hat{i}}] \} \nonumber\\
&+& i K_s c_s \sum_{x, i < j} \bar{\psi}_x 
\sigma_{ij} \hat{F}_{ij}(x) \psi_x 
+ i K_s c_t \sum_{x, i} \bar{\psi}_x 
\sigma_{0i} \hat{F}_{0i}(x) \psi_x,
\label{FNAL}
\end{eqnarray}
where $K_{s,t}$ and $c_{s,t}$ are the spatial and
temporal hopping parameters and the clover
coefficients, respectively.  
The hopping parameters $K_{s,t}$ are related to the bare quark mass
$m_0 = a_t m_{q0}$ through 
\begin{equation}
a_t m_{q0} \equiv 1/(2K_t) - 3r_s/\zeta - 1, 
\;\;\;\;\;\; \zeta \equiv K_t/K_s.
\label{mqzeta}
\end{equation}
The form Eq.~(\ref{FNAL}) on the anisotropic lattice is the
same as that on the isotropic lattice in
Ref.~\cite{El-Khadra:1997mp}. 
Note however that Ref.~\cite{El-Khadra:1997mp} uses the inverse of our 
definition for $\zeta$. 
We refer to their definition as $\zeta_{\rm F} \equiv
K_s/K_t = 1/\zeta$. 
Using Eq.~(\ref{mqzeta}) one can convert $\{m_{q0},\zeta\}$
to $\{K_s,K_t\}$. 
In our convention, the relation between 
$\{\nu, \omega, \omega_0\}$ and $\{\zeta, c_s, c_t\}$
is given by
\begin{equation}
\zeta = \xi_0/\nu, \;\;\; c_s = \omega/\nu, \;\;\; c_t =
\xi_0\omega_0/\nu 
\label{mass2hop}
\end{equation}
or equivalently, 
\begin{equation}
\nu = \xi_0/\zeta, \;\;\; \omega = c_s\nu, \;\;\; \omega_0
= c_t\nu /\xi_0. 
\label{hop2mass}
\end{equation}

Following Ref.~\cite{El-Khadra:1997mp}, we call the quark action 
Eq.~(\ref{massform}) as the ``mass form'' and
Eq.~(\ref{FNAL}) as the ``hopping parameter form''.

\subsection{Tree level tuning of bare parameters for arbitrary mass}

\label{sec:massivetree}

To derive the tree level value of bare parameters,
we follow the Fermilab method and 
calculate the lattice Hamiltonian\cite{El-Khadra:1997mp}. 
After some algebra (see appendix \ref{sec:hamiltonian} for details), 
we obtain the lattice Hamiltonian Eq.~(\ref{contlikeH}).
Using the FWT transformation Eq.~(\ref{FWT}),
we then transform it to the non-relativistic form,
in which the upper components of the Dirac spinor
completely decouple from the lower ones ({\it i.e.}
eliminate ${\bf \gamma}\cdot{\bf D}$ and ${\bf
\alpha}\cdot{\bf E}$). The transformed Hamiltonian is
given by 
\begin{equation}
\frac{1}{a_t}\hat{H}^{U} = \hat{\bar{\Psi}}\left( m_1 + \gamma_0
A_0 - \frac{{\bf D}^2}{2m_2} - \frac{i{\bf\Sigma}\cdot{\bf B}}{2m_B}
- \gamma_0\frac{[{\bf \gamma}\cdot{\bf D},{\bf
\gamma}\cdot{\bf E}]}{8m_E^2} + \cdots \right)\hat{\Psi} 
\label{NRHamiltonian}
\end{equation}
with 
\begin{eqnarray} 
a_tm_1 &=& \log (1+m_0),\label{polemass}\\
\frac{1}{a_tm_2} &=& \frac{2{\zeta_{\rm F}^{\prime}}^2}{m_0(2+m_0)}
+ \frac{r_s^{\prime}\zeta_{\rm F}^{\prime}}{1+m_0}, \label{kineticmass}\\
\frac{1}{a_tm_B} &=& \frac{2{\zeta_{\rm F}^{\prime}}^2}{m_0(2+m_0)}
+ \frac{c_s^{\prime}\zeta_{\rm F}^{\prime}}{1+m_0},\label{magneticmass}\\
\frac{1}{(a_tm_E)^2} &=& 4{\zeta_{\rm F}^{\prime}}^2 \left
[ \frac{(1+m_0)^2}{m_0^2 (2+m_0)^2} + (c_t-1)\frac{1}{m_0(2+m_0)}\right],
\label{electronicmass}
\end{eqnarray}
where $\zeta_{\rm F}^{\prime}$, $r_s^{\prime}$ and $c_s^{\prime}$ 
are defined in Eq.~(\ref{dashedparam}).
The ${\bf\Sigma}\cdot{\bf B}$ term gives the leading order
contribution to the hyperfine splitting,
while $[{\bf \gamma}\cdot{\bf D},{\bf
\gamma}\cdot{\bf E}]$ term yields the fine structure splitting.

The matching condition 
$\hat{H}^{U} = \hat{H}_{\rm NR} + O(a_s^2)$ 
is equivalent to
\begin{equation}
m_1 = m_2 = m_B = m_E = m_q.
\label{matching}
\end{equation}
This yields the tree level value of bare parameters for
massive quark:
\begin{equation}
\xi_0\zeta_{\rm F} = \nu = \sqrt{\left(\frac{\xi_0 r_s
m_0(2+m_0)}{4(1+m_0)}\right)^2+\frac{m_0(2+m_0)}{2\log (1+m_0)}} - 
\frac{\xi_0 r_s m_0(2+m_0)}{4(1+m_0)},
\label{treezeta}
\end{equation}
\begin{equation}
c_s = r_s \,\,\,\,\,\, (\omega = r_s\nu),
\label{treecB}
\end{equation}
\begin{equation}
c_t = \frac{(\xi_0\zeta_{\rm F})^2-1}{m_0(2+m_0)} + \frac{\xi_0^2
r_s\zeta_{\rm F}}{1+m_0} + \frac{(\xi_0 r_s)^2m_0(2+m_0)}{4(1+m_0)^2}.
\label{treecE}
\end{equation}
We note that $c_s$ is independent of the quark mass, 
while $\nu$ and $c_t$ have complicated
mass dependences.
The term $\xi_0 m_0 \simeq a_s m_{q0}$ seems to exist
in Eq.~(\ref{treezeta}) and (\ref{treecE}).
To see this explicitly, we expand $\nu$ and $c_t$ in $m_0$. This gives
\begin{equation}
\nu = 1 + \frac{1}{2}(1-\xi_0 r_s)m_0 +
\frac{1}{24}(-1 +6\xi_0 r_s +3(\xi_0 r_s)^2)m_0^2
+ O(m_0^3),
\label{nu_smallm}
\end{equation}
\begin{equation}
c_t = \frac{1+\xi_0 r_s}{2} + 
\frac{1}{12}(-2 -3\xi_0 r_s +3(\xi_0 r_s)^2)m_0
+ O(m_0^2).
\label{ct_smallm}
\end{equation}
The $a_s m_{q0}$ term, which is $O(1)$ for heavy quarks
at currently accessible lattice
spacings of $a_s^{-1} \approx 2$ GeV,
appears in $\nu$ and $c_t$ even at the tree level.
Since $\xi_0 m_0 = a_s m_{q0}$ is always multiplied with 
the spatial Wilson coefficient $r_s$ 
in Eqs.~(\ref{nu_smallm}) and (\ref{ct_smallm}), 
one can eliminate 
the $a_s m_{q0}$ term at the tree level by choosing
\begin{equation}
r_s = 1/\xi_0.
\label{rs_xiinv}
\end{equation}
However, this choice has the disadvantage that the mass splitting between
unphysical doubler states and the physical state
decreases as $\xi_0$ increases.
Moreover, the hopping terms in the quark action are no longer proportional to 
the $1\pm \gamma_\mu$ projection operators.
It is also doubtful that, beyond the tree level,
the $a_s m_{q0}$ term can be still 
eliminated by this choice.

If one adopts the conventional choice
\begin{equation}
r_s = 1,
\label{rs_1}
\end{equation}
the $a_s m_{q0}$ term remains, but 
the unphysical doubler states decouple.
This choice also has the practical merit
that the quark action has the full projection property, so that
the coding is easier and the computational cost is lower.

\begin{figure}[t]
\centerline{\epsfxsize=9.5cm \epsfbox{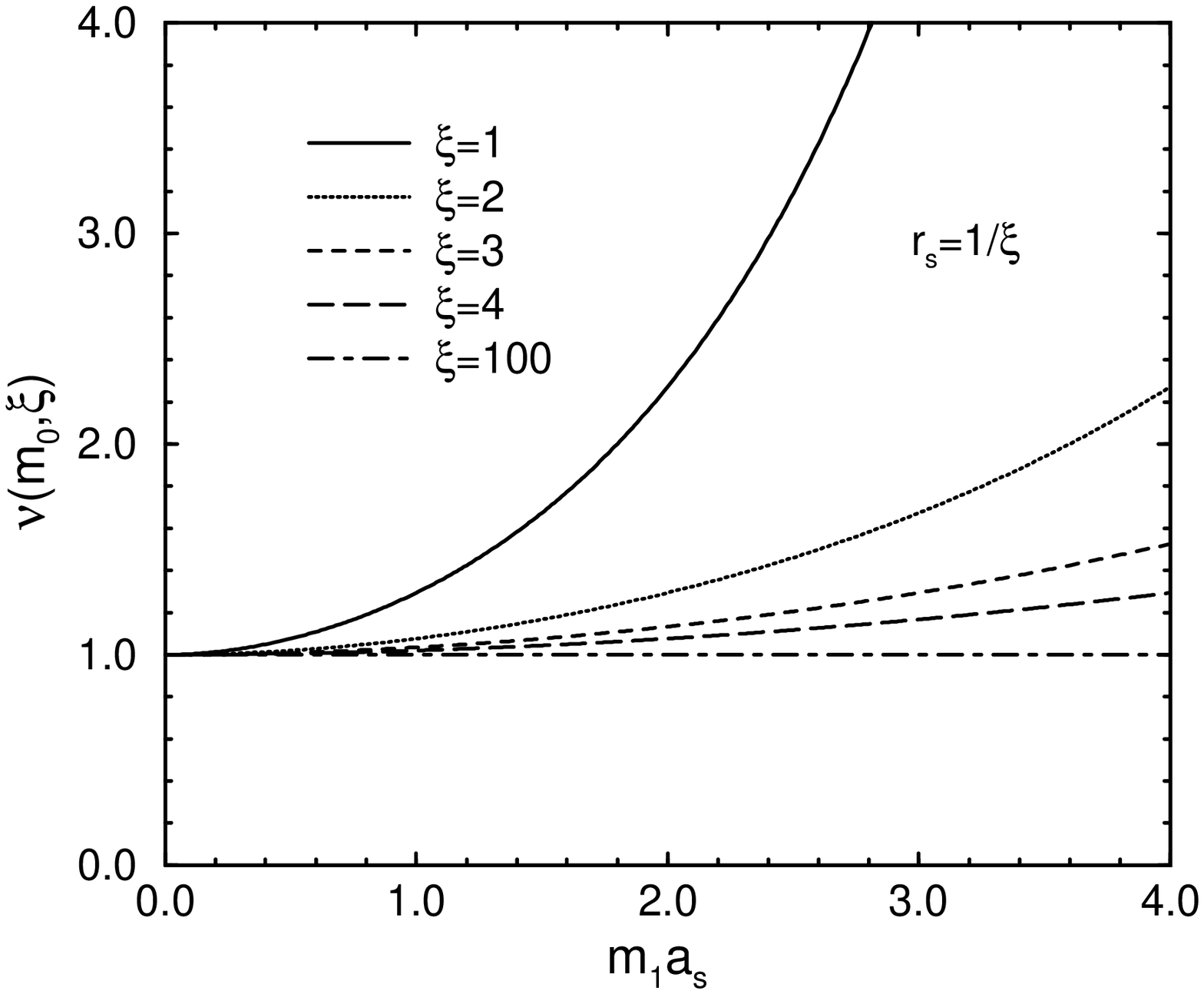}
            \epsfxsize=9.5cm \epsfbox{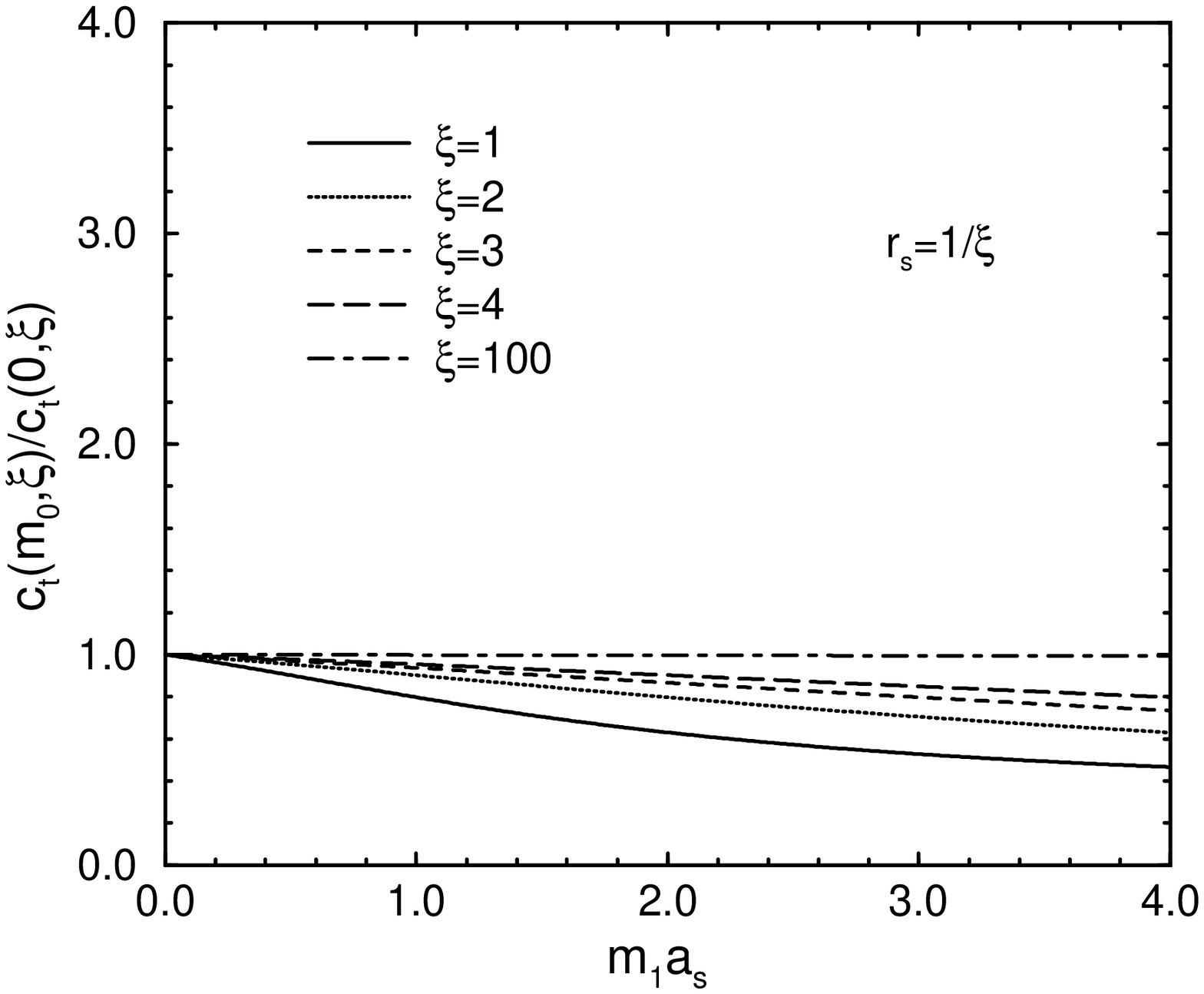}}
\caption{Tree level full mass dependences of $\nu$ and $c_t$
for $r_s=1/\xi=1/\xi_0$. Horizontal axis is the pole mass in
spatial lattice units $m_1a_s = \xi\log (1+m_0)$.
Vertical axis is normalized to be 1 in the massless limit.
}
\label{fig:zeta_ce_tree_rxiinv}
\end{figure}

\begin{figure}[t]
\centerline{\epsfxsize=9.5cm \epsfbox{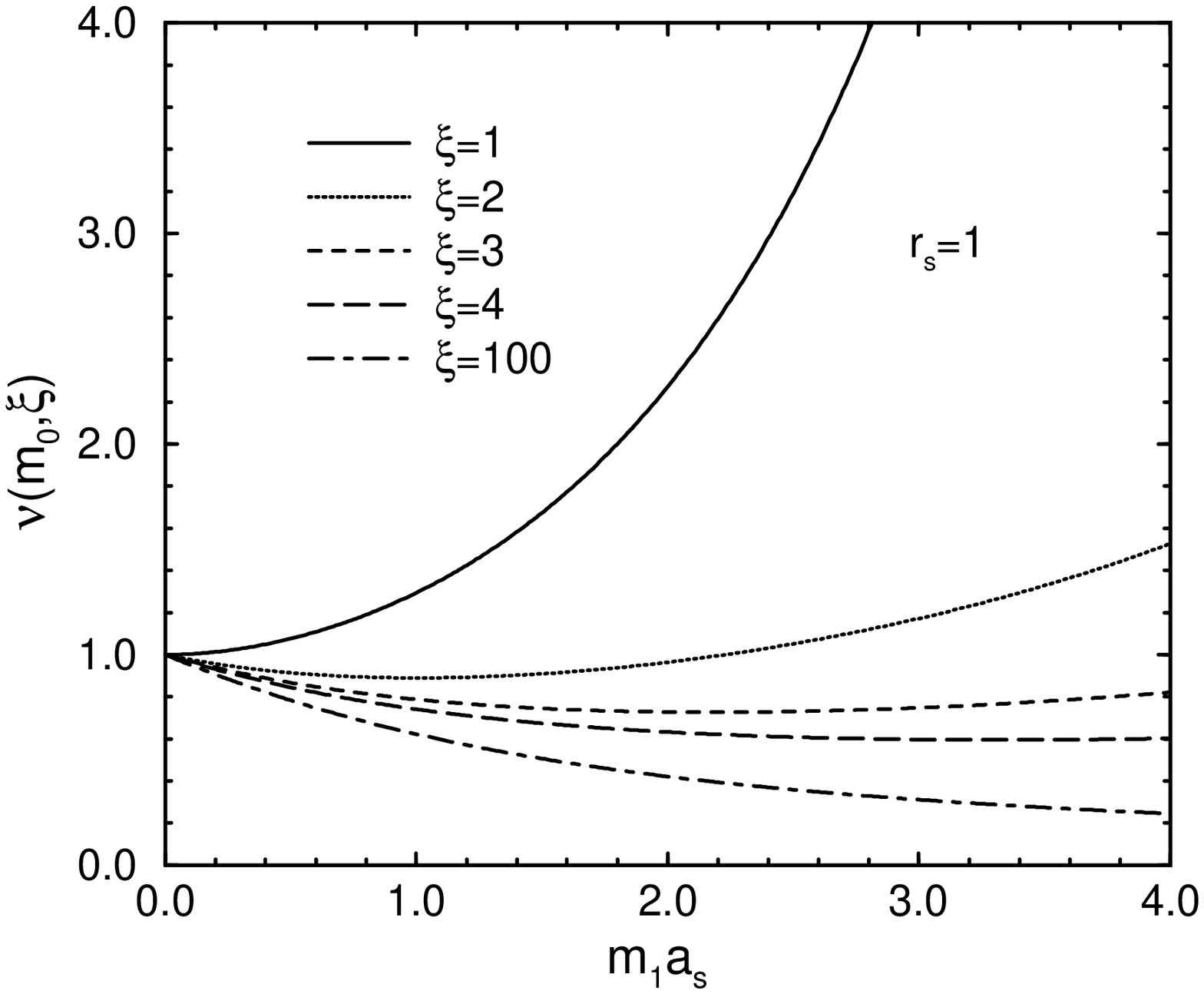}
            \epsfxsize=9.5cm \epsfbox{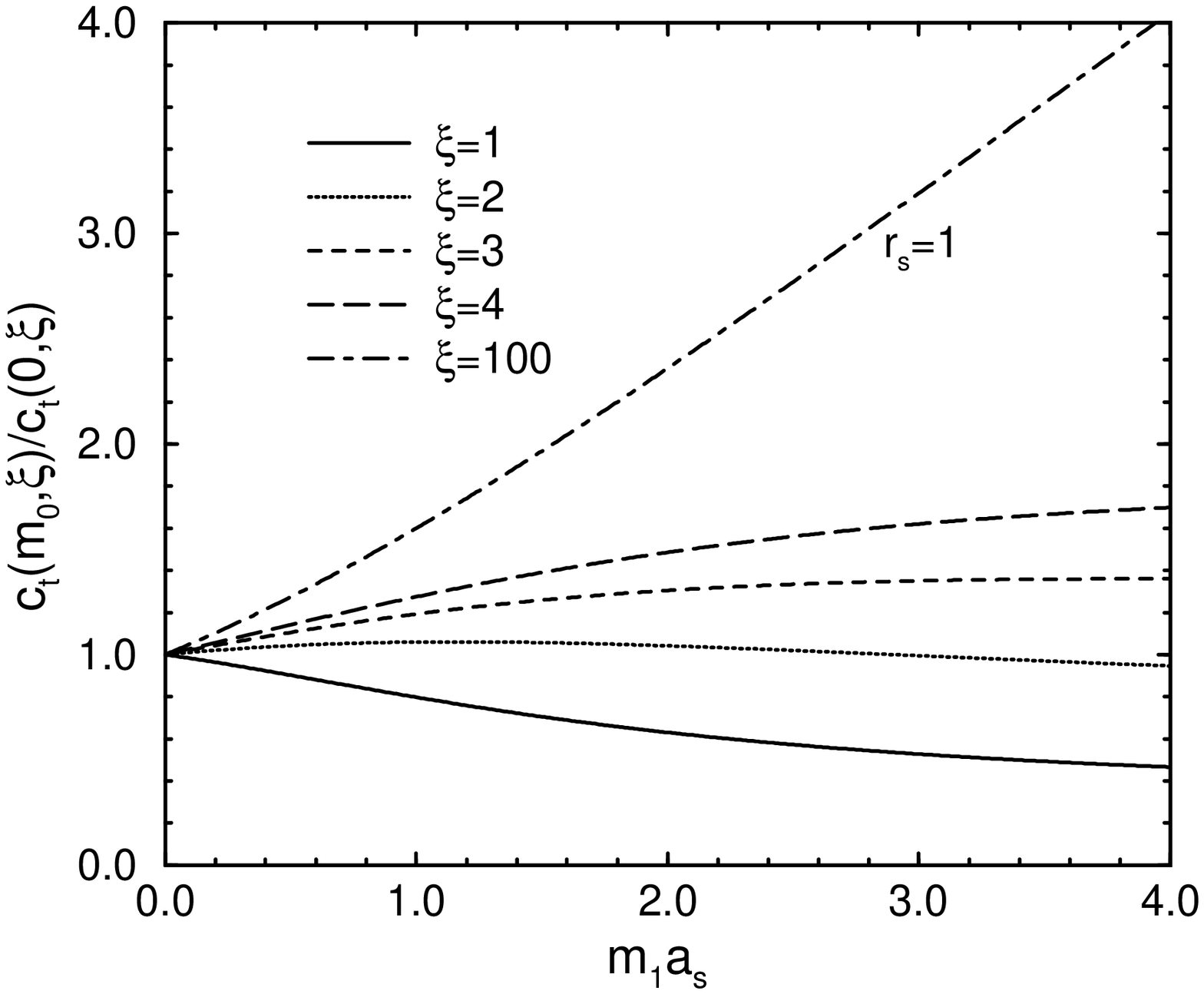}}
\caption{The same as Fig.~\ref{fig:zeta_ce_tree_rxiinv}, but for $r_s=1$.}
\label{fig:zeta_ce_tree_r1}
\end{figure}

The tree-level full mass dependences of $\nu$ and $c_t$ for 
$r_s = 1/\xi_0$ and $r_s = 1$ are shown in
Figs.~\ref{fig:zeta_ce_tree_rxiinv} and \ref{fig:zeta_ce_tree_r1}. 
In order to compare at the same $a_s$, we choose $m_1a_s$
as the horizontal axis 
instead of $m_1a_t$ where $m_1$ is the pole mass.
Since $a_s^{-1}\gsim 1$ GeV and $m_1 \le m_{\rm bottom}\sim 4.5$ GeV
in current typical simulations, we plot results for $m_1a_s \le 4$. 

For $r_s = 1/\xi_0$ shown in Fig.~\ref{fig:zeta_ce_tree_rxiinv},  
both $\nu$ and $c_t$ are monotonic
functions in mass, and they converge to their 
massless values as $\xi_0$ increases at any fixed values of $m_1a_s$. 
Hence, the $a_s m_{q0}$ dependence can be controlled by increasing $\xi_0$.
At $\xi =100$ the mass dependences of $\nu$ and $c_t$ 
completely disappear with the cost that the physical and unphysical states
are almost degenerate. In actual simulations with $r_s = 1/\xi_0$,
taking $2 \le \xi_0 \ll \infty$ to decouple unphysical doublers,
one is allowed to use the massless values for $\nu$ and $c_t$,
since their mass dependences are monotonic and very weak. 
In this case mass dependent parameter tuning 
can be avoided even at $a_s m_0 \sim 1$.

For $r_s = 1$, on the other hand, the mass dependences of
$\nu$ and $c_t$ are complicated and non-negligible even for large
$\xi_0$. Indeed
$\nu$ and $c_t$ do not converge to their massless values as $\xi_0$ increases
at fixed $m_1a_s$, as shown in Fig.~\ref{fig:zeta_ce_tree_r1}.
The deviation from the massless values at
$\xi_0=2$ is smaller than the one at $\xi_0=1$, but it becomes
larger again as $\xi_0$ increases. Therefore, 
taking $\xi_0=2$-3 in simulations with $r_s = 1$,
one needs to perform a mass dependent parameter tuning.

For both choices of $r_s$, 
it is better to use a moderate value of $\xi_0$, rather than
excessively large values. 
In our numerical study of the charmonium spectra,
we adopt the choice $r_s = 1$, and make a mass dependent parameter 
tuning, due to the practical reasons mentioned above.

Finally we show the tree level value of the parameters in the massless limit.
By taking $a_tm_{q0} \rightarrow 0$ in
Eqs.~(\ref{treezeta})-(\ref{treecE}), one obtains
\begin{equation}
\nu =1, \;\;\; \omega = r_s, \;\;\; \omega_0 = \frac{1+\xi_0 r_s}{2\xi_0},
\label{masslesstree_mass}
\end{equation}
in the mass form, or 
\begin{equation}
\zeta=\xi_0, \;\;\; c_s = r_s, \;\;\; c_t = \frac{1+\xi_0 r_s}{2}.
\label{masslesstree_hopping}
\end{equation}
in the hopping parameter form.
Note that there is an ambiguity in the tree level value of $a_s/a_t$,
since $\xi_0=\xi$ at the tree level but $\xi_0\not=\xi$ in the
simulation. Fortunately, this
ambiguity almost disappears after the tadpole improvement,
as shown in the next subsection.

\subsection{Tadpole improvement}\label{sec:TI}

In this section we apply the tadpole improvement \cite{Lepage:1993xa}
to the parameters of the anisotropic lattice action 
at tree level in order to partially include higher order
corrections.
One first rewrites the lattice action using a more
continuum-like link variable $\tilde{U}_{i,0} =
U_{i,0}/u_{s,t}$, where $u_{s,t} = \langle U_{i,0} \rangle$ is
the expectation value of the spatial or temporal link variable,
{\it i.e.} one replaces
\begin{equation}
U_{i,0} \rightarrow u_{s,t}\tilde{U}_{i,0},
\label{tadpoleU}
\end{equation}
and then repeats the tree-level calculations. 
We will show below how the tree-level values of bare parameters 
are modified.

\subsubsection{Gauge action}

By the replacement of Eq.~(\ref{tadpoleU}), the anisotropic gauge
action Eq.~(\ref{gaugeact}) becomes
\begin{equation}
S_g  \rightarrow 
- \sum \frac{6}{\tilde{g}^2}\left[\frac{1}{\tilde{\xi_0}} \tilde{P}_{ss'} +
\tilde{\xi_0} \tilde{P}_{st}  
+ {\rm const.\ independent\ of\ \tilde{U}_{\mu}} \right] ,
\label{tadpolegauge}
\end{equation}
where $\tilde{P}_{\mu 
\nu} = \frac{1}{3} \mbox{Re}\ \mbox{Tr}\  \tilde{U}_{\mu
\nu}$, and $\tilde{g}^2$ and $\tilde{\xi_0}$ are given by
\begin{equation}
\tilde{g}^2 = \frac{g^2}{u_s^3 u_t} \simeq
\frac{g^2}{\sqrt{\langle P_{ss'}\rangle \langle
P_{st}\rangle}}\ ,
\;\;\;\;\;\;\;\;\;\;
\tilde{\xi_0} = \frac{u_t}{u_s}\xi_0 \simeq
\sqrt{\frac{\langle P_{st}\rangle}{\langle P_{ss'}\rangle}}\xi_0. 
\label{tigauge}
\end{equation}
Requiring space-time symmetry for the action
Eq.~(\ref{tadpolegauge}) in the classical limit, one obtains
the tree-level tadpole-improved value of the anisotropy (denoted
by an index `TI'),
\begin{equation}
\xi^{\rm TI} = \tilde{\xi_0} = (u_t/u_s)\xi_0.
\label{TIxi0}
\end{equation}
In practice $\xi^{\rm TI}$ in Eq.~(\ref{TIxi0}) agrees with
the renormalized anisotropy $\xi$ 
within a few \% accuracy at $g^2 \sim 1$. 
Therefore one can replace the factor $(u_t/u_s)\xi_0$ by $\xi$  
in the following equations. This simplifies the tree level
expression. Moreover, the arbitrariness for the choice of
anisotropy disappears.

\subsubsection{Fermion action}

When the fermion action is rewritten in terms of $\tilde{U}_{i}$ and
$\tilde{U}_{0}$ instead of $U_{i}$ and $U_{0}$,
the action keeps the same form with
\begin{equation}
\tilde{K_s} = u_s K_s, \;\;\;\;\; \tilde{K_t} = u_t K_t,
\end{equation}
\begin{equation}
\tilde{c_s} = u_s^3 c_s, \;\;\;\;\; 
\tilde{c_t} = u_s u_t^2 c_t.
\end{equation}
Then $\zeta = K_t/K_s$ and the bare quark mass $a_t m_{q0} = 1/2K_t -
(1+3r_s/\zeta)$ 
are modified to 
\begin{equation}
\tilde{\zeta} = \tilde{K_t}/\tilde{K_s} = (u_t/u_s)\zeta,
\end{equation}
\begin{eqnarray}
a_t \tilde{m}_{q0} &=& \frac{1}{2\tilde{K_t}} -
(1+3r_s/\tilde{\zeta})
\label{TIbearemass}\nonumber \\
&=& \frac{a_t m_{q0}}{u_t} + \frac{1}{u_t} -1 + (3r_s/\zeta)
\frac{1-u_s}{u_t}.
\end{eqnarray}
Using parameters with the tilde, one can repeat the derivation in the
previous subsection. For a massless quark, one obtains
\begin{equation}
\tilde{\zeta} = \tilde{\xi_0} \simeq \xi, \;\;\;\; 
\tilde{c_{s}} = r_s, 
\;\;\;\; \tilde{c_{t}} = \frac{1+\tilde{\xi_0}r_s}{2} \simeq 
\frac{1+\xi r_s}{2}.
\end{equation}
Therefore, tadpole-improved (TI) tree-level estimates are 
\begin{equation}
\zeta^{\rm TI} = 
(u_s/u_t)\tilde{\xi_0} = \xi_0,
\label{TIzeta}
\end{equation}
which indicates that non-perturbative $\zeta$ at
$\tilde{m}_{q0} \sim 0$ is closer to $\xi_0$ than to $\xi$, and
\begin{equation}
c_s^{\rm TI} = \frac{r_s}{u_s^3}, \;\;\;\; c_t^{\rm TI} = 
\frac{1}{u_s u_t^2}\frac{1+(u_t/u_s)\xi_0 r_s}{2} \simeq
\frac{1}{u_s u_t^2}\frac{1+\xi r_s}{2}.
\label{TIcst}
\end{equation}
As can be seen in Eqs.~(\ref{TIzeta}) and (\ref{TIcst}),
the tadpole improvement eliminates the uncertainty of choice
of anisotropy ({\it i.e.} whether to chose $\xi_0$ or $\xi$) at tree level.
Converting to the $\{\nu, \omega, \omega_0\}$ convention,
one obtains 
\begin{equation}
\nu^{\rm TI} = 1, \;\;\;\; \omega^{\rm TI} =
\frac{r_s}{u_s^3}, \;\;\;\;  \omega_0^{\rm TI}
= \frac{1}{u_s^2 u_t}\frac{1+(u_t/u_s)\xi_0 r_s}{2(u_t/u_s)\xi_0}.
\end{equation}
Note that $\nu^{\rm TI}$ is normalized to 1 since $\nu$ equals 
$\xi_0/\zeta$ and 
not $\xi/\zeta$, hence the former definition is practically more
convenient than the latter one.
Note also that tadpole factors in $c_t^{\rm TI}$ and
$\omega_0^{\rm TI}$ are different because $\omega_0$ equals
$c_t\nu /\xi_0$ and not $c_t\nu /\xi$.

Similarly, for massive quarks, tadpole-improved tree-level
estimates become
\begin{equation}
1/\zeta^{\rm TI} = \frac{u_t}{u_s}\left\{\sqrt{\left(\frac{r_s
\tilde{m_0}(2+\tilde{m_0})}{4(1+\tilde{m_0})}\right)^2+\frac{\tilde{m_0}(2+ 
\tilde{m_0})}{2(u_t/u_s)^2\xi_0^2\log (1+\tilde{m_0})}} - \frac{r_s
\tilde{m_0}(2+\tilde{m_0})}{4(1+\tilde{m_0})}\right\} 
\label{TImasszeta}
\end{equation}
with $\nu^{\rm TI}=\xi_0/\zeta^{\rm TI}$, and 
\begin{equation}
c_s^{\rm TI} = \frac{r_s}{u_s^3} , 
\label{TImasscB}
\end{equation}
\begin{eqnarray}
c_t^{\rm TI} &=& \frac{1}{u_s
u_t^2}\left\{\frac{(\nu^{\rm
TI})^2-1}{\tilde{m_0}(2+\tilde{m_0})} +
\left(\frac{u_t}{u_s}\right) \frac{\xi_0
r_s \nu^{\rm TI}}{1+\tilde{m_0}} +
\left(\frac{u_t}{u_s}\right)^2 \frac{(\xi_0
r_s)^2\tilde{m_0}(2+\tilde{m_0})}{4(1+\tilde{m_0})^2}\right\} ,
\label{TImasscE}
\end{eqnarray}
where $\tilde{m_0} = a_t \tilde{m}_{q0}$.

\section{Simulations}\label{sec:simulations}

We proceed to calculate the charmonium spectrum in the
quenched approximation as our 
first numerical study using the anisotropic lattice. 
In this section we describe the 
computational details of our quenched charmonium calculation.

\subsection{Choice of simulation parameters}
\label{sec:choiceparam}

For the gauge sector, we use the anisotropic Wilson gauge action
given in Eq.~(\ref{gaugeact}). Throughout this paper, we employ $\xi 
=3$, where $\xi$ is the renormalized anisotropy.
In order to achieve $\xi =3$, we tune
the bare anisotropy $\xi_0$, using the parameterization of $\eta
\equiv \xi/\xi_0$ given by Klassen\cite{Klassen:1998ua}:
\begin{equation}
\eta(\beta,\xi) = 1 + \left( 1 - \frac{1}{\xi} \right) 
                      \frac{\hat{\eta}_{1}(\xi)}{6} \,
                      \frac{1+a_{1}g^{2}}{1+a_{0}g^{2}} g^{2},
\label{xirenormalization}
\end{equation}
where  $a_{0}= -0.77810$, $a_{1} = -0.55055$ and
\begin{equation}
\hat{\eta}_{1}(\xi) = \frac{1.002503 \xi^{3} + 0.39100 \xi^{2} 
                            + 1.47130 \xi - 0.19231}
            {\xi^{3} + 0.26287 \xi^{2} + 1.59008 \xi -0.18224}.
\end{equation}

We perform simulations in the quenched approximation,
at four values of gauge coupling $\beta = 5.70$,
5,90, 6.10 and 6.35. These couplings correspond to $a_s = 0.07-$0.2 fm 
and $a_t m_{\rm charm} = 0.16-$0.48 for $m_{\rm
charm} = 1.4$ GeV. The spatial lattice size $L$ is chosen so that 
the physical box size is about 1.6 fm, while the temporal lattice
size $T$ is always set to be $T=2 \xi L = 6L$. 

For the charm quark, we use the anisotropic clover quark action
Eq.~(\ref{FNAL}) with the conventional choice of the spatial Wilson 
coefficient, $r_s =1$, as mentioned in Sec.\ref{sec:massivetree}.
We take two values for the bare quark mass $m_0=(m_0^1,m_0^2)$ at 
each $\beta$ in order to
interpolate (or extrapolate) results in $m_0$ to  
the charm quark mass $m_0^{\rm charm}$.
The charm quark mass $m_0^{\rm charm}$ is fixed from the experimental value
of the spin averaged $1S$ meson mass.
In this procedure, we use both the pole mass $M_{\rm pole}$ and 
kinetic mass $M_{\rm kin}$ for the $1S$ meson.
For $\zeta$,
the ratio of the hopping parameters,
we adopt both the tree-level tadpole-improved value $\zeta^{\rm TI}$,
and a non-perturbative value $\zeta^{\rm NP}$
determined from the meson dispersion relation. 
We describe our method of tuning $\zeta$ in detail in 
Sec.\ref{sec:mztuning}.
For the spatial clover coefficient $c_{s}$,
we employ the tree-level tadpole-improved value for massive quarks
Eq.~(\ref{TImasscB}). Note that $c_{s}$ has no mass dependence at the
tree level. 
On the other hand, we adopt the tree-level tadpole-improved value in 
the massless limit Eq.~(\ref{TIcst}) for the temporal
clover coefficients $c_{t}$. 
We discuss possible systematic errors arising from our choice
of the parameters $\zeta$ and $c_{s,t}$ in Sec.\ref{sec:syserr}.
The tadpole factors $u_{s,t}$ in Eqs.~(\ref{TIcst}) and (\ref{TImasscB}) are
estimated by the mean plaquette
prescription:
\begin{equation}
u_s  = \langle P_{ss'} \rangle^{1/4},\;\;\;\;\;\; u_t =1 .
\label{TIfactor}
\end{equation}
If we adopted the alternative definition 
$u_t  = (\langle P_{st}\rangle / \langle P_{ss'} \rangle )^{1/2}$
instead, $u_t$ would be greater than 1.
We use $\xi$ instead of $(u_t/u_s)\xi_0$
in Eq.~(\ref{TIcst}).

Gauge configurations are generated by a 5-hit pseudo heat bath update 
supplemented by four over-relaxation steps.
These configurations are then fixed to the Coulomb gauge 
at every 100-400 sweeps.
On each gauge fixed configuration, we invert the
quark matrix by the BiCGStab algorithm to obtain the quark 
propagator. 
We always perform the iteration of the BiCGStab inverter by
$T$ times, where $T$ is the temporal lattice size.
By changing the stopping condition for the 
quark propagator, we have checked that 
this criterion is sufficient to achieve the desired numerical accuracy.
We accumulate 400-1000 configurations for hadronic
measurements. 

Our simulation parameters are compiled in Tables 
\ref{tab:simparam} and \ref{tab:simparam2}.
In Table \ref{tab:simparam3}, we compare
some of the parameters used in our simulation 
(labeled by ``set A'')
with those in the previous
studies by Klassen (``set B'' and ``set D'')\cite{Klassen:1999fh,Klassen} 
and by Chen (``set C'')\cite{Chen:2001ej} 
for later references. 

\subsection{Meson operators}

In this work, we calculate all of S- and P-state meson
masses of charmonia, namely 
$^1S_0$($\eta_c$), 
$^3S_1$($J/\psi$), $^1P_1$($h_c$),
$^3P_0$($\chi_{c0}$), $^3P_1$($\chi_{c1}$) and
$^3P_2$($\chi_{c2}$).
For this computation, we measure 
the correlation function of the operators
which have the same quantum number as one of above particles.
In Table \ref{tab:SPope} we give the operators for the S- and P-state
mesons. There are two types of operators, those of the form  
$\bar{\psi} \Gamma \psi$ and of $\bar{\psi} \Gamma \Delta\psi$, 
where $\Gamma$ represents a combination of $\gamma$ matrices
and $\Delta$ the spatial lattice derivative. 
We call them the $\Gamma$ operator and the $\Gamma \Delta$ operator,
respectively. The latter appears only for the P-state mesons.
Note that there are two lattice representations for the $^3P_2$ state 
(E and T representations) due to breaking of rotational symmetry. 

We measure the correlation functions of the $\Gamma$ operators
\begin{equation}
 C^{ss'}_{\rm state}(t) = \sum_{\bf x} 
\langle 
\bar{\psi}_{{\bf x},t} \Gamma \psi_{{\bf x},t} 
\cdot 
\sum_{\bf y_0,z_0} \bar{\psi}_{{\bf z_0},0} \Gamma \psi_{{\bf y_0},0} 
f^{s'}_{\bf x_0-z_0} f^{s}_{\bf x_0-y_0} 
\rangle,
\end{equation}
where $f^{s}_{\bf x}$ is a source smearing function, and we always adopt
a point sink. We employ the point source ($s=0$) with
$f^{s=0}_{\bf x}=\delta_{\bf x,0}$ and an exponentially smeared source ($s=1$)
with $f^{s=1}_{\bf x}=A_s e^{-B_s \mid {\bf x}\mid}$, where 
$A_s$ and $B_s$ are smearing parameters.
Therefore we have three source combinations, $ss'=00$, 01 and 11, for 
the $\Gamma$ operators.
The smearing parameters $A_s$ and $B_s$ at each $\beta$ are chosen so
that the effective mass of the $1S$ meson for  
$ss'=01$ has a wide plateau. 

To obtain the correlation functions of the $\Gamma \Delta$ operators,
we measure 
\begin{equation}
 C^{ss'}_{ijkl}(t) = \sum_{\bf x} 
\langle 
\bar{\psi}_{{\bf x},t} \Gamma_i \Delta_j \psi_{{\bf x},t} 
\cdot 
\sum_{\bf y_0,z_0} \bar{\psi}_{{\bf z_0},0} \Gamma_k \psi_{{\bf y_0},0} 
f^{s'=2}_{l, \bf x_0-z_0} f^{s}_{\bf x_0-y_0} 
\rangle,
\end{equation}
where 
$
\Delta_i \psi_{{\bf x},t} = \psi_{{\bf x}+\hat{i},t}- \psi_{{\bf x}-\hat{i},t}
$
is the discretized derivative at the sink, and we employ a 
smeared derivative source ($s=2$) given by
\begin{equation}
f^{s=2}_{i,{\bf x}} = A_s e^{-B_s \mid
{\bf x}+\hat{i}\mid} - A_s e^{-B_s \mid {\bf x}-\hat{i}\mid}
\;\;\;\;\;\; (i=1,2,3)
\end{equation}
with $A_s$ and $B_s$ the same as those for $s=1$.
For the $^3P_0$ state, for example, we calculate
$C^{ss'}_{\rm ^3P_0} = \sum_{i,j=1}^3 C^{ss'}_{iijj}$ with 
$\Gamma_i = \gamma_i$.
For the $\Gamma \Delta$ operators, 
we have two source combinations, $ss'= 02$ and 12. 
In total,
S-state mesons have $ss'=00$, 01 and 11 source combinations,
and P-state mesons have 00, 01, 11, 02 and 12 source
combinations except for $^3P_2$.
Since there is no $\Gamma$ operator for $^3P_2$,
it has only 02 and 12 source combinations.

To calculate the dispersion relation of S-state mesons,
we measure correlation functions for four
lowest non-zero momenta,
\begin{equation}
a_s{\bf p} = (2\pi /L) \times \{(1,0,0),\ (1,1,0),\
          (1,1,1),\ (2,0,0)\},
\end{equation}
in addition to those at rest.
Correlation functions with the same value of $| {\bf p} |$ but different
orientations are averaged to increase the statistics.

\subsection{Tuning bare quark mass $m_0$ and fermion anisotropy $\zeta$}
\label{sec:mztuning}

Let us describe our method of tuning $\zeta$ and
$m_0$ in detail. 
We determine the input parameters $m_0 (=m_0^1,m_0^2)$ and $\zeta (=\zeta^{\rm
TI},\zeta^{\rm NP})$ as follows. First we fix $\zeta = \xi =
3$ and choose $m_0^1$ and $m_0^2$ where the 1S meson mass 
roughly agrees with the experimental value. Then we determine
both the tree-level tadpole-improved value 
$\zeta^{\rm TI}$ and the nonperturbative value $\zeta^{\rm NP}$
at $m_0=m_0^1$ and $m_0^2$.

To obtain $\zeta^{\rm TI}$ at fixed $m_0$ , we use
Eqs.~(\ref{TIbearemass}) and (\ref{TImasszeta}). 
We replace the factor $u_t/u_s$ in Eq.~(\ref{TImasszeta}) with $\xi /\xi_0$,
using Eq.~(\ref{TIxi0}). 
On the other hand, $\zeta^{\rm NP}$ is obtained by demanding 
that the relativistic dispersion relation is
restored at small momenta for the 1S meson.
The dispersion relation on a lattice is given by
\begin{eqnarray}
E(p)^2 &=& E(0)^2 + c^2 {\bf p}^2 + O(a_s^2 p^4)\\
       &=& M_{\rm pole}^2 + \frac{M_{\rm pole}}{M_{\rm kin}} 
{\bf p}^2 + O(a_s^2 p^4),
\label{EPrel}
\end{eqnarray}
where $c$ is called the `speed of light', and 
$M_{\rm pole}$ and $M_{\rm kin}$ 
are the pole and kinetic masses of the 1S meson.
Throughout this paper, a capital letter $M$
denotes the meson mass, while a small one $m$ the quark mass.
Generally $c$ is not equal to one due to lattice artifacts.
We extract the speed of light $c$ by fitting $E(p)^2$ linearly in $p^2$ 
for three or four lowest momenta, 
since the linearity of $E(p)^2$ in $p^2$ is well satisfied.
We identify $\zeta^{\rm NP}$ with a point where
$c =1$ 
or equivalently $M_{\rm pole}=M_{\rm kin}$ for the 1S meson.
To determine $\zeta^{\rm NP}$, we perform preparatory simulations
and calculate $c$ for 
$\zeta = 2.8$, 3.0 and 3.2 at $m_0=m_0^1$ and
$m_0^2$ using 100-200 gauge configurations. 
Then we find $\zeta =\zeta^{\rm NP}$, where $c=1$, from an
interpolation of $\zeta$. As shown in Table \ref{tab:simparam2}, the speed of
light $c$ at $\zeta^{\rm NP}$ is indeed equal to 1 within
1\%, which is roughly the size of the statistical error.

Production runs for the charmonium spectrum described in
Sec.\ref{sec:choiceparam} are performed at
$m_0=(m_0^1,m_0^2)$ and $\zeta =(\zeta^{\rm TI},\zeta^{\rm
NP})$ for each $\beta$. Accidentally, for $\beta =5.90$ and
6.35, $\zeta^{\rm TI}= \zeta^{\rm NP}$ 
holds within our numerical accuracy,
so we use the same data for the analysis at these $\beta$.

Finally we linearly
interpolate or extrapolate results at $m_0=(m_0^1,m_0^2)$ to those at 
$m_0 = m_0^{\rm charm}$, 
with fixed $\zeta$ ($=\zeta^{\rm TI}$ or $\zeta^{\rm NP}$).
As already mentioned,
we identify $m_0^{\rm charm}$
with a point where the spin-averaged 1S meson mass 
$M_{\rm lat}(1\bar{S})$ in units
of a physical quantity $Q_{\rm lat}$ is equal to the 
corresponding experimental value:
\begin{equation}
\frac{M_{\rm lat}(1\bar{S})}{Q_{\rm lat}} = 
\frac{M_{\rm exp}(1\bar{S})}{Q_{\rm exp}}
\label{tunem0}
\end{equation}
with $M_{\rm exp}(1\bar{S}) = 3067.6$ MeV for charmonium.
In this work, we adopt the Sommer scale $r_0$ and
the spin-averaged
mass splittings $\dM(1\bar{P}-1\bar{S}) \equiv M(1\bar{P}) - M(1\bar{S})$
and $\dM(2\bar{S}-1\bar{S}) \equiv M(2\bar{S}) - M(1\bar{S})$
as the scale quantity $Q$.
The spin-averaged masses are defined by  
\begin{eqnarray}
M(n\bar{S}) &=& [3M(n^3S_1)+M(n^1S_0)]/4, \label{spinaverageS}\\
M(n\bar{P}) &=& [3M(n^1P_1)+5M(n^3P_2)+3M(n^3P_1)+M(n^3P_0)]/12
\label{spinaverageP}
\end{eqnarray}
with $n (=1,2,\cdots)$ the radial quantum number. 
The experimental values of the mass splittings $\dM (1\bar{P}-1\bar{S})$ and
$\dM(2\bar{S}-1\bar{S})$ are 457.9 MeV and 595.4 MeV, respectively.
The experimental values of $r_0$ is not known, and 
we use a phenomenological estimate $r_0=0.50$ fm.
For the definition of the lattice meson mass $M_{\rm lat}$ 
in Eq.~(\ref{tunem0}),
we have two choices in the case of $\zeta = \zeta^{\rm TI}$:
one is the pole mass $M_{\rm pole}$ and the other is the kinetic
mass $M_{\rm kin}$.
On the other hand, in the case of $\zeta = \zeta^{\rm NP}$,
$M_{\rm pole} = M_{\rm kin}$ should hold by definition.
In practice, there can be small deviations due to the statistical error. 
Therefore we have 4(=2$\times$2) choices for $(M_{\rm lat},\zeta)$ in total.

\subsection{Mass fitting}
\label{sec:massfitting}

From meson correlation functions
we extract the meson mass (energy) by standard $\chi ^2$
fitting with a multi-hyperbolic-cosine ansatz 
(termed $n_{\rm fit}$-cosh fit below)
\begin{equation}
C^{ss'}_{\rm state}(t) 
= \sum_{i=0}^{n_{\rm fit}-1} A_i^{ss'} 
\cosh [(\frac{T}{2}-t) M_i],
\label{fitansatz}
\end{equation}
where $ss'$ represents the source combination (00, 01, etc.),
$t$ is the time separation from the source, and 
$n_{\rm fit}$ is the number of states included in the fit.

We determine the mass of the ground state and the first radial excited
state for each particle, and the mass splittings such as $\dM(1P-1S)$
and $\dM(2S-1S)$, 
from a 2-cosh fit using several
correlation functions with different source combinations simultaneously.
Here we use the correlation functions of $ss'=00$, 01 and 11 sources for
S-states, while 00, 11, 02 and 12 sources are used for 
P-states except for $^{3}P_{2}$.
For $^{3}P_{2}$, we use the 
correlation functions of 02 and 12 sources.
The 2-cosh fit for each S-state always gives the 
ground state mass consistent with that from the 1-cosh fit.
On the other hand, for the P-state, the 2-cosh fit is preferred over the
1-cosh fit because the $1P$ mass 
from the 1-cosh fit using the correlation function of
11 and 12 sources occasionally disagrees by a few $\sigma$, 
due to excited state contaminations. 
To determine the mass of the first excited state accurately,
it is better to adopt results from the 3-cosh fit.
However, we do not perform the 3-cosh fit systematically 
because of the instability of it, and adopt 
results from the 2-cosh fit for the first excited state mass.
This may cause an overestimation of the first excited state mass
due to a contamination from higher excited states.

To determine the spin-averaged $1S$ mass and the $1S$ energy at 
${\bf p \neq 0}$, and the spin mass splittings such as 
$\dM(1^3S_1-1^1S_0)$ and $\dM(1^3P_1-1^3P_0)$, we perform a 1-cosh fit
($n_{\rm fit}=1$) using the source combination which gives the
widest plateau in the effective mass. 
We use the 01 source for the S-state, and the 12 source
for the P-state. 
We always 
check that the spin mass splitting from a simultaneous 2-cosh fit 
mentioned above
agrees with that from the 1-cosh fit within 1-2 $\sigma$.
We also check that the splitting $\dM(1^3P_1-1^3P_0)$ from a 1-cosh fit using
the 11 source agrees with that using the 12 source.

In these analyses, we perform both the uncorrelated
fit, and the correlated fit which takes account of the
correlation between different time slices and different
sources. The uncorrelated fit 
is always stable and gives
$\chi^2/N_{DF} \lsim 0.5 \ (Q \sim 1)$.
The correlated fit with 1-cosh ansatz is also stable and
produces results consistent with those from the uncorrelated fit.
However, the correlated 2-cosh fit is often unstable, either failing
to invert the covariance matrix or giving large $\chi^2/N_{DF} \gg 1$ even
if it converges. Therefore we adopt the uncorrelated fit for
our final analysis.

The fitting range $[ t_{\rm min},t_{\rm max} ]$ for the final
analysis is determined as follows.
From an inspection of
the effective mass plot, we determine $t_{\rm max}$ which roughly has the
same physical length independent of $\beta$.
We repeat the 1- and 2-cosh fits for each $\beta$, 
varying $t_{\rm min}$ with fixed $t_{\rm max}$, and
find a range of $t_{\rm min}$ where the ground state
mass and the first excited state mass (for 2-cosh fit) are
stable against $t_{\rm min}$. We also check that it has reasonable
value of $\chi^2/N_{DF}$.
The final $t_{\rm min}$ is then chosen from the region accepted
above so that its physical length is roughly equal
independent of $\beta$.

\begin{figure}[t]
\centerline{\epsfxsize=9.5cm \epsfbox{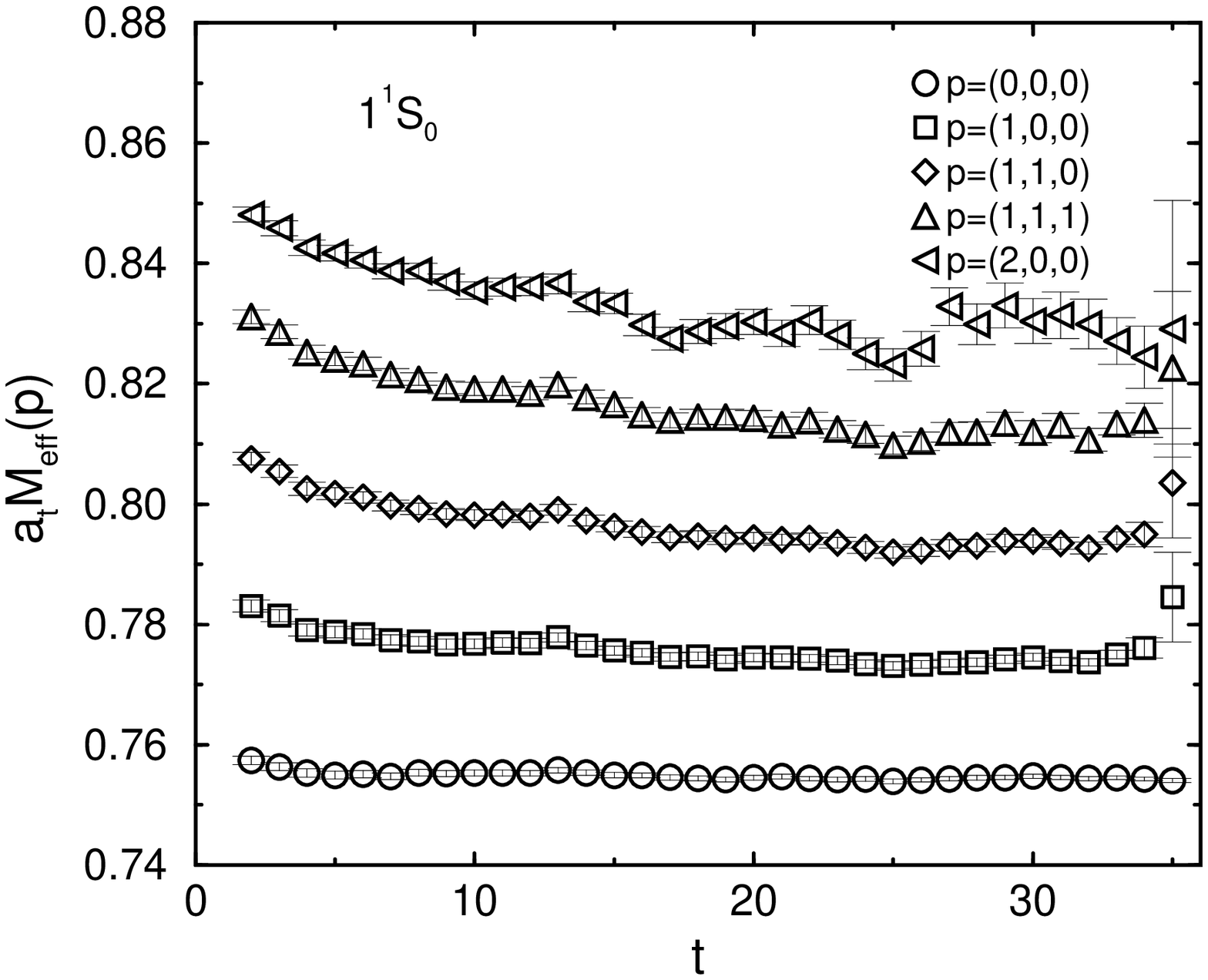}
            \epsfxsize=9.5cm \epsfbox{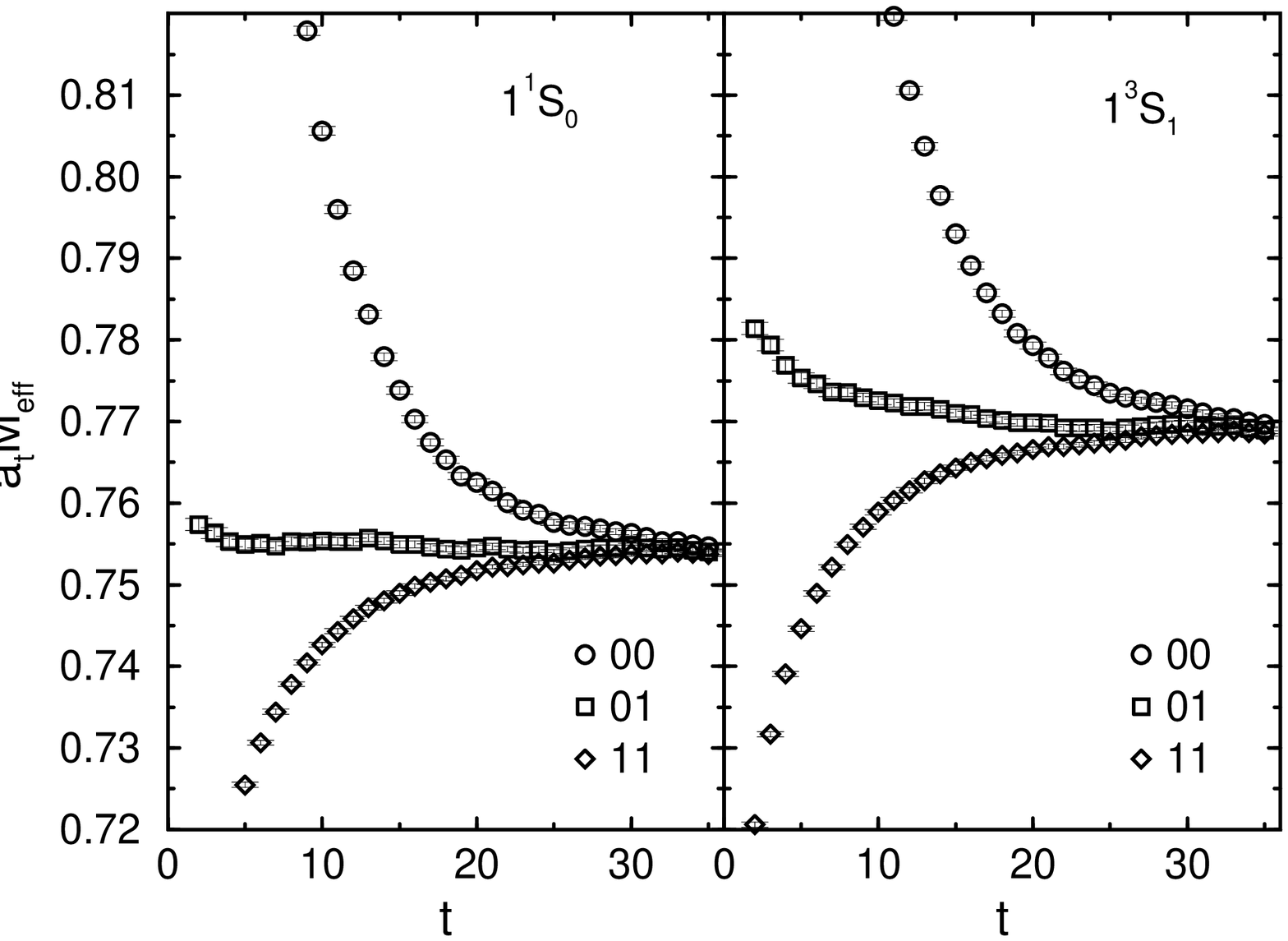}}
\caption{S-state effective masses at $\beta=5.90$, $a_tm_{q0}=0.144$ and
         $\zeta = 2.99$. The left figure shows the 
         $1^1S_0$ masses at ${\bf p} \neq 0,$ while the right
         shows the $1^1S_0$ and $1^3S_1$ masses for the source $ss'=00$, 01 and 11.}
\label{fig:SeffMb590mq144}
\end{figure}

\begin{figure}[t]
  \begin{center}
    \leavevmode
    \epsfxsize=0.65 \hsize
    \epsffile{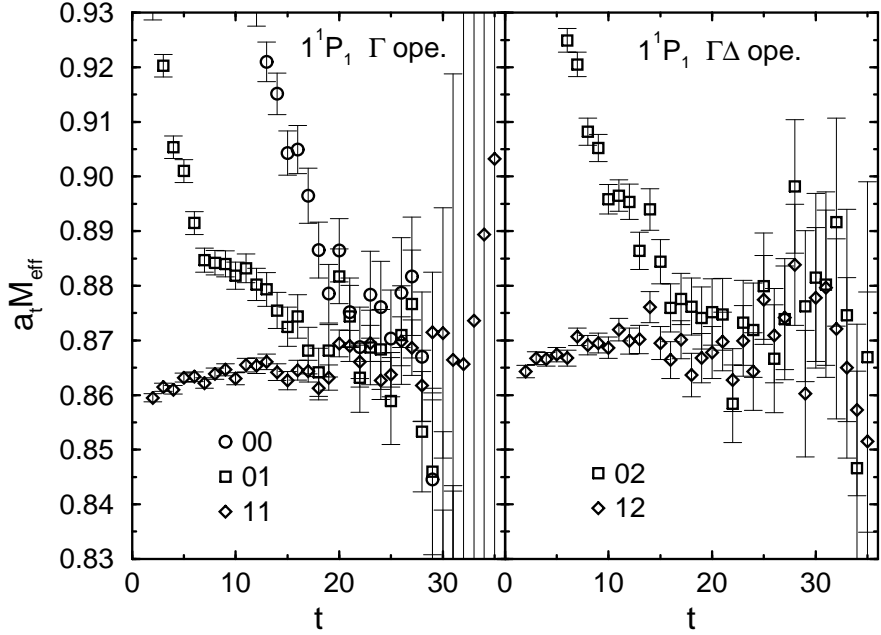}
    \caption{P-state ($1^1P_1$)
     effective masses at $\beta=5.90$, $a_tm_{q0}=0.144$ and 
             $\zeta = 2.99$.
The left figure shows the masses from the $\Gamma$ operator,
while the right shows those from the $\Gamma\Delta$ operator.}
    \label{fig:PeffMb590mq144}
   \end{center}
\end{figure}

\begin{figure}[t]
\centerline{\epsfxsize=9.5cm \epsfbox{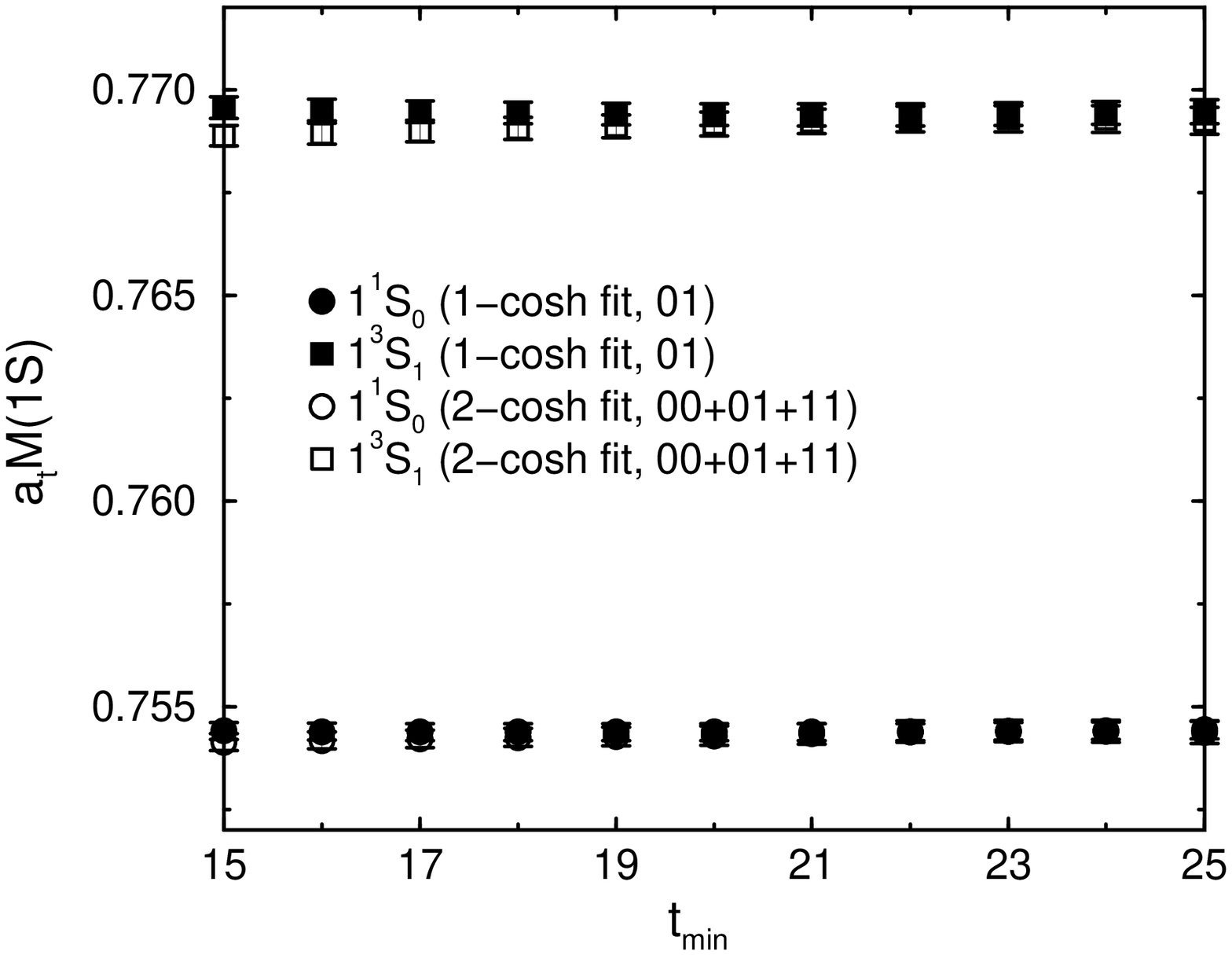}
            \epsfxsize=9.5cm \epsfbox{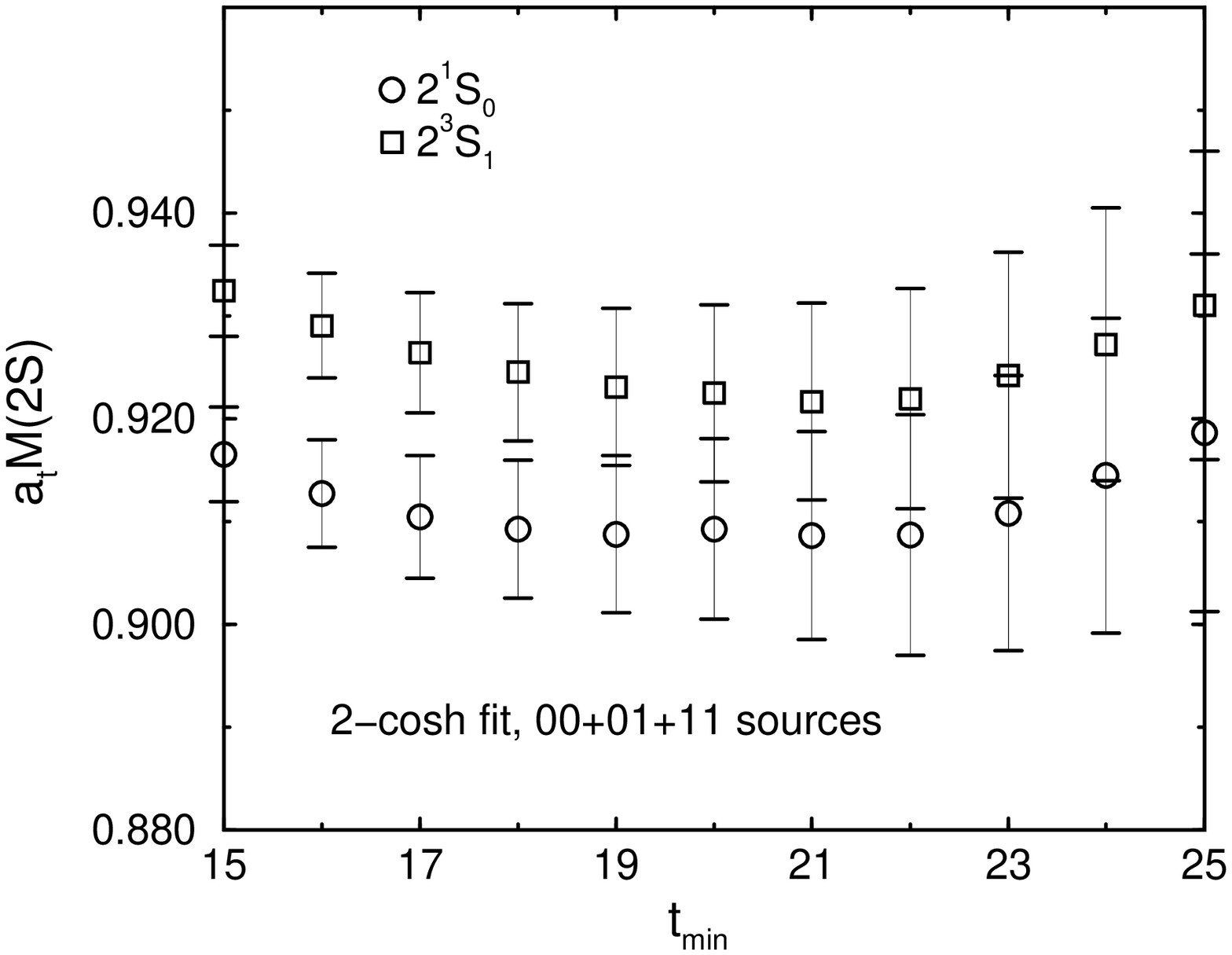}}
\centerline{\epsfxsize=9.5cm \epsfbox{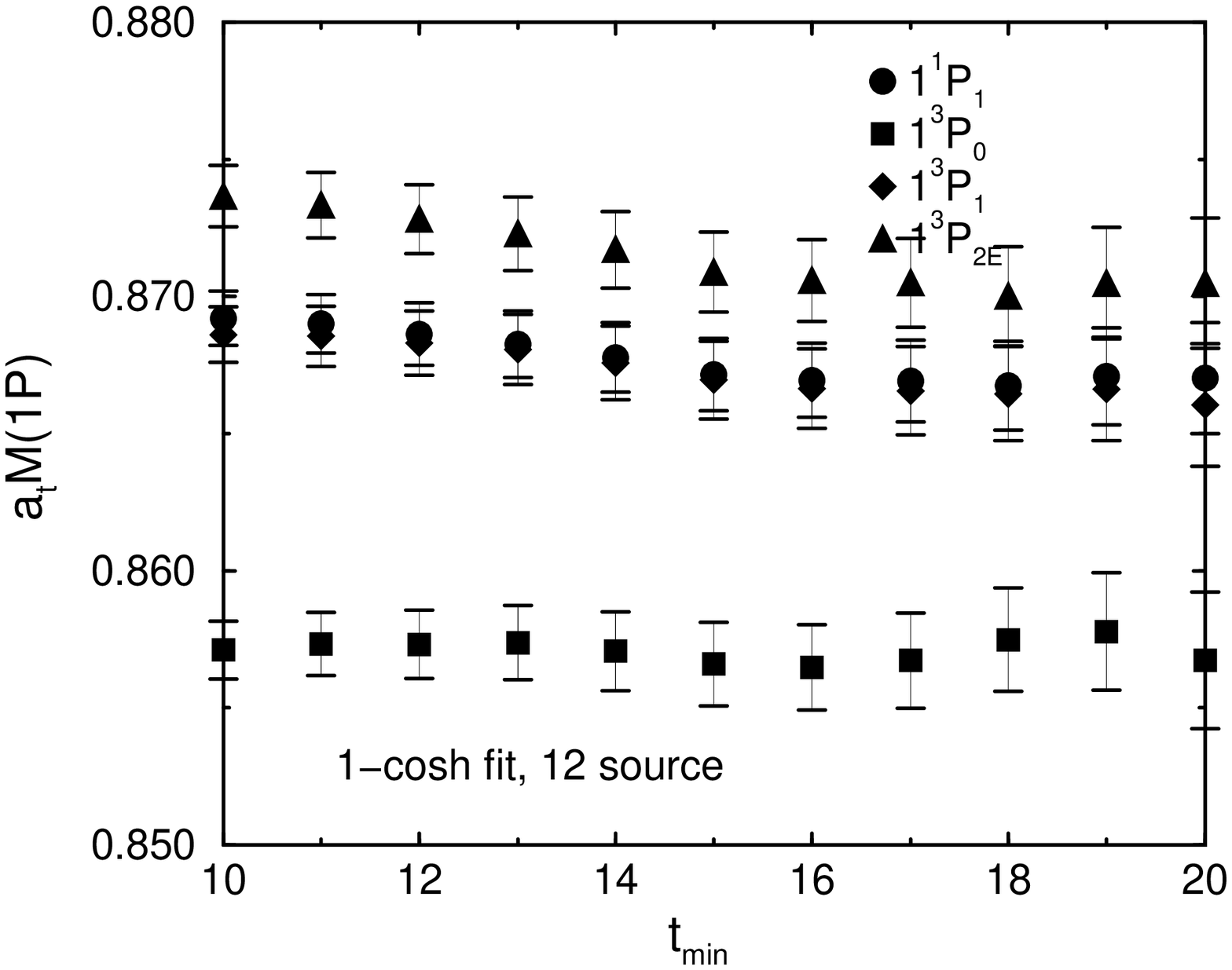}
            \epsfxsize=9.5cm \epsfbox{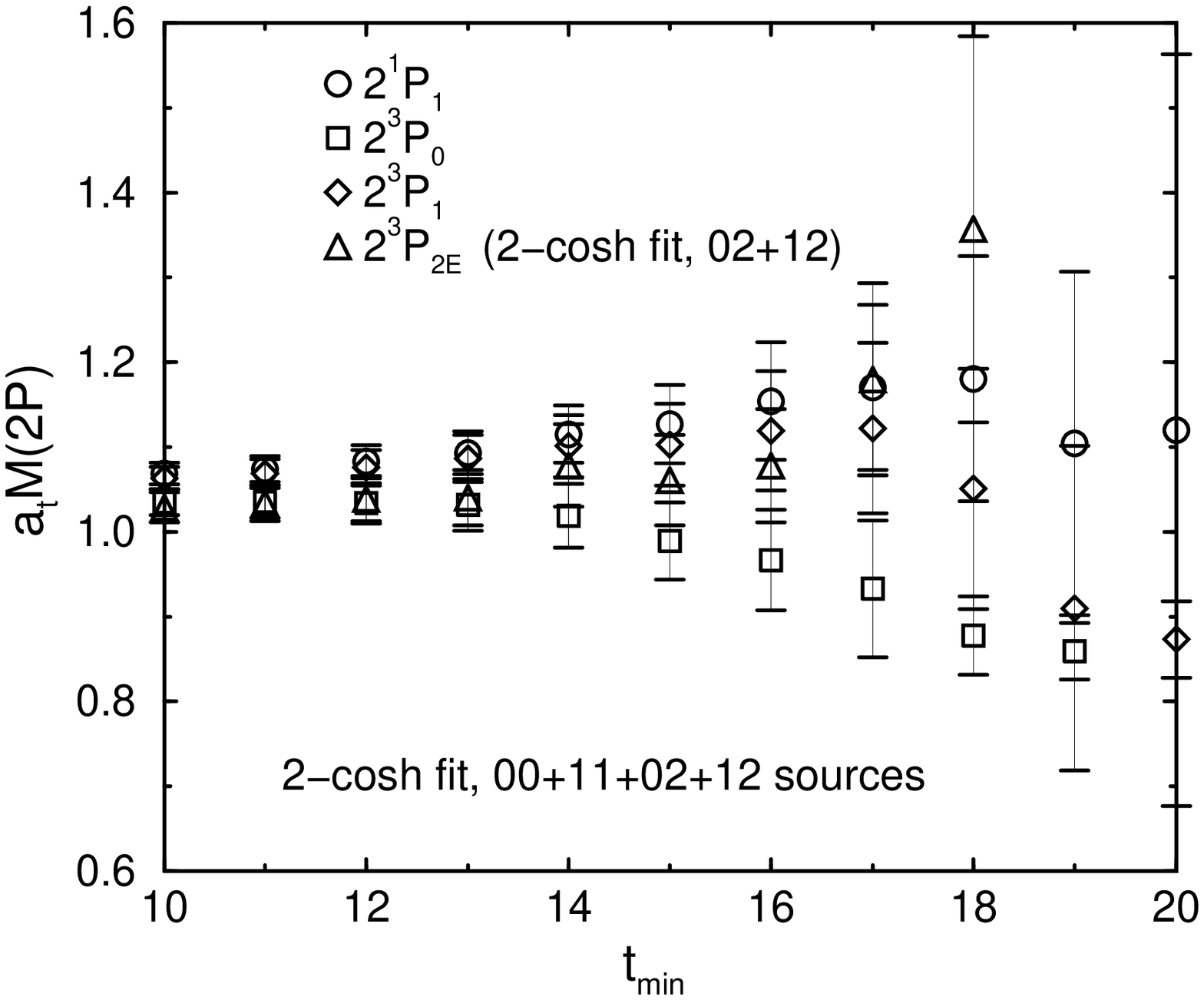}}
\caption{Fit range ($t_{\rm min}$) 
dependence of masses at $\beta=5.90$, $a_tm_{q0}=0.144$ and 
             $\zeta = 2.99$. The legend denotes the state (fit
 ansatz, quark source).}
\label{fig:tminb590mq144}
\end{figure}

\begin{figure}[t]
  \begin{center}
    \leavevmode
    \epsfxsize=0.65 \hsize
     \epsffile{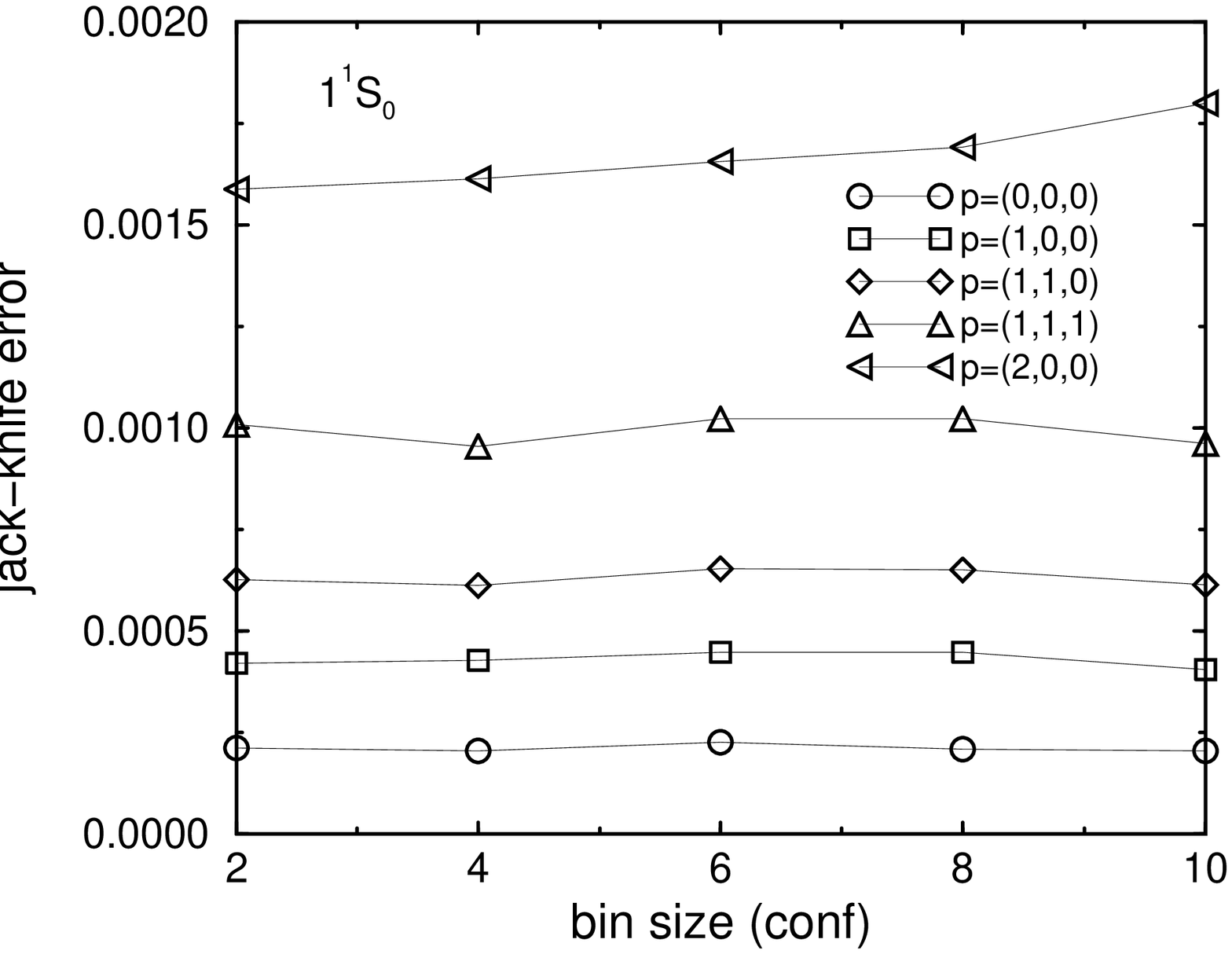}
    \caption{Bin size dependence of jack-knife error of 
    $a_tM(1^1S_0)$ with ${\bf p = 0}$ and ${\bf p \neq 0}$
    at $\beta=6.10$, $a_tm_{q0}=0.024$ and $\zeta = 2.88$.}
    \label{fig:bindeppi}
   \end{center}
\end{figure}

\begin{figure}[h]
  \begin{center}
    \leavevmode
    \epsfxsize=0.65 \hsize
    \epsffile{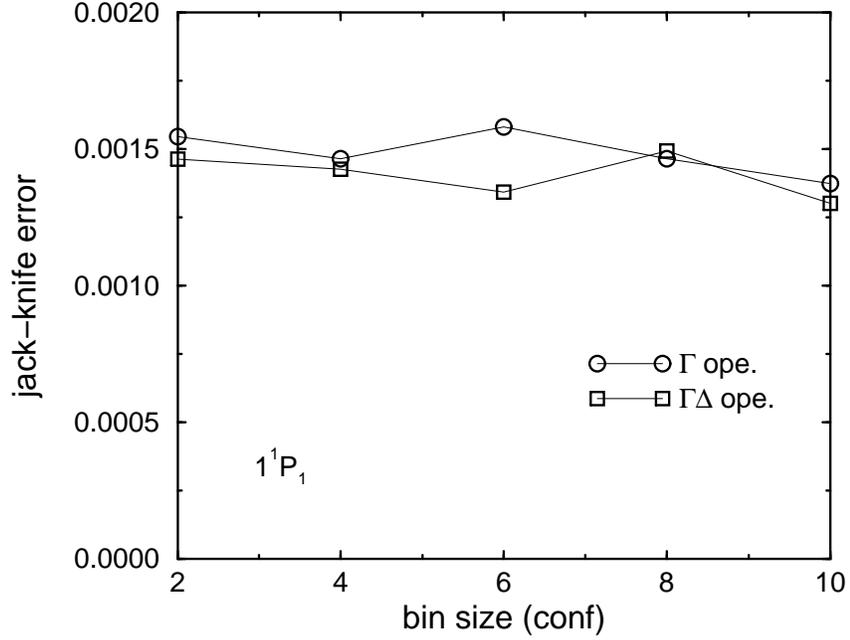}
    \caption{Bin size dependence of jack-knife error of 
$a_tM(1^1P_1)$ at $\beta=6.10$, $a_tm_{q0}=0.024$ and $\zeta = 2.88$.
}
    \label{fig:bindep1P1}
   \end{center}
\end{figure}

Typical examples of the effective mass plot and $t_{\rm min}$-dependence
of the fitted mass are shown in
Figs.~\ref{fig:SeffMb590mq144}-\ref{fig:PeffMb590mq144}, and in
Fig.~\ref{fig:tminb590mq144}, respectively.
Our final fitting ranges are summarized in Table
\ref{tab:frange}.
Statistical errors of masses and mass splittings are
estimated by the jack-knife method.
The typical bin size dependences of jack-knife errors for the
ground state masses are shown in
Figs.~\ref{fig:bindeppi} and \ref{fig:bindep1P1}. 
We always adopt a bin size of 10 configurations, {\it i.e.}
1000-4000 sweeps.

\subsection{Scaling violation and the continuum limit} \label{sec:syserr} 

We discuss scaling violation for our action and 
how the results 
at finite $a_s$ are extrapolated to the continuum limit
$a_s \rightarrow 0$.
Since we use the anisotropic Wilson gauge action with nonperturbatively
tuned $\xi_0$, the scaling violation from the gauge sector
starts at $O((a_s\Lambda_{\rm QCD})^2)$.

For the quark sector, we use the anisotropic clover quark
action with tadpole-improved 
clover coefficients $c_{s,t}$, and either the tadpole-improved value 
$\zeta^{\rm TI}$ or
nonperturbative value $\zeta^{\rm NP}$ for $\zeta$.
Since we adopt the tree-level tadpole-improved value of $c_{s}$ for massive
($a_sm_q \ge 0$) quarks, the scaling violation arising from the choice of
$c_{s}$ is $O((a_s\Lambda_{\rm QCD})^2)$ and $O(\alpha a_s\Lambda_{\rm QCD})$.
On the other hand, for $c_{t}$, we adopt the tree-level tadpole-improved
value correct only in the massless ($a_sm_q = 0$) limit, which 
generates an additional $O(a_s\Lambda_{\rm QCD} \cdot a_sm_q) = O(a_s^2\Lambda_{\rm QCD}m_q)$ error.
Recall that the $a_sm_q$ (not only $a_tm_q$) 
dependence of the parameter remains with our
choice of the spatial Wilson coefficient $r_s=1$ at the tree level, 
as discussed in Sec.\ref{sec:ansioQCDact}. 
In the case of $\zeta =\zeta^{\rm NP}$, therefore, the scaling violations
are $O((a_s\Lambda_{\rm QCD})^2)$ and $O(a_s^2\Lambda_{\rm QCD}m_q)$
at leading order, and $O(\alpha a_s\Lambda_{\rm QCD})$
at next-to-leading order. 
The size of these errors are estimated to be 
$O((a_s\Lambda_{\rm QCD})^2)=7-$1\%, $O(a_s^2\Lambda_{\rm QCD}m_q)=37-$4\%
and $O(\alpha a_s\Lambda_{\rm QCD})=4-$1\% for $\beta =5.70-$6.35
corresponding to $a_s^{-1} \approx 1.0-$2.8 GeV.
Here we took $\Lambda_{\rm QCD}=250$ MeV 
($\simeq \Lambda_{\rm \overline{MS}}^{N_f=0}$) and $m_q=1.4$ GeV 
($\simeq m_{\rm charm}$), and the renormalized coupling
constant $\alpha$ is estimated from Eq.~(\ref{tigauge}). 
It is expected that the $O(\alpha a_s\Lambda_{\rm QCD})$ errors are
largely eliminated by the tadpole improvement.

When the tree level tadpole improved value $\zeta^{\rm TI}$ is used
instead of $\zeta^{\rm NP}$,
we have additional
$O(\alpha)$ and $O(\alpha a_sm_q)$ 
errors, since the kinetic term is a dimension four operator.
The size of the additional errors is estimated to be 
$O(\alpha)=15-$12\% and $O(\alpha a_sm_q)=22-$6\%. 
Again we expect that the dominant
part of this error is eliminated by the tadpole improvement.

In this work we adopt an $a_s^2$-linear extrapolation for the
continuum limit, because the leading order scaling violation is always 
$O((a_s\Lambda_{\rm QCD})^2,a_s^2\Lambda_{\rm QCD}m_q)$
irrespective of the choice of $\zeta$.
We also perform an $a_s$-linear extrapolation 
to estimate systematic errors.
In practice we use results at three finest lattice spacings {\it i.e.}
$\beta =5.90$-6.35 ($a_sm_q \le 1$) for the continuum extrapolation,
excluding results at $\beta = 5.70$ ($a_sm_q > 1$), 
which appear to have larger 
discretization errors as expected from the naive order estimate.
Performing such extrapolations for all sets of 
$M_{\rm lat} = (M_{\rm pole},M_{\rm kin})$ and $\zeta =(\zeta^{\rm
TI},\zeta^{\rm NP})$, we adopt the choice which 
shows the smoothest scaling behavior for the final value, and 
use others to estimate the systematic errors.

\section{Results}\label{sec:results}

Now we present our results of the quenched charmonium spectrum
obtained with the anisotropic quark action.
In this section,  
we first compare results of 
$\zeta^{\rm NP}$ with $\zeta^{\rm TI}$.
Second, we determine the lattice scale, and study the 
effect of $(M_{\rm lat},\zeta)$ tuning.
We then show the results of charmonium masses and mass splittings,
and estimate their continuum limit.

\subsection{Dispersion relation and $\zeta^{\rm NP}$}

\begin{figure}[t]
\centerline{\epsfxsize=9.5cm \epsfbox{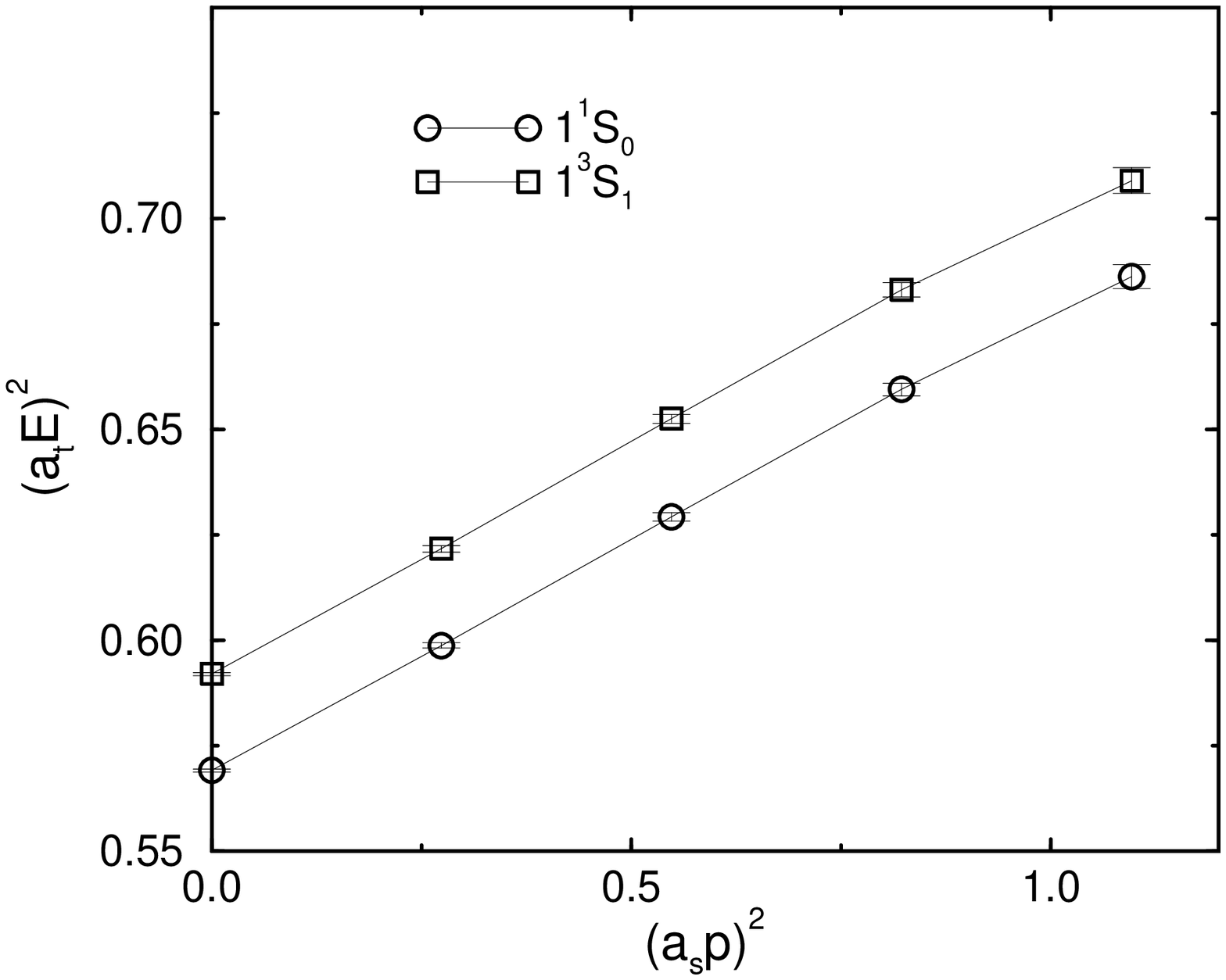}
            \epsfxsize=9.5cm \epsfbox{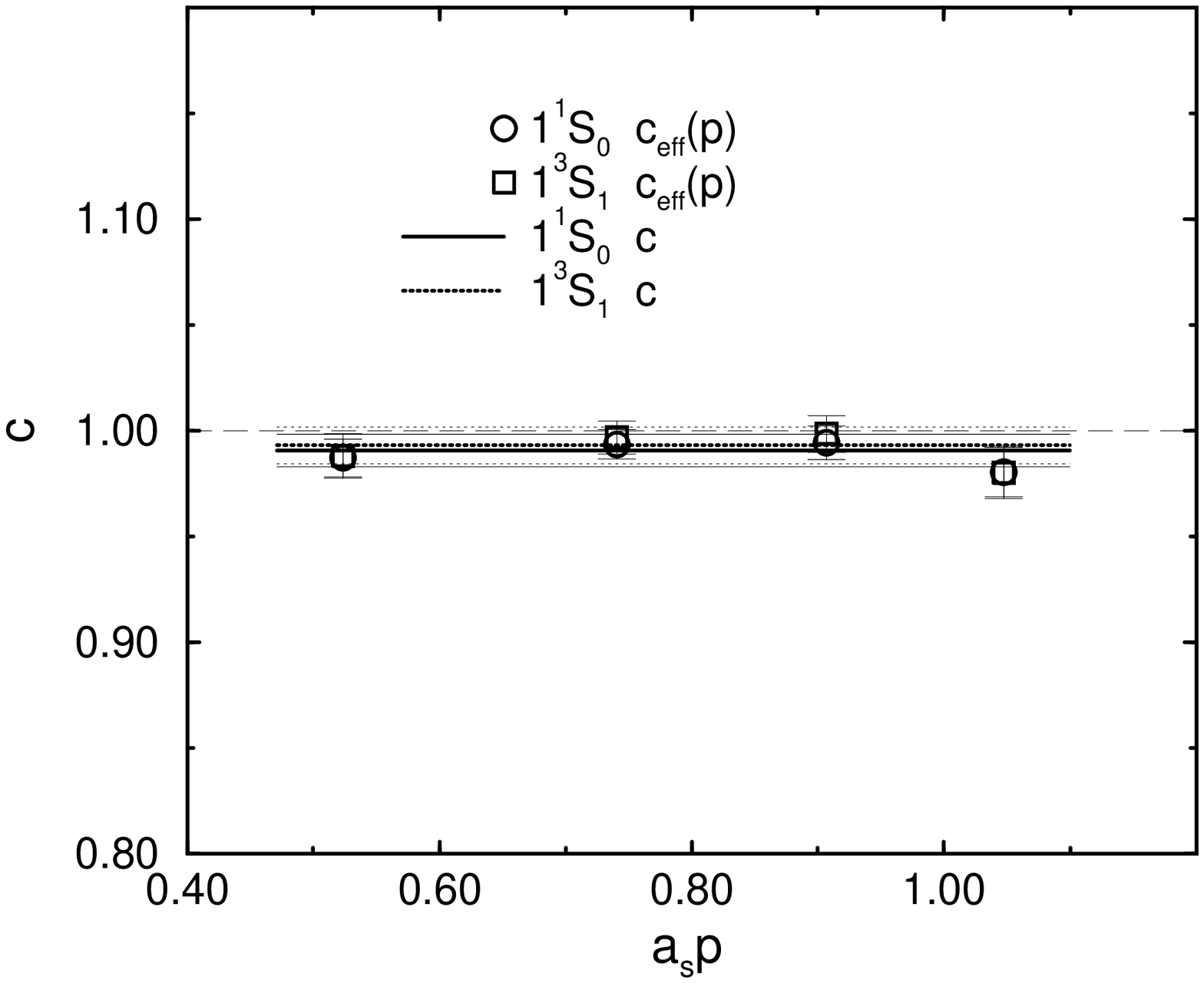}}
    \caption{Dispersion relation (left) and speed of light (right) of S-state 
             at $\beta=5.90$, $a_tm_{q0}=0.144$ and $\zeta = 2.99$.
On the right, we show the effective speed of light $c_{\rm eff}({\bf p})$ and 
the speed of light $c$ from the fit. 
}
    \label{fig:ceNPZ}
\end{figure}

In Fig.~\ref{fig:ceNPZ}, we plot a typical example of the 
dispersion relation and the speed of light. 
As shown in the left figure, 
the linearity of $E^2$ in $p^2$ is satisfied well.
Indeed the `effective speed of light', defined by
\begin{equation}
c_{\rm eff}({\bf p}) = \sqrt{\frac{E({\bf p})^2-E({\bf 0})^2}{{\bf p}^2}},
\label{effc}
\end{equation}
has a wide plateau as shown in the right figure. 
Therefore we employ the linear fit in $p^2$ to extract the
speed of light $c$ from $E^2$.
This figure also illustrates that the speed of light $c$ 
for $^1S_0$ agrees well with that for $^3S_1$ within errors.
This is indeed the case for all data points as observed in 
Table \ref{tab:simparam2}.
The speed of light $c$ seems universal 
for all mesons as pointed out in Ref.\cite{Chen:2001ej}.

The nonperturbative value of $\zeta$, $\zeta^{\rm NP}$, is
obtained by demanding that the speed of light $c$ is equal to
1 within 1\%.
On the other hand, the tree-level tadpole-improved value, $\zeta^{\rm TI}$, 
gives $c$ deviating from 1 by $2-$4\% {\it i.e.} $2-$4 $\sigma$ at most,
which is much smaller than the size of the $O(\alpha,\alpha a_sm_q)$
error ($12-$15\%, $6-$22\%) estimated in the previous section.
This suggests that $O(\alpha,\alpha a_sm_q)$ 
errors associated with $\zeta^{\rm TI}$ 
are almost eliminated by the tadpole improvement,
as expected.

\begin{figure}[t]
  \begin{center}
    \leavevmode
    \epsfxsize=0.65 \hsize
     \epsffile{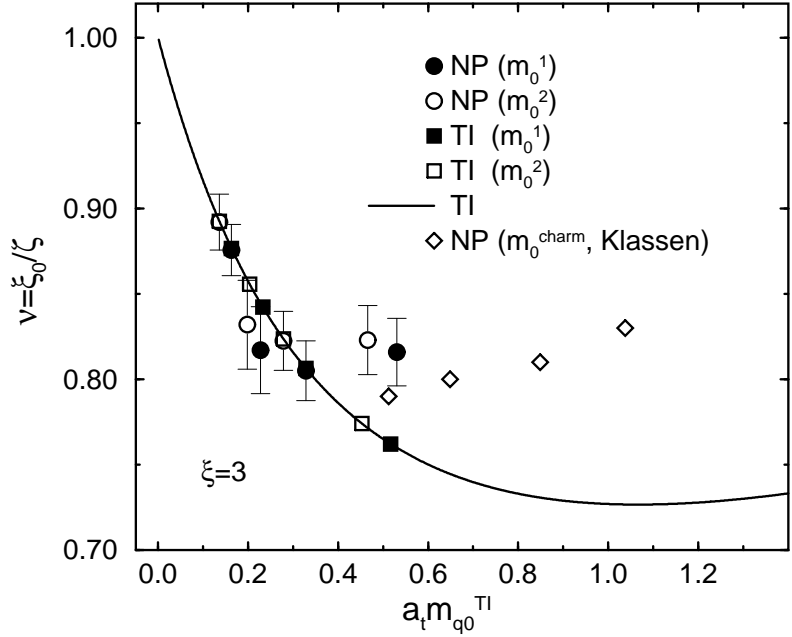}
    \caption{The tadpole improved bare mass 
$\tilde{m_0} \equiv a_tm_{q0}^{\rm TI}$
versus $\nu = \xi_0/\zeta$ at $\xi =3$. `TI' and `NP' denote the 
tree level tadpole improved value and nonperturbative value
respectively. Circles and squares are our data 
at $m_0=m_0^1, m_0^2 \ (\approx m_0^{\rm charm})$ for $\beta =5.7$-6.35.
The error bars for the circles denote the statistical uncertainty of 
$\nu^{\rm NP} = \xi_0/\zeta^{\rm NP}$.
We also plot Klassen's data at $m_0=m_0^{\rm charm}$ for $\beta =5.5$-5.8
as open diamonds.
}
    \label{fig:tunedzeta}
   \end{center}
\end{figure}

In Fig.~\ref{fig:tunedzeta}, 
$\nu^{\rm NP} = \xi_0/\zeta^{\rm NP}$ and 
$\nu^{\rm TI} = \xi_0/\zeta^{\rm TI}$ at $m_0=m_0^1$ and $m_0^2$
are plotted as a function of $\tilde{m_0} = a_tm_{q0}^{\rm TI}$.
We find that $\nu^{\rm NP}$ 
(circles) and $\nu^{\rm TI}$ 
(squares and solid line) agree within errors at $\tilde{m_0}\le 0.3$
but deviate from each other at $\tilde{m_0}\simeq 0.5$ ($\beta=5.7$).
The latter is one of the reasons why we exclude this point in the continuum
extrapolation.
One also notices that the slope of $\nu$ approaching the value $\nu =1$
in the continuum limit is steep, and in addition, the difference
$\nu^{\rm NP}-\nu^{\rm TI}$ for our data
does not have a smooth dependence in $a_tm_{q0}^{\rm TI}$.
As discussed in Sec.\ref{sec:hfsrevisit}, these features of $\nu^{\rm NP}$
bring complications in
the scaling behavior of the hyperfine splitting.

\subsection{Lattice scale}

In this work, we determine the lattice spacing
via the Sommer scale $r_0$\cite{Sommer:1994ce}, 
the $1\bar{P} - 1\bar{S}$ meson mass splitting, 
and the $2\bar{S} - 1\bar{S}$ splitting.
We compare the results obtained with these different scales, in order to
estimate the quenching errors.

\subsubsection{Scale from the Sommer scale $r_0$}

\begin{figure}[t]
\centerline{\epsfxsize=9.5cm \epsfbox{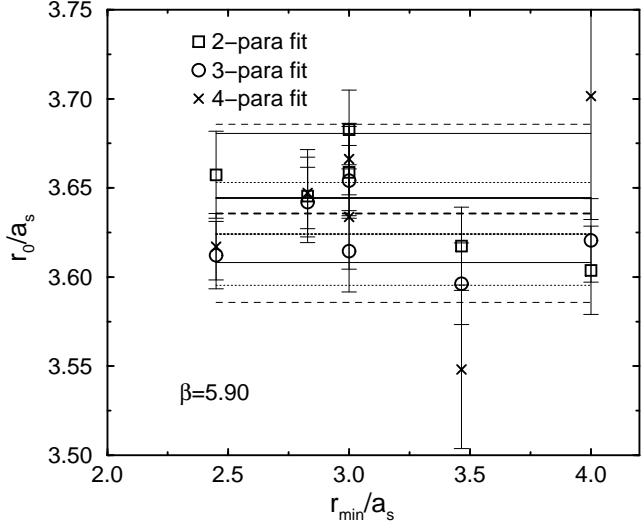}
            \epsfxsize=9.5cm \epsfbox{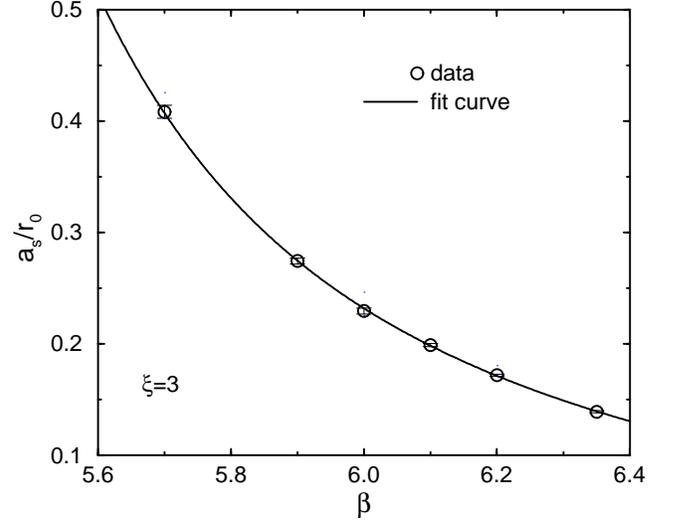}}
    \caption{Results of $r_0/a_s$.
The left figure shows typical fit range ($r_{\rm min}$) dependence of $r_0/a_s$
and its averaged value.
The right is the result of $a_s/r_0$ as a function of $\beta$
 and its fit curve Eq.~(\ref{alltonfitform}).}
    \label{fig:r0fit}
\end{figure}

In order to calculate the static quark potential
needed for the extraction of $r_0$,
additional pure gauge simulations listed in Table \ref{tab:R0} are performed. 
Using $La_s \geq 1.4$ fm lattices, 
we measure the smeared Wilson loops at
every 100-200 sweeps at six values of $\beta$ in the range $\beta =5.70-$6.35. 
Details of the smearing method \cite{Bali:1992ab,Bali:1993ru} 
are the same as those in Ref.\cite{Aoki:1999ff}.
We determine the potential $V(\hat{r})$ at each $\beta$ from 
a correlated fit with the ansatz
\begin{equation}
W(\hat{r},\hat{t})
 = C(\hat{r})e^{a_tV(\hat{r}) \cdot \hat{t}},
\label{Vfit}
\end{equation}
where $\hat{r}=r/a_s$ and $\hat{t}=t/a_t$ are the spatial and temporal
extent of the Wilson loop in lattice units.
The fitting range of $\hat{t}$ is chosen 
by inspecting the plateau of the effective potential
$
a_tV_{\rm eff}(\hat{r},\hat{t}) 
 = \log ( W(\hat{r},\hat{t}) / W(\hat{r},\hat{t}+1)).
$
A correlated fit to $V(\hat{r})$ is then performed
with the ansatz
\begin{equation}
a_tV(\hat{r}) = a_tV_{0} + (a_ta_s\sigma) \hat{r} - (e/\xi) \frac{1}{\hat{r}}
+ a_t\delta V, \,\,\,\,\,\,\, a_t\delta V = 
                         l \left( \frac{1}{\hat{r}}
                       - \left[\frac{1}{\hat{r}}\right]  \right),
\label{potfit}
\end{equation}
where $\sigma$ is the string tension and $\left[\frac{1}{\hat{r}}\right]$ 
is the lattice Coulomb term from one-gluon exchange 
\begin{equation}
\left[\frac{1}{\hat{r}}\right] 
 = 4\pi \int_{-\pi}^{\pi} \frac{d^{3}\mathbf{k}}{(2\pi)^{3}}
                         \frac{\cos (\mathbf{k} \cdot \hat{r})}
                              {4 \sum_{i=1}^{3} \sin^{2} 
                               (k_{i}a_s/2)}_.
\label{GXchg}
\end{equation}
We extract $r_0/a_s$ from the condition that 
$r^2 \frac{d(V - \delta V)}{dr}|_{r=r_0} = c$, 
{\it i.e.}
\begin{equation}
r_0/a_s = \sqrt{\frac{c-e}{\xi a_ta_s\sigma}}
\label{r0sigma}
\end{equation}
with $c=1.65$.
The error of $r_0/a_s$ is estimated by adding 
the jack-knife error with bin size 5 and the variation over the fitting
range of $\hat{r}$.
Keeping to the ansatz Eq.~(\ref{potfit}), we attempt three different fits:
(i) 2-parameter fit with $e=\pi /12$ and $l=0$ fixed,
(ii) 3-parameter fit with $e=\pi /12$ fixed,
and (iii) 4-parameter fit.
We check that $r_0/a_s$ from these three fits agree well within errors
(See Fig. \ref{fig:r0fit}).
We adopt $r_0/a_s$ from the 2-parameter fit as our final value.
Results of $r_0/a_s$ at each $\beta$ are summarized in Table \ref{tab:R0}.

Next we fit $r_0/a_s$ as a function of $\beta$
with the ansatz proposed by Allton \cite{Allton:1996kr},
\begin{equation}
(a_s/r_0)(\beta) =  
                         f(\beta) ~ ( \,
  1 + c_2           \, \hat{a}(\beta)^2
    + c_4           \, \hat{a}(\beta)^4  \, 
                            )/ c_0 \, , \qquad
     \hat{a}(\beta) \equiv { f(\beta) \over f(\beta_1)} \, ,
\label{alltonfitform} 
\end{equation}
where $\beta_1=6.00$ and 
$f(\beta)$ is the two-loop scaling function of SU(3) gauge theory,
\begin{equation} 
 f(\beta = 6/g^2) \, \equiv \,
(b_0 g^2)^{-{b_1\over 2b_0^2}} \, \exp(-{1\over 2b_0 g^2}) \, , \qquad
  b_0 = {11\over (4\pi)^2 } \, , \quad  b_1 = {102\over (4\pi)^4} \, ,
\label{f2loop} 
\end{equation}
and $c_n (n=2,4)$ parameterize deviations from the two-loop scaling.
From this fit, we obtain 
that
\begin{equation}
                 c_0  =  0.01230(29),\,\,\,\, 
                 c_2  =  0.163(54),\,\,\,\,
                 c_4  =  0.053(22)
\label{alltonrst}
\end{equation}
with $\chi^2/N_{DF}=0.51$.
As shown in Fig.~\ref{fig:r0fit}, the fit curves reproduce the data very 
well. We use Eq.~(\ref{alltonrst}) in our later analysis.
Finally, we obtain $a_s$ from the input of
$r_0 = 0.50$ fm. The values of $a_s$ at each $\beta$
are given in Table \ref{tab:simparam}.

\subsubsection{Scale from charmonium mass splittings}

\begin{figure}[t]
    \leavevmode
    \epsfxsize=0.65 \hsize
\centerline{\epsfxsize=9.5cm \epsfbox{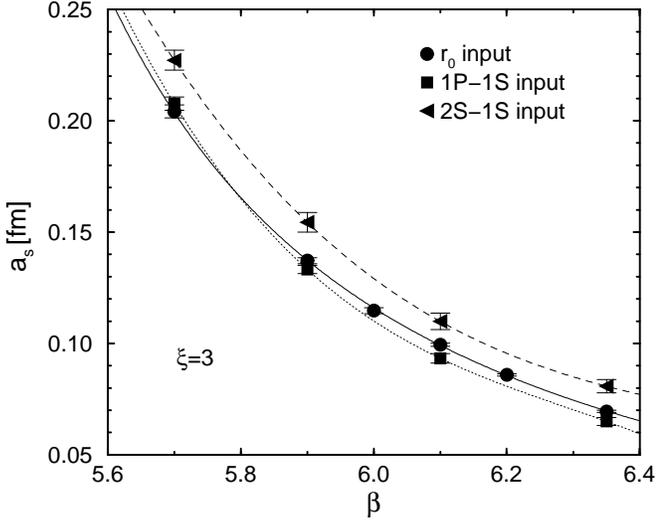}
            \epsfxsize=9.5cm \epsfbox{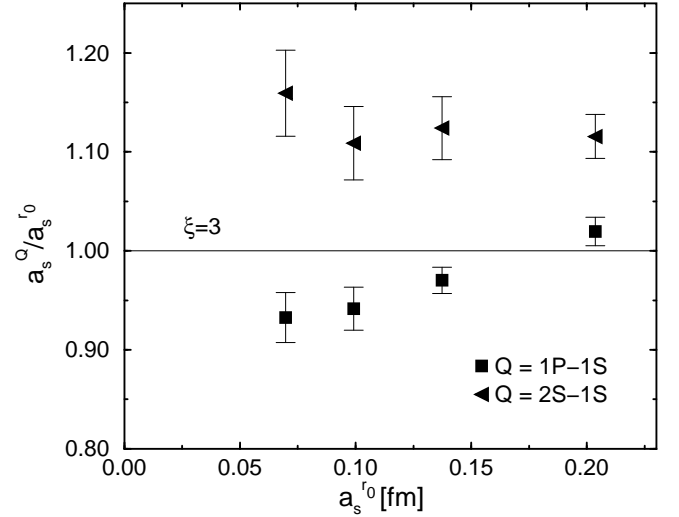}}
    \caption{The left-hand side shows $\beta$-dependence of 
the lattice spacing. The solid line is the fit curve Eq.~(\ref{alltonfitform}), 
while dotted and dashed lines are spline interpolations
to square and triangle symbols respectively.
On the right-hand side $a_s^{1\bar{P}-1\bar{S}}/a_s^{r_0}$ and 
$a_s^{2\bar{S}-1\bar{S}}/a_s^{r_0}$ as a function of $a_s^{r_0}$ are plotted.
}
    \label{fig:beta-as}
\end{figure}

The quarkonium $1P - 1S$ and $2S - 1S$ splittings are often used 
to set the scale in 
heavy quark simulations since the experimental values are well 
determined and they are roughly independent of quark mass for 
charm and bottom.
Here we take the spin average for $1S$, $1P$ and $2S$ masses, so that
the most of the uncertainties from the spin splitting cancel out.
The lattice spacing at  $m_0 = m_0^{\rm charm}$ is given by
\begin{equation}
a_s^Q = \xi \hat{Q}_{\rm lat}/Q_{\rm exp} \;\;\;\;\;\;  
(Q=\dM(1\bar{P}-1\bar{S}),\ \dM(2\bar{S}-1\bar{S})),
\label{asfromsplitting}
\end{equation}
where $\hat{Q}_{\rm lat}$ denotes the value in the temporal lattice unit.
We use the data of $(M_{\rm pole},\zeta^{\rm TI})$ and
check that other choices do not change $a_s^Q$ sizably.
In Table \ref{tab:m0ch-as} we summarize the values of 
$m_0^{\rm charm}$ and $a_s^Q$ for all $Q$ including $r_0$,
and plot the $\beta$-dependence of $a_s^Q$ in Fig.~\ref{fig:beta-as}.
We observe that
$a_s^{1\bar{P}-1\bar{S}} < a_s^{r_0} < a_s^{2\bar{S}-1\bar{S}}$
holds for $\beta =5.70-$6.35.
To show this explicitly, on the right we also plot the ratio
$a_s^{1\bar{P}-1\bar{S}}/a_s^{r_0}$ and 
$a_s^{2\bar{S}-1\bar{S}}/a_s^{r_0}$ as a function of $a_s^{r_0}$.
Deviations from unity are about $-$5\% for
$a_s^{1\bar{P}-1\bar{S}}/a_s^{r_0}$,  
+($10-$15)\% for $a_s^{2\bar{S}-1\bar{S}}/a_s^{r_0}$ and hence
+($10-$25)\% for $a_s^{2\bar{S}-1\bar{S}}/a_s^{1\bar{P}-1\bar{S}}$
at our simulation points.
The major source of discrepancy among the 
lattice spacings from different observables
is the quenching effect. 
Another source is the uncertainty of input value of $r_0 =0.50$ fm,
which is only a phenomenological estimate.
Other systematic errors are expected for $a_s^{2\bar{S}-1\bar{S}}$
for the following reasons.
Our fitting for $2S$ masses may be contaminated by 
higher excited states.
In addition, the lattice size $\sim$1.6 fm
may be too small to avoid finite size effects for $2S$ masses. 
On the other hand, the fitting for $1P$ masses are more reliable,  
and we have checked that 
the finite size effects are negligible for $\dM(1\bar{P}-1\bar{S})$ in
preparatory simulations (see also Ref.\cite{Chen:2001ej}).
For these reasons, 
we consider the scale $a_s^{1\bar{P}-1\bar{S}}$ to be the best choice 
for physical results on the spectrum. We present the
results for three scales in the following, however, to show the dependence
of the spectrum on the choice of the input for the lattice spacing.
In order to make a comparison with the results 
by Klassen and Chen, who employ $r_0$ to set the scale,
we use the results with $a_s^{r_0}$.

\subsection{Effect of $(M_{\rm lat},\zeta)$ tuning}\label{sec:effecttune}

\begin{figure}[t]
\centerline{\epsfxsize=9.5cm \epsfbox{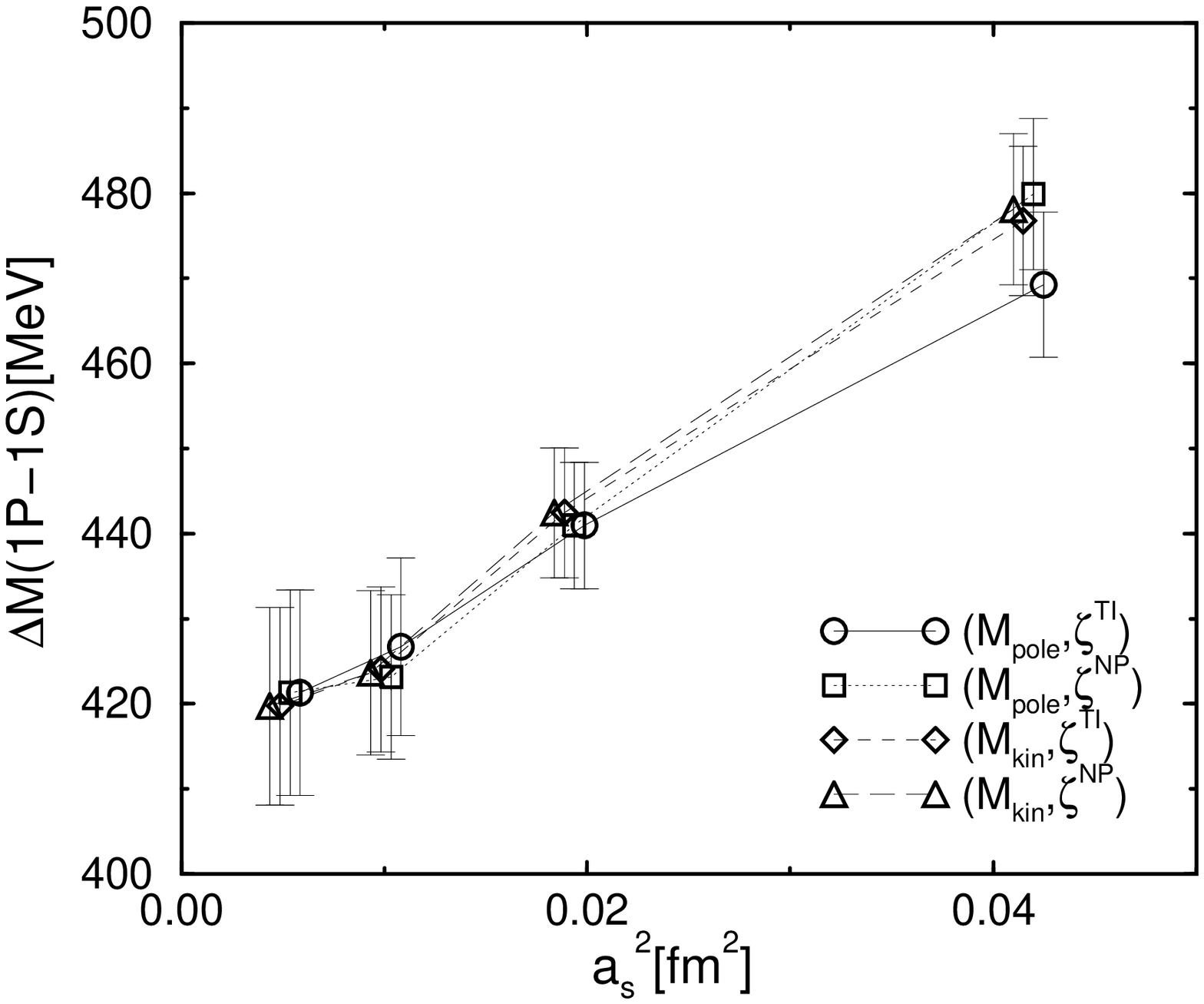}
            \epsfxsize=9.5cm \epsfbox{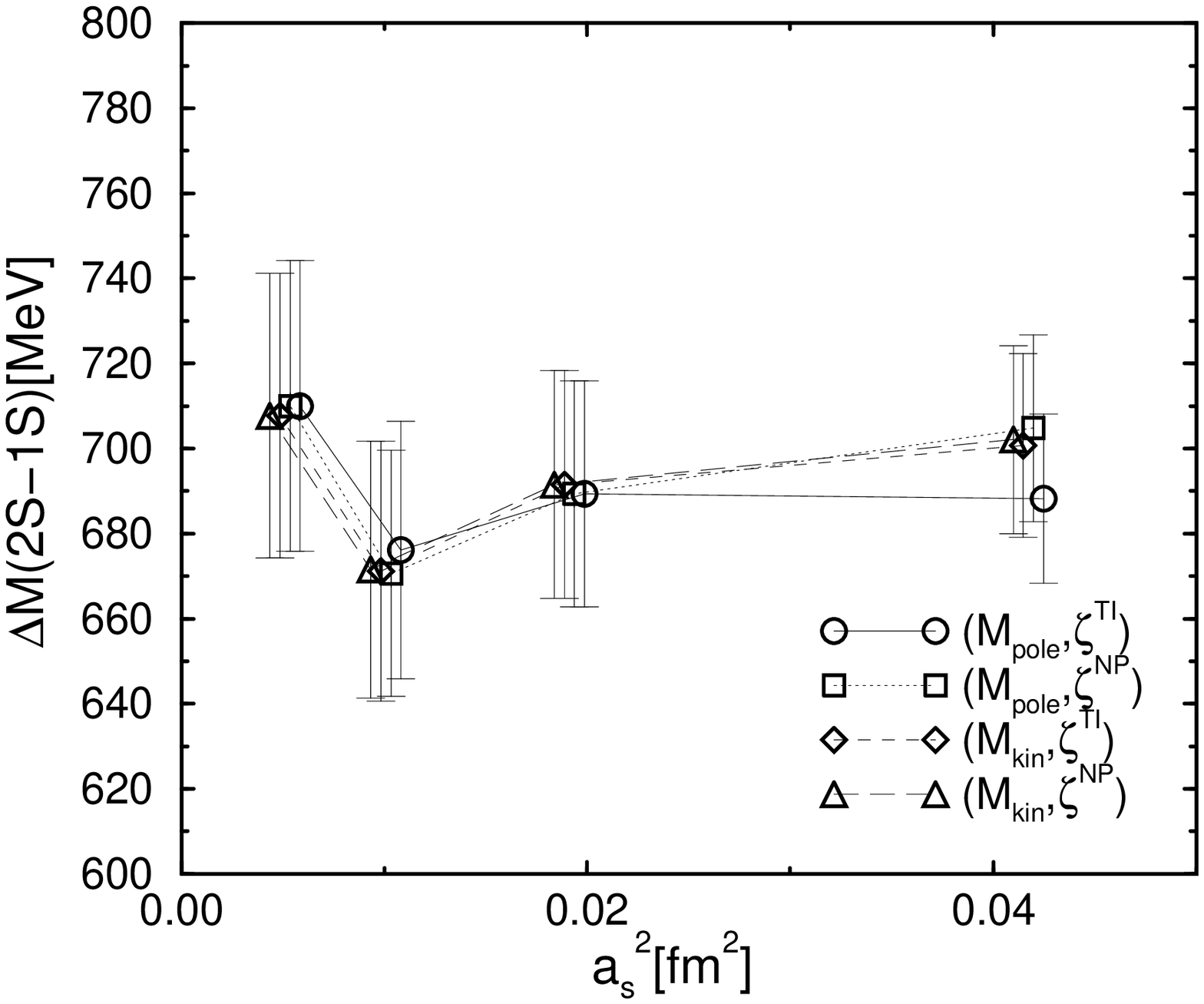}}
\centerline{\epsfxsize=9.5cm \epsfbox{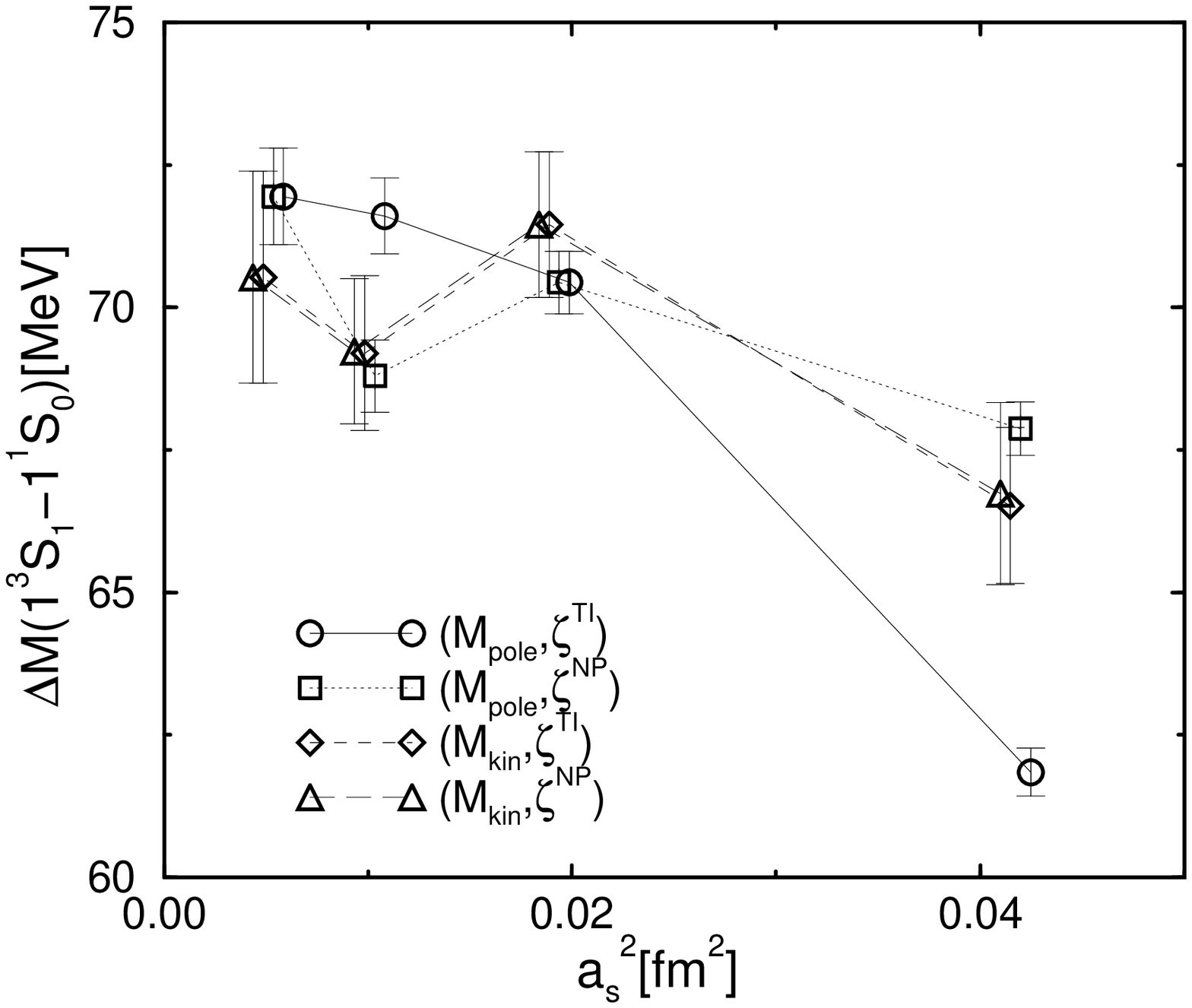}
            \epsfxsize=9.5cm \epsfbox{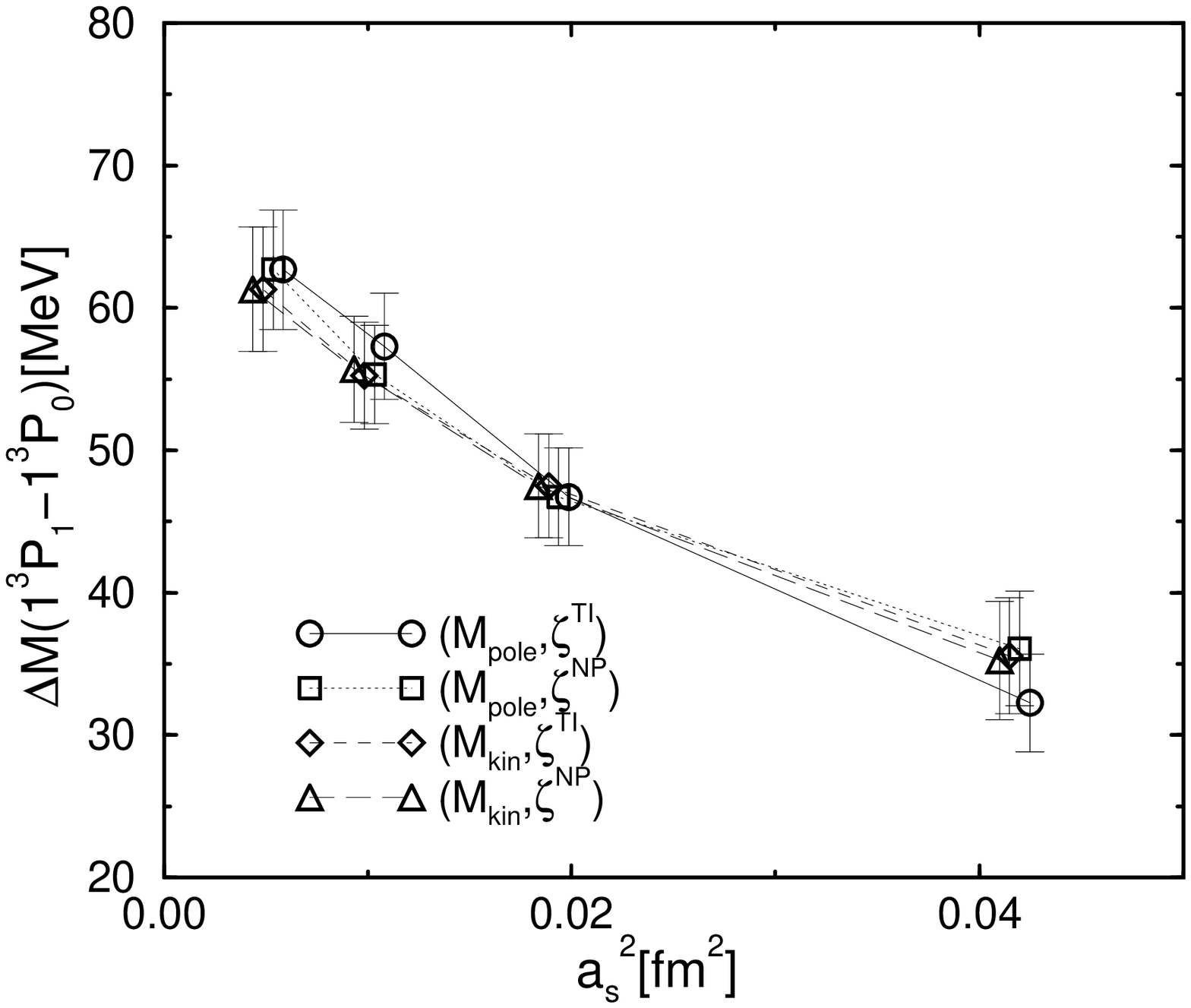}}
\caption{Comparison of results for various $(M_{\rm lat},\zeta)$ tunings.
The scale is set by $r_0$. The data points are slightly shifted along the
 horizontal axis for the distinguishability.}
\label{fig:Comptuning}
\end{figure}

In Fig.~\ref{fig:Comptuning}, we plot the results of 
spin-averaged mass splittings and spin mass splittings for 
each choice of $(M_{\rm lat},\zeta)$.
The upper two figures show the spin-averaged splittings 
$\dM(1\bar{P}-1\bar{S})$ and $\dM(2\bar{S}-1\bar{S})$, 
while the lower two show the S-state hyperfine splitting
$\dM(1^3S_1-1^1S_0)$ and the P-state fine structure $\dM(1^3P_1-1^3P_0)$.
Numerical values for each choice at $\beta = 6.1$
are given in Table \ref{tab:comptuning}.
Here we set the scale with $r_0$ because it has the smallest
statistical error.

For all of mass splittings in Fig.~\ref{fig:Comptuning}, 
the results for 
$(M_{\rm pole},\zeta^{\rm NP}) \simeq (M_{\rm kin},\zeta^{\rm NP})$ 
well agree with those for $(M_{\rm kin},\zeta^{\rm TI})$,
suggesting that the mass splittings are independent of the choice of
$\zeta$ whenever the $M_{\rm kin}$ tuning is adopted. 
This can be understood as follows\cite{El-Khadra:1997mp}.
Setting the measured kinetic mass to the experimental value 
$M_{\rm kin}=M_{\rm exp}$ for the meson roughly corresponds to 
setting $m_2 = m_{\rm charm}$ for the quark, where 
the kinetic mass for the quark $m_2$ is given by Eq.~(\ref{kineticmass})
at the tree level.
Since the spin-averaged splitting 
is dominated by $m_2$, 
setting $m_2 = m_{\rm charm}$ for each $\zeta$ results in the same value
for this splitting. 
With our choice of the spatial clover coefficient $c_s = r_s$, 
$m_B=m_2$ also holds independent of $\zeta$ at the tree level.
Hence the spin splitting takes approximately the same value
because it is dominated by
the magnetic mass $m_B$ given by Eq.~(\ref{magneticmass}).

As a result, we practically have only two choices for $(M_{\rm lat},\zeta)$,
{\it i.e.} $(M_{\rm pole},\zeta^{\rm TI})$ and 
$(M_{\rm pole},\zeta^{\rm NP}) \simeq (M_{\rm kin},\zeta^{\rm NP}) 
\simeq (M_{\rm kin},\zeta^{\rm TI})$.
As observed in Fig.~\ref{fig:Comptuning}, however,
the results for $(M_{\rm pole},\zeta^{\rm TI})$ agree with 
those for the other choices at three finest $a_s$, within a few $\sigma$ for
the hyperfine splitting and 1$\sigma$ for other mass splittings.
This shows that the choice $(M_{\rm pole},\zeta^{\rm TI})$
is as acceptable as any other, with our numerical accuracy, 
for the lattices we adopted.
Since the hyperfine splitting for the choice $(M_{\rm pole},\zeta^{\rm TI})$
has a smoother lattice spacing dependence (at $\beta \ge 5.9$) and 
a smaller error than that for other choices in Fig.~\ref{fig:Comptuning},
we decide to use the data with $(M_{\rm pole},\zeta^{\rm TI})$
for the continuum extrapolations.
The results for other choices are used to estimate the systematic errors.
A slight bump in the lattice spacing dependence of the 
hyperfine splitting for $(M_{\rm pole},\zeta^{\rm NP})$ is 
in part ascribed to the statistical error of $\zeta^{\rm NP}$ itself,
as discussed in Sec.\ref{sec:hfsrevisit}.

\subsection{The charmonium spectrum}

\begin{figure}[t]
  \begin{center}
    \leavevmode
    \epsfxsize=9.5cm
    \epsffile{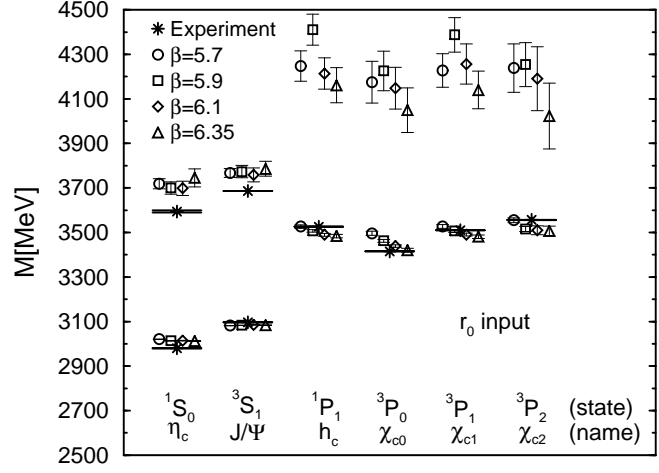}
   \end{center}
\centerline{\epsfxsize=9.5cm \epsfbox{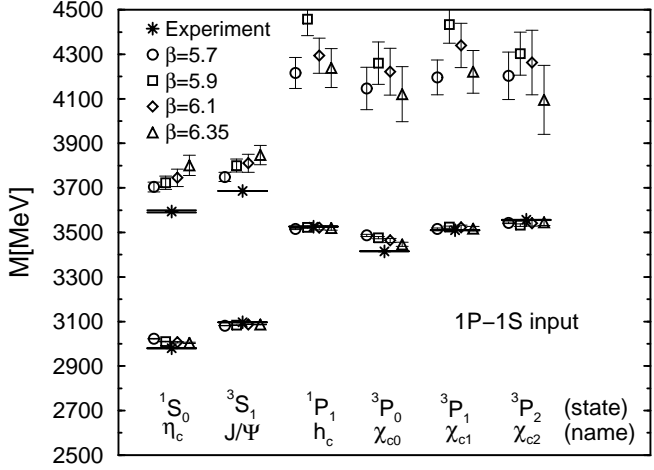}
            \epsfxsize=9.5cm \epsfbox{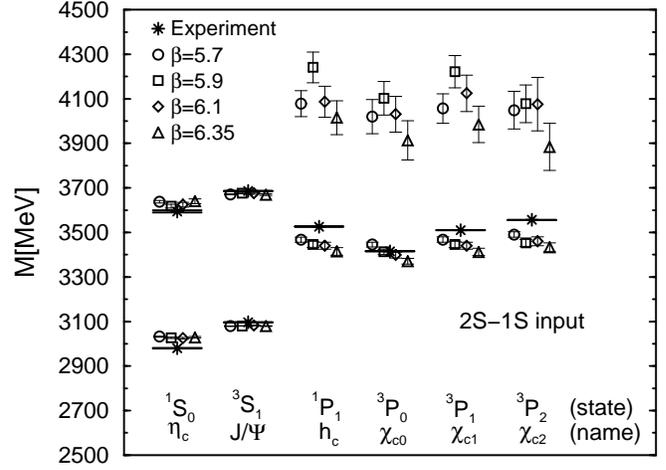}}
\caption{Charmonium spectrum at finite $\beta$. The scale is fixed from
$r_0$, $\dM(1\bar{P}-1\bar{S})$ and $\dM(2\bar{S}-1\bar{S})$.}
\label{fig:spectrum}
\end{figure}

The results for charmonium spectrum, 
obtained for $(M_{\rm pole},\zeta^{\rm TI})$, for the three choices of 
scale are plotted in Fig.~\ref{fig:spectrum} together with 
the experimental values, and numerical values are listed in 
Tables \ref{tab:cc_r0}-\ref{tab:cc_2s1s}.
As observed in Fig.~\ref{fig:spectrum}, the gross features of 
the mass spectrum are consistent with the experiment.
For example, the splittings among the $\chi_c$ states are 
resolved well and 
with the correct ordering ($\chi_{c0} < \chi_{c1} < \chi_{c2}$).
Statistical errors for the 1S, 1P and 2S state masses are of 1 MeV, 
10 MeV and 30 MeV, respectively.
When we set the scale from the $1\bar{P}-1\bar{S}$ ($2\bar{S}-1\bar{S}$)
splitting, the spin structure and the 
$2\bar{S}-1\bar{S}$ ($1\bar{P}-1\bar{S}$) splittings are 
predictions from our simulations.

\subsection{S-state hyperfine splitting}\label{sec:hfs}

We now discuss our results for the S-state hyperfine splitting 
$\dM(1^3S_1-1^1S_0)$,
which is the most interesting quantity in this work.
The hyperfine splitting (HFS), arising from 
the spin-spin interaction between quarks,
is very sensitive to the choice of the clover term, as noticed from 
Eqs.~(\ref{NRHamiltonian}) and (\ref{magneticmass}).
Since the clover term also controls the lattice discretization error
of the fermion sector,
the calculation of the HFS is a good testing ground for
the lattice quark action.

\begin{figure}[t]
  \begin{center}
    \leavevmode
    \epsfxsize=0.65 \hsize
    \epsffile{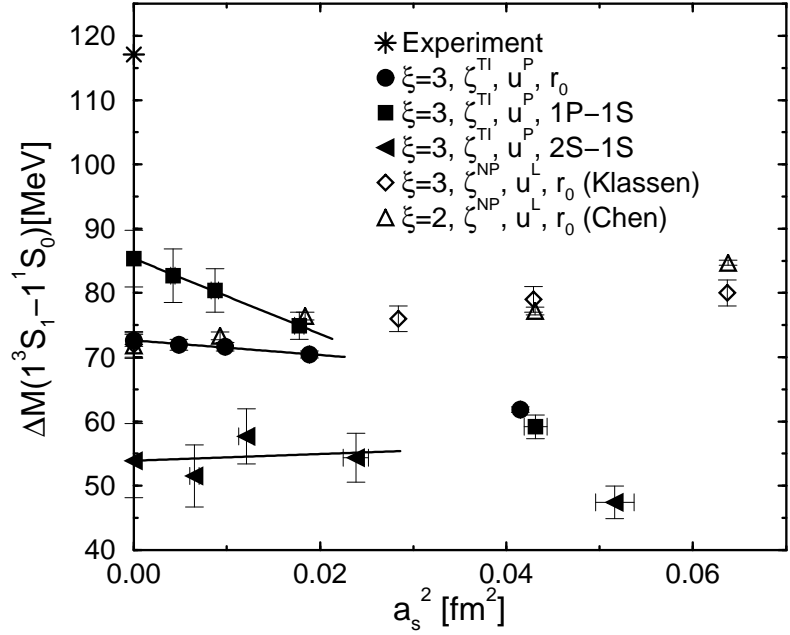}
    \caption{S-state hyperfine splitting $\dM(1^3S_1-1^1S_0)$. 
Results obtained with $\tilde{c_s} = u_s^3 c_s = 1$ are collected here.
Our results are shown by filled symbols for each input, while
results by Klassen (set B) and Chen (set C)
with the $r_0$ input are shown by open
symbols. In the legend, we give the choice of the anisotropy $\xi$,
$\zeta$ tuning, tadpole factor 
and scale input.
These captions also apply to the figures that follow.}
    \label{fig:hfs}
   \end{center}
\end{figure}

In Fig.~\ref{fig:hfs} we plot our results for the S-state HFS 
with $(M_{\rm pole},\zeta^{\rm TI})$ for each scale input by filled symbols.
From the $a_s^2$-linear continuum extrapolation using 3 points at 
$\beta =5.90$-6.35, we obtain 
\begin{equation}
\dM(1^3S_1-1^1S_0) \  = \ \left\{ \begin{array}{ll}
72.6(0.9)(+1.2)(-3.8) \; {\rm MeV} & \mbox{($r_0$ input)} \\
85.3(4.4)(+5.7)(-2.5) \; {\rm MeV} & \mbox{($1\bar{P} - 1\bar{S}$ input)} \\
53.9(5.8)(-1.5)(-2.0) \; {\rm MeV} & \mbox{($2\bar{S} - 1\bar{S}$ input)} \\
117.1(1.8) \; {\rm MeV} & \mbox{(experiment)} \\
\end{array} \right. ,
\label{hfscontfinal}
\end{equation}
where the first error is the statistical error.
The second error represents the ambiguity in the
continuum extrapolation, estimated as the difference
between the $a_s^2$-linear and the $a_s$-linear fits.
The third error is the systematic error associated with 
the choice of $(M_{\rm lat},\zeta)$. We estimate it from the maximum  
difference at the continuum limit between 
the choice of $(M_{\rm pole},\zeta^{\rm TI})$ and the other three choices.
Our estimate of the S-state HFS is smaller than the experimental 
value by 27 \% if the $1\bar{P}-1\bar{S}$ splitting is used to set the
scale. A probable source for this 
large deviation is quenching effects.

In this figure, we also plot previous anisotropic results
by Klassen (set B in Table \ref{tab:simparam3})\cite{Klassen} 
and Chen (set C)\cite{Chen:2001ej} at $\xi =2$ and 3
with the {\it same} choice of the clover coefficients $c_{s,t}$
and using $r_0$ to set the scale.
The difference between our simulation and theirs is 
the choice of $\zeta$ and  the tadpole factor for $c_{s,t}$, 
as noted in Table \ref{tab:simparam3}.
We use $\zeta^{\rm TI}$ and the tadpole factor estimated from the
plaquette $u^P$, while they used
$\zeta^{\rm NP}$ and tadpole estimate from the mean link 
in the Landau gauge $u^L$.
As shown in this figure,
our result in the continuum limit with $r_0$ input
agrees with the results by Klassen\cite{Klassen} and Chen\cite{Chen:2001ej}.
The results with a {\it different} choice 
of the clover coefficients $c_{s,t}$ by Klassen (set D) 
will be shown in Sec.\ref{sec:hfsrevisit}, where
we will study the effect of $c_{s}$ to the HFS.

\subsection{P-state fine structure}

\begin{figure}[t]
  \begin{center}
    \leavevmode
    \epsfxsize=0.65 \hsize
    \epsffile{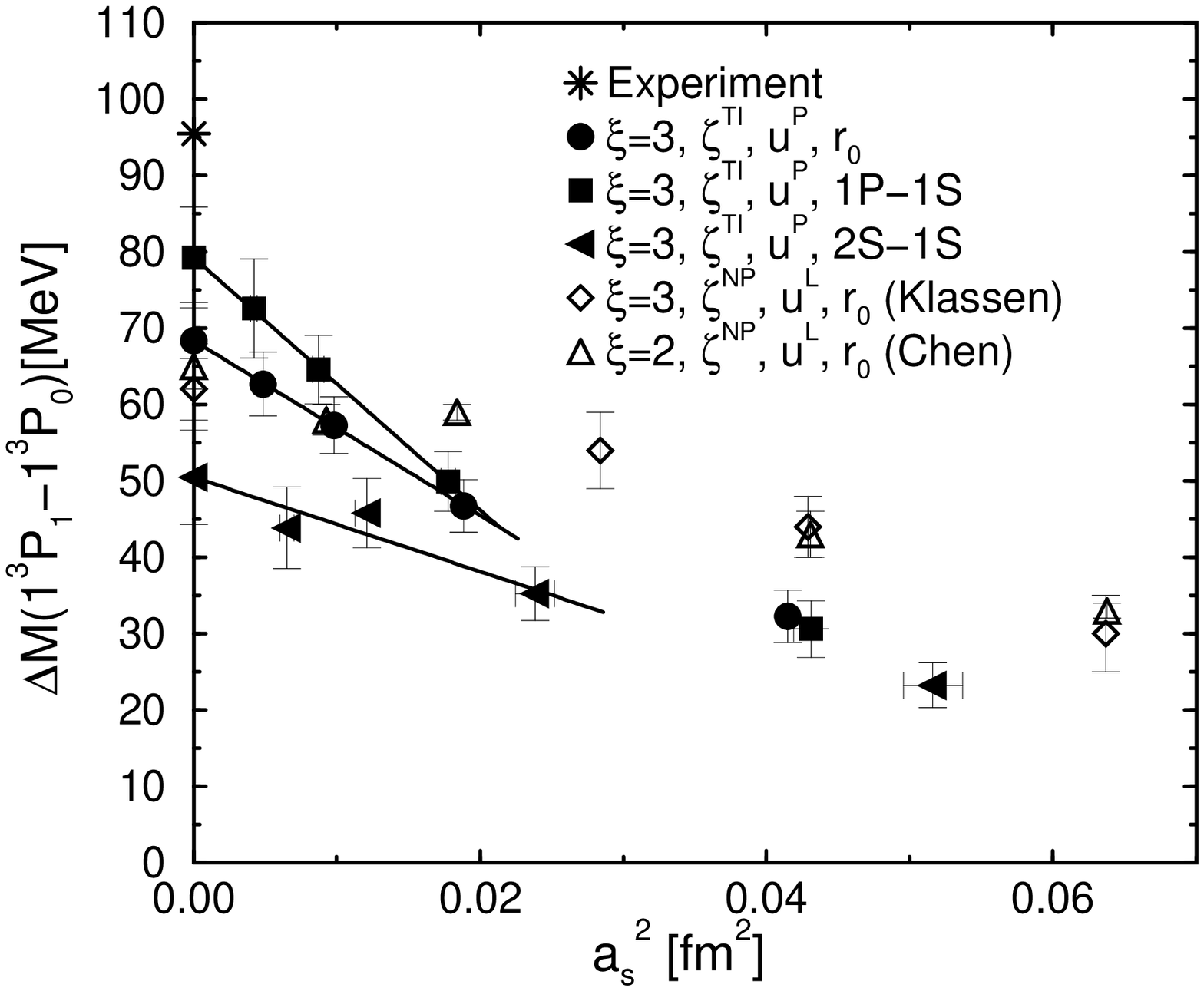}
    \caption{P-state fine structure splitting
    $\dM(1^3P_1-1^3P_0)$. }
    \label{fig:Psp}
   \end{center}
\end{figure}

\begin{figure}[t]
  \begin{center}
    \leavevmode
    \epsfxsize=0.65 \hsize
    \epsffile{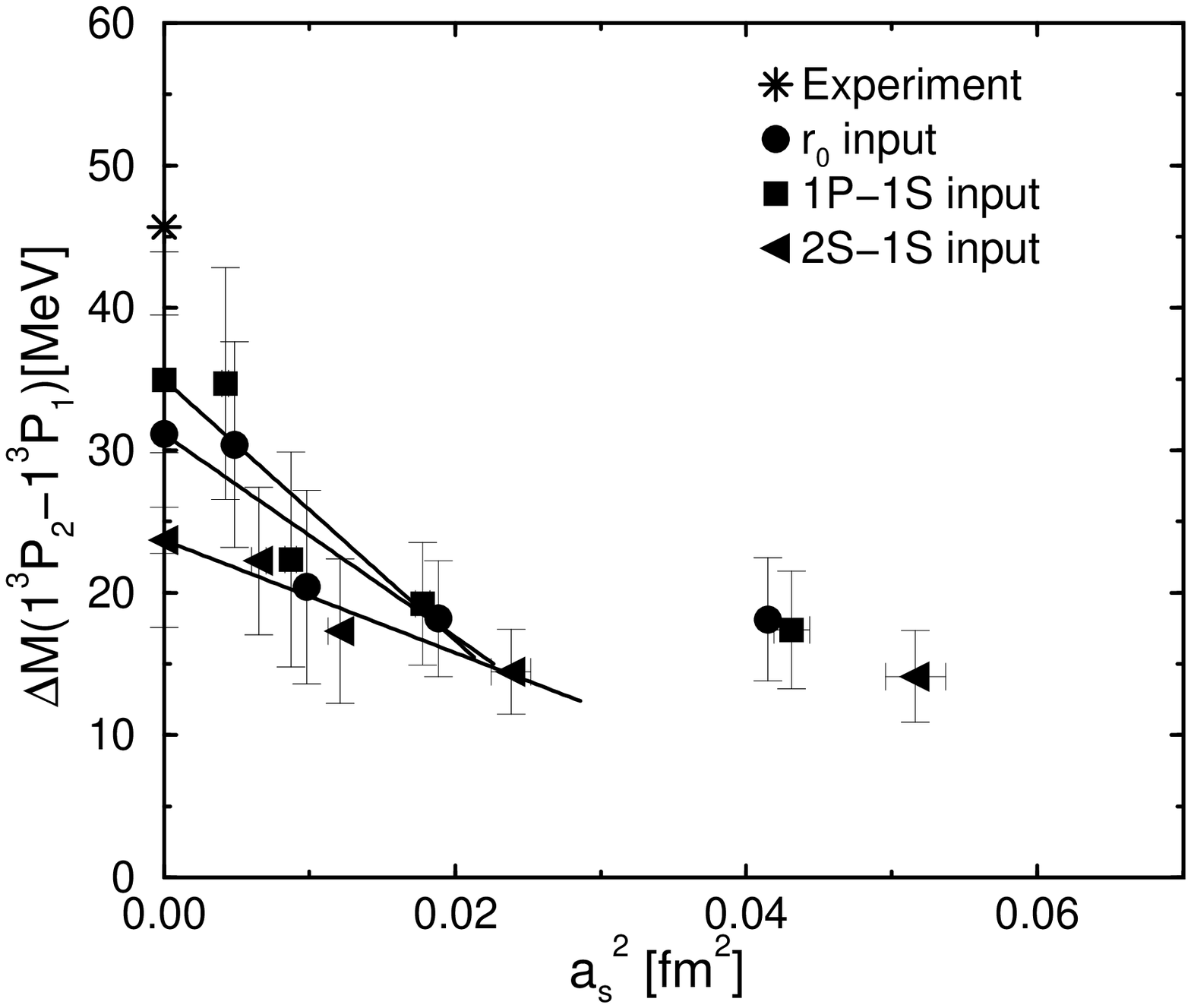}
    \caption{P-state fine structure splitting
    $\dM(1^3P_2-1^3P_1)$. }
    \label{fig:Psp2}
   \end{center}
\end{figure}

Results for the P-state fine structure are shown in Figs.~\ref{fig:Psp} and
\ref{fig:Psp2}.
The value of the P-state fine structure in the continuum limit and  
the systematic errors 
are estimated in a similar manner to the case of the S-state HFS.
For $1^3P_1-1^3P_0$ splitting, we obtain
\begin{equation}
\dM(1^3P_1-1^3P_0) \ = \ \left\{ \begin{array}{ll}
68.4(5.0)(+11.8)(-3.0) \; {\rm MeV} & \mbox{($r_0$ input)} \\
79.2(6.6)(+16.5)(-2.4) \; {\rm MeV} & \mbox{($1\bar{P} - 1\bar{S}$ input)} \\
50.5(6.2)(+7.9)(-2.2) \; {\rm MeV} & \mbox{($2\bar{S} - 1\bar{S}$ input)}. \\
95.5(0.8) \; {\rm MeV} & \mbox{(experiment)} \\
\end{array} \right.
\label{Psp1cont}
\end{equation}
Note that the systematic errors from the choice of the fit
ansatz (second error) are rather large here, due to the large scaling violation
seen in Fig.~\ref{fig:Psp}.
The result with the $1\bar{P} - 1\bar{S}$ input
yields a 17\% (2.5$\sigma$) smaller value than the experiment.
Our result with the $r_0$ input is consistent with the previous results by 
Klassen\cite{Klassen} and Chen\cite{Chen:2001ej}.

For $1^3P_2-1^3P_1$ splitting, we obtain
\begin{equation}
\dM(1^3P_2-1^3P_1) \ = \ \left\{ \begin{array}{ll}
31.1(8.4)(+8.1)(-1.0) \; {\rm MeV} & \mbox{($r_0$ input)} \\
35.0(9.0)(+9.6)(-0.7) \; {\rm MeV} & \mbox{($1\bar{P} - 1\bar{S}$ input)} \\
23.7(6.1)(+5.6)(-0.8) \; {\rm MeV} & \mbox{($2\bar{S} - 1\bar{S}$ input)} \\
45.7(0.2) \; {\rm MeV} & \mbox{(experiment)} \\
\end{array} \right. ,
\label{Psp2cont}
\end{equation}
where we use the result from
the E representation operator for $^3P_2$. As observed in 
Tables \ref{tab:cc_r0}-\ref{tab:cc_2s1s}, the mass difference 
$\dM(1^3P_{2T}-1^3P_{2E})$ is always consistent with zero, suggesting that
the rotational invariance for this quantity is 
restored well in our approach.
The value of $\dM(1^3P_2-1^3P_1)$ 
is smaller than the experimental one by 23\% ($1\sigma$) 
with the $1\bar{P} - 1\bar{S}$ input.
There is no lattice result from the anisotropic 
relativistic approach to be compared with.

\begin{figure}[h]
  \begin{center}
    \leavevmode
    \epsfxsize=0.65 \hsize
    \epsffile{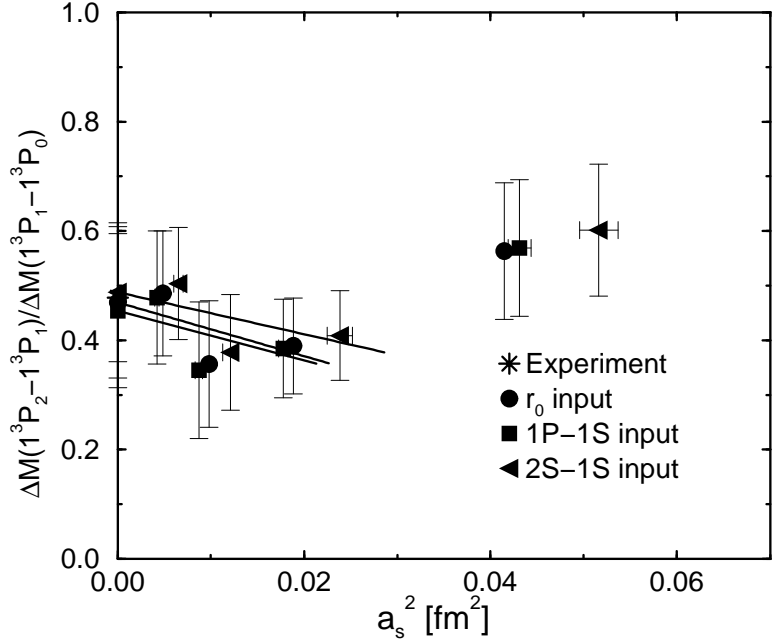}
    \caption{Fine structure ratio
    $\dM(1^3P_2-1^3P_1)/\dM(1^3P_1-1^3P_0)$. }
    \label{fig:ratP}
   \end{center}
\end{figure}

Next we consider the ratio of the two fine structures,
$\dM(1^3P_2-1^3P_1)/\dM(1^3P_1-1^3P_0)$.
In Fig.~\ref{fig:ratP}, we plot the lattice spacing dependence
of this ratio. As shown in this figure, the scaling violation of 
the ratio is smaller than that for the individual splittings
(Figs.~\ref{fig:Psp} and \ref{fig:Psp2}).
Moreover, results  are always consistent with the experimental value
within errors. Presumably this is in part due to 
a cancellation of systematic errors such as the discretization effect
and the quenching effect in the ratio.
Our continuum estimate of this ratio is 
\begin{equation}
\frac{\dM(1^3P_2-1^3P_1)}{\dM(1^3P_1-1^3P_0)} \  
= \ \left\{ \begin{array}{ll}
0.47(14)(+06) \;  & \mbox{($r_0$ input)} \\
0.45(14)(+05) \;  & \mbox{($1\bar{P} - 1\bar{S}$ input)} \\
0.49(13)(+06) \;  & \mbox{($2\bar{S} - 1\bar{S}$ input)} \\
0.48(00) \;  & \mbox{(experiment)} \\
\end{array} \right. .
\label{ratPcont}
\end{equation}
Our results agrees well with the experimental value.
We omit
the systematic error arising from the choice of $(M_{\rm lat},\zeta)$, which
is found to be much smaller than others.

\begin{figure}[t]
  \begin{center}
    \leavevmode
    \epsfxsize=0.65 \hsize
    \epsffile{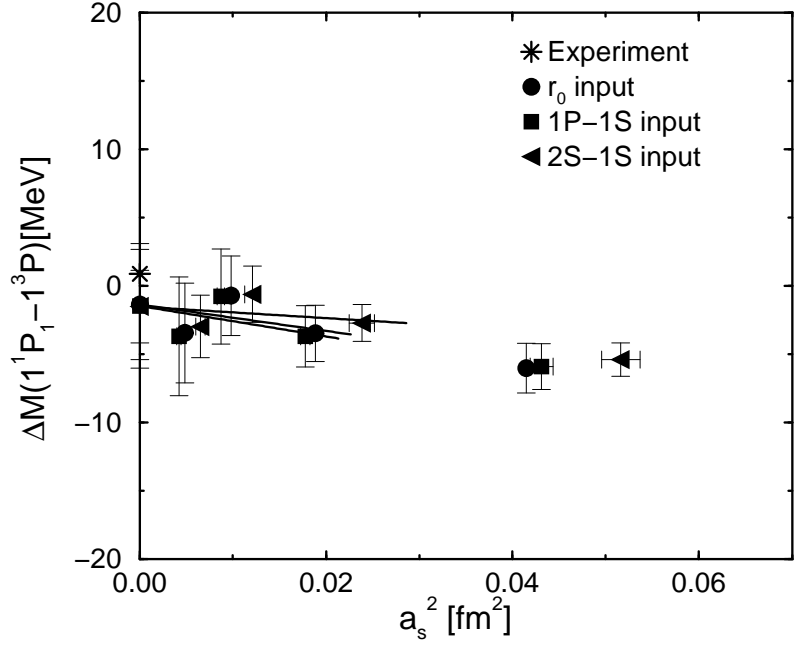}
    \caption{Splitting $\dM(1^1P_1-1^3P)$. }
    \label{fig:1P13PavNL}
   \end{center}
\end{figure}

Another interesting quantity is the P-state hyperfine splitting,
$\dM(1^1P_1-1^3P)$, where $M(1^3P) \equiv [5M(1^3P_2)+3M(1^3P_1)+M(1^3P_0)]/9$.
This should be much smaller than the S-state hyperfine splitting
because the P-state wavefunction vanishes at the origin. 
The lattice spacing dependence is shown in Fig.~\ref{fig:1P13PavNL}
and the continuum estimate is
\begin{equation}
\dM(1^1P_1-1^3P) \  
= \ \left\{ \begin{array}{ll}
-1.4(4.0)(+0.6) \; {\rm MeV} & \mbox{($r_0$ input)} \\
-1.5(4.6)(+0.7) \; {\rm MeV} & \mbox{($1\bar{P} - 1\bar{S}$ input)} \\
-1.5(2.6)(+0.3) \; {\rm MeV} & \mbox{($2\bar{S} - 1\bar{S}$ input)} \\
+0.9(0.3) \; {\rm MeV} & \mbox{(experiment)} \\
\end{array} \right. .
\label{hfsPcont}
\end{equation}
The sign is always negative at finite $a_s$ and in the continuum limit, 
but within errors the continuum value is consistent with the 
experimental value. 
We do not observe sizable differences between results using
different scale inputs for this quantity.

\subsection{$1P - 1S$ splitting}

\begin{figure}[t]
  \begin{center}
    \leavevmode
    \epsfxsize=0.65 \hsize
    \epsffile{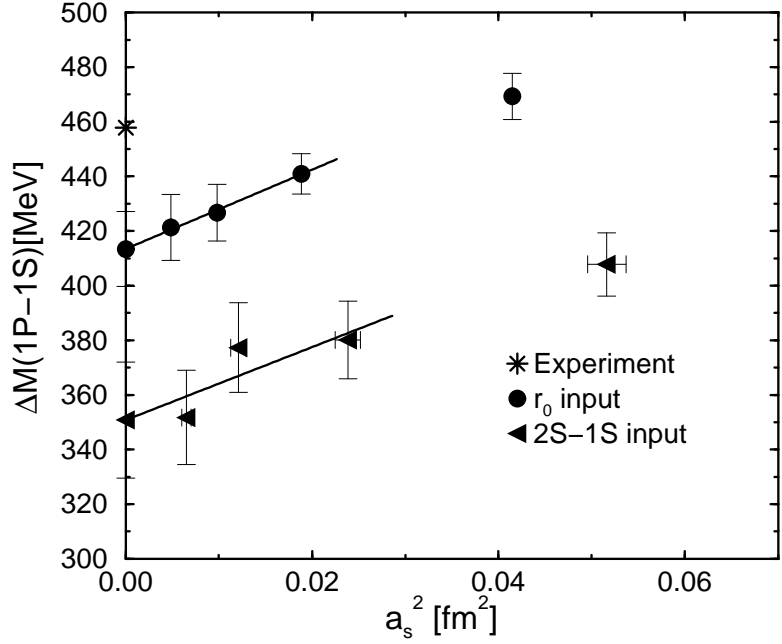}
    \caption{Spin averaged $1\bar{P}-1\bar{S}$ splitting. In the
   figures, we always omit the bar for the spin average.}
    \label{fig:PSavdifE}
   \end{center}
\end{figure}

\begin{figure}[t]
\centerline{\epsfxsize=9.5cm \epsfbox{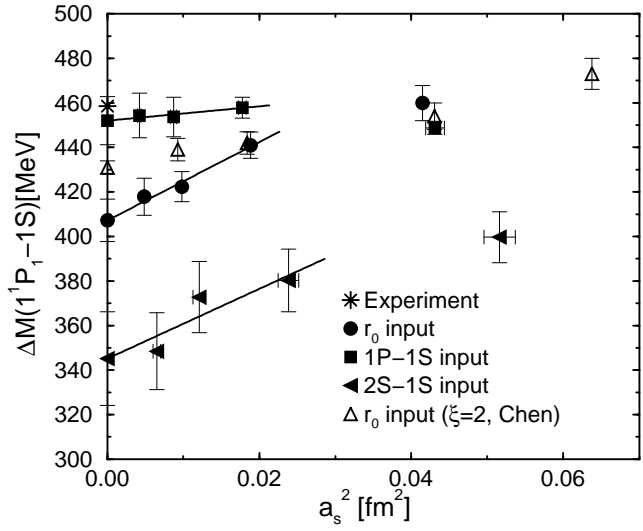}
            \epsfxsize=9.5cm \epsfbox{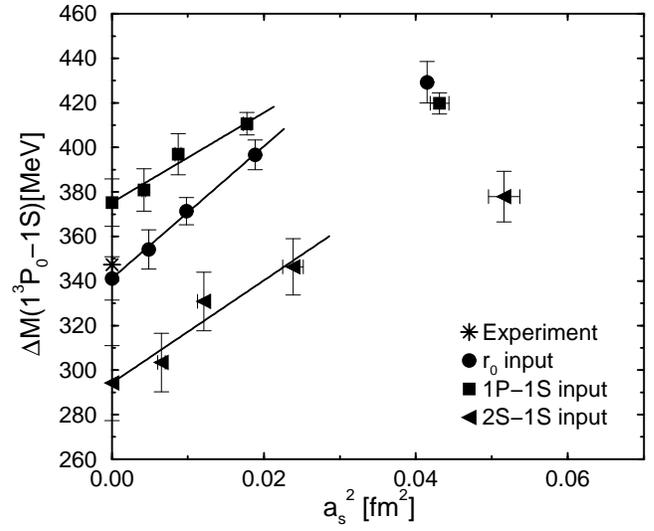}}
\centerline{\epsfxsize=9.5cm \epsfbox{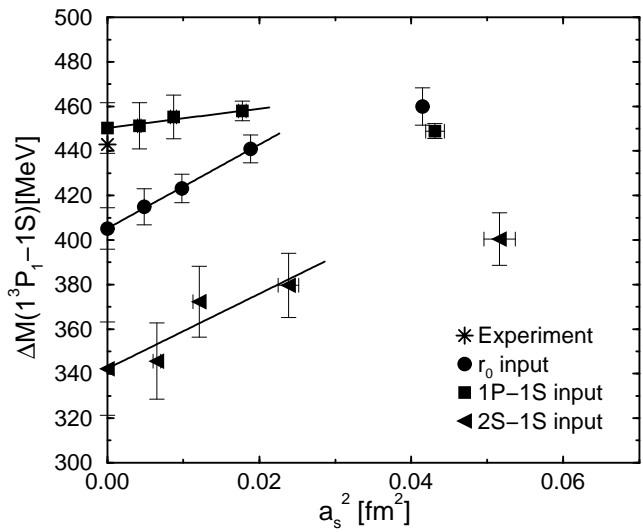}
            \epsfxsize=9.5cm \epsfbox{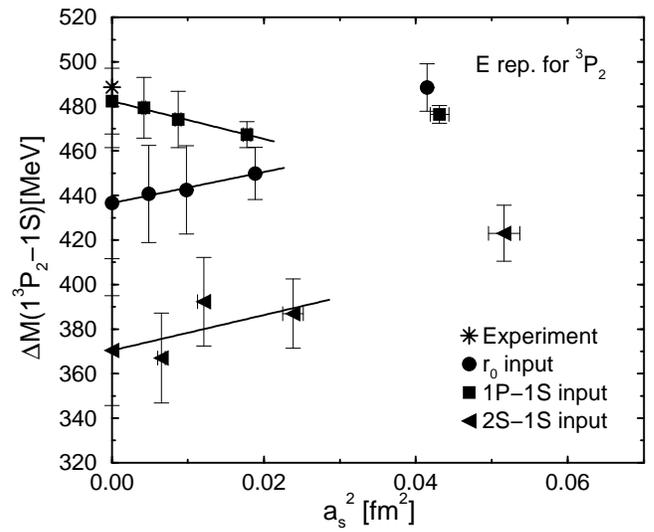}}
\caption{Spin dependent $1P-1\bar{S}$ splittings. }
\label{fig:PSdif}
\end{figure}

The mass splittings between the orbital (radial) exited state
and the ground state such as the $1P - 1S$ ($2S - 1S$) splitting 
are dominated by the kinetic term 
in the non-relativistic Hamiltonian Eq.~(\ref{NRHamiltonian}).
Since the dependence on the choice of $(M_{\rm lat},\zeta)$ is small compared 
to the statistical error, as shown in Fig.~\ref{fig:Comptuning}, 
we ignore the systematic error from 
the choice of $(M_{\rm lat},\zeta)$ in this and next subsections.
Results of the spin-averaged and spin-dependent $1P - 1S$ splittings
are shown in Figs.~\ref{fig:PSavdifE} and \ref{fig:PSdif}.
In the continuum limit, the spin-averaged $1P - 1S$ splitting is 
\begin{equation}
\dM(1\bar{P} - 1\bar{S}) \  
= \ \left\{ \begin{array}{ll}
413(14)(-15) \; {\rm MeV} & \mbox{($r_0$ input)} \\
351(21)(-20) \; {\rm MeV} & \mbox{($2\bar{S} - 1\bar{S}$ input)} \\
458(01) \; {\rm MeV} & \mbox{(experiment)} \\
\end{array} \right. .
\label{PSavdifEcont}
\end{equation}
The spin-dependent $1P - 1S$ splitting
deviates from the experimental value by 0-10\% (1-5$\sigma$) with
the $r_0$ input and 15-25\% (3-5$\sigma$) with the $2S - 1S$ input,
as shown in Fig.~\ref{fig:PSdif}.
The result of the $1^1P_1 - 1\bar{S}$ splitting with the $r_0$ input
agrees with the result by Chen within a few $\sigma$ in the continuum limit.

\subsection{$2S - 1S$ and $2P - 1P$ splittings}

\begin{figure}[t]
  \begin{center}
    \leavevmode
    \epsfxsize=0.65 \hsize
    \epsffile{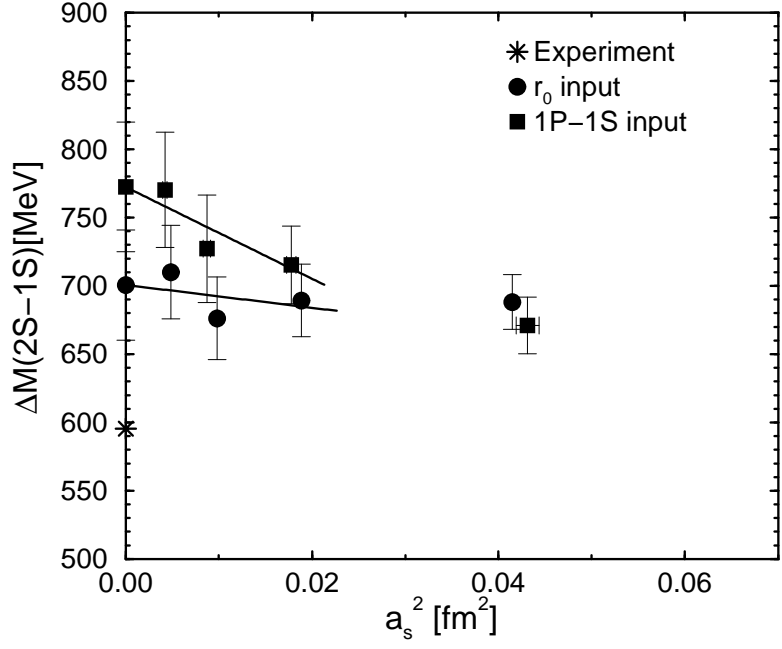}
    \caption{Spin averaged $2\bar{S}-1\bar{S}$ splitting. }
    \label{fig:EXdifSav}
   \end{center}
\end{figure}

\begin{figure}[t]
\centerline{\epsfxsize=9.5cm \epsfbox{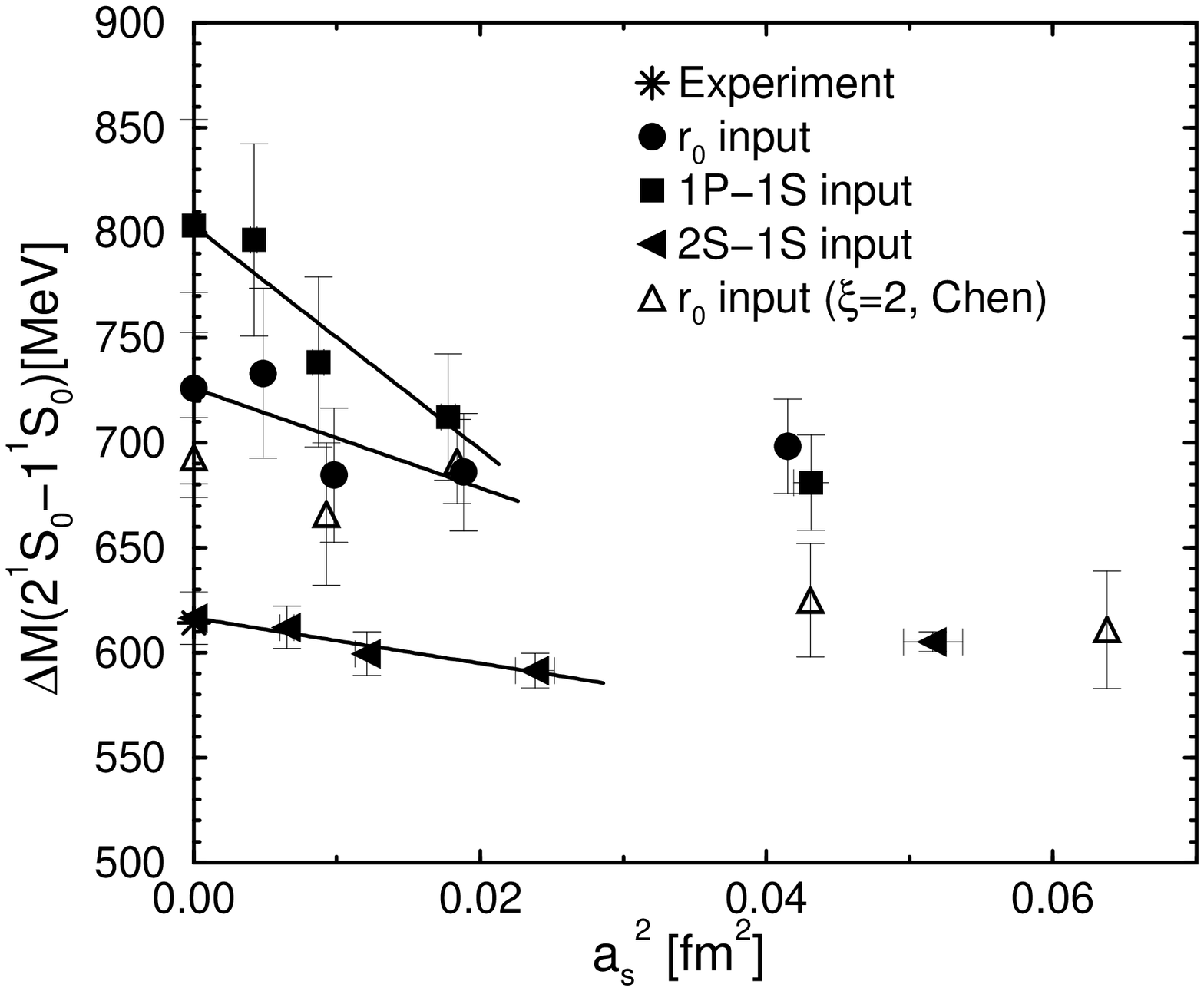}
            \epsfxsize=9.5cm \epsfbox{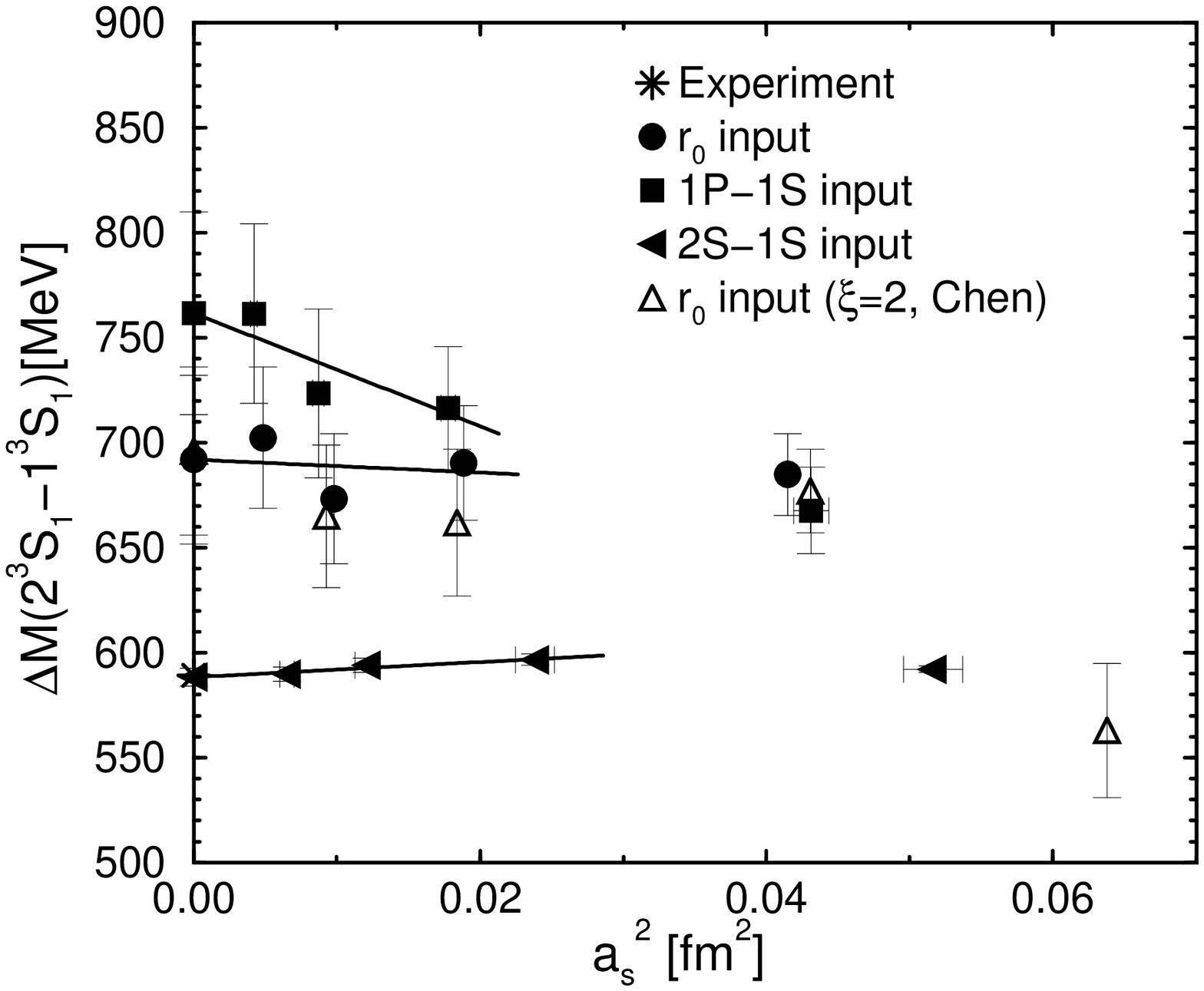}}
\caption{Spin dependent $2S-1S$ splittings. }
\label{fig:EXdifS}
\end{figure}

\begin{figure}[t]
\centerline{\epsfxsize=9.5cm \epsfbox{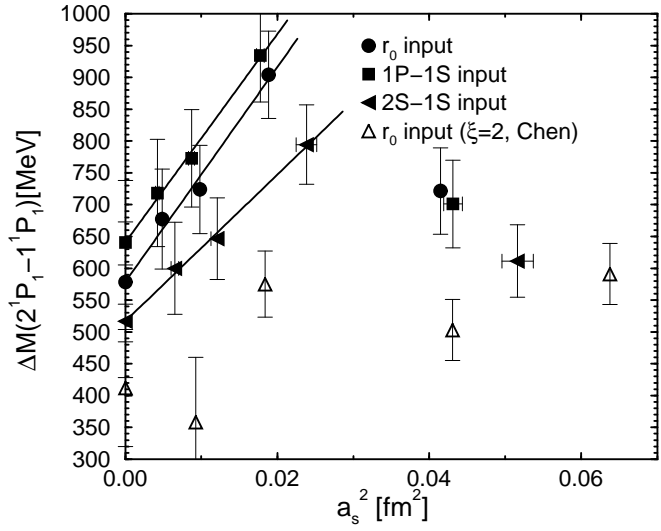}
            \epsfxsize=9.5cm \epsfbox{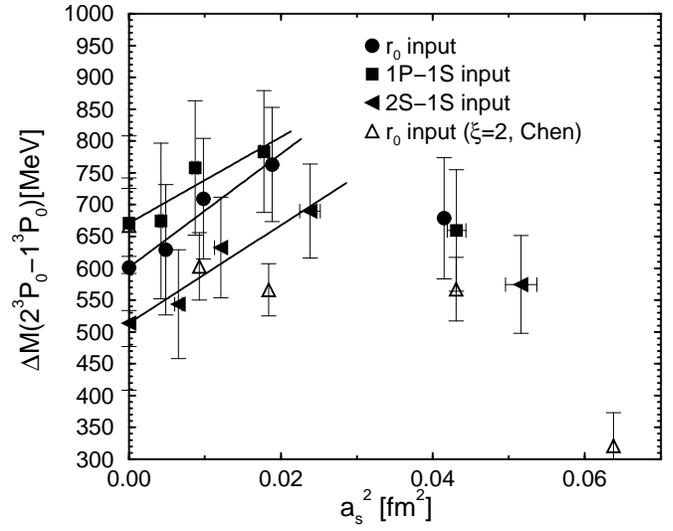}}
\centerline{\epsfxsize=9.5cm \epsfbox{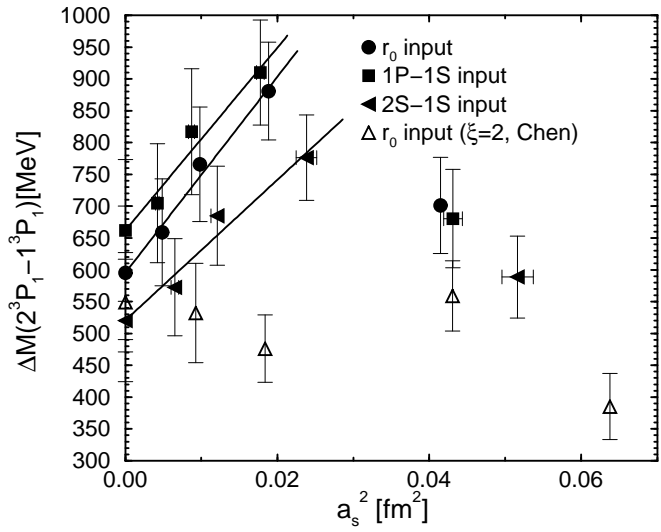}
            \epsfxsize=9.5cm \epsfbox{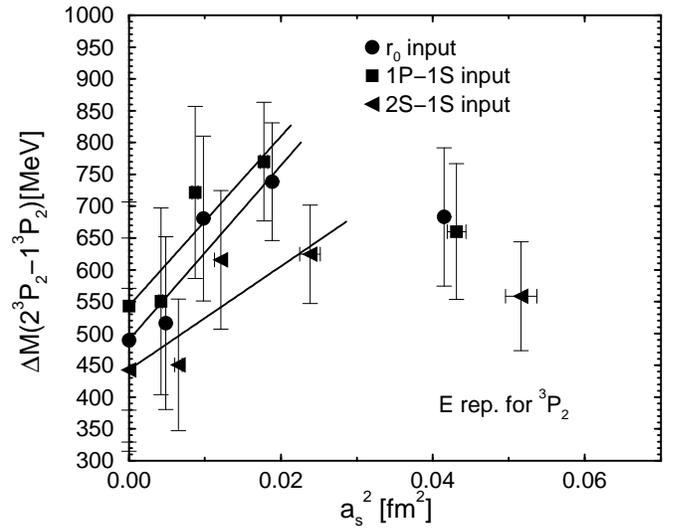}}
\caption{Spin dependent $2P-1P$ splittings.}
\label{fig:EXdifP}
\end{figure}

In Figs.~\ref{fig:EXdifSav} and \ref{fig:EXdifS}, we show 
the results of the spin-averaged and spin-dependent $2S - 1S$ splittings.
In the continuum limit, these splittings
deviate from the experimental values by 
$\sim$20\% (2.5$\sigma$) with the $r_0$ input and $\sim$30\% (4$\sigma$) 
with the $1\bar{P} - 1\bar{S}$ input.
For the spin-averaged $2S - 1S$ splitting, we obtain
\begin{equation}
\dM(2\bar{S} - 1\bar{S}) \  
= \ \left\{ \begin{array}{ll}
701(40)(+13) \; {\rm MeV} & \mbox{($r_0$ input)} \\
772(47)(+35) \; {\rm MeV} & \mbox{($1\bar{P} - 1\bar{S}$ input)} \\
595(01) \; {\rm MeV} & \mbox{(experiment)} \\
\end{array} \right. .
\label{EXdifSav}
\end{equation}
Besides quenching effects, possible sources of the deviations are 
finite size effects and the mixing of the $2S$ with
higher excited states.
Figure \ref{fig:EXdifP} shows the result for $2P - 1P$ splittings.
Note that there is no experimental value for this splitting
at present.
Our results of $2S - 1S$ and $2P - 1P$ splittings are 
consistent with previous results by Chen.
We also calculate mass splittings such as 
$\dM(2^3S_1-2^1S_0)$ and $\dM(2\bar{P}-2\bar{S})$, but these  
suffer from large statistical and systematic errors.
We leave accurate determinations of the excited state masses 
for future studies.

\subsection{The charmonium spectrum in the continuum limit}\label{sec:cc_cont}

\begin{figure}[h]
  \begin{center}
    \leavevmode
    \epsfxsize=0.90 \hsize
    \epsffile{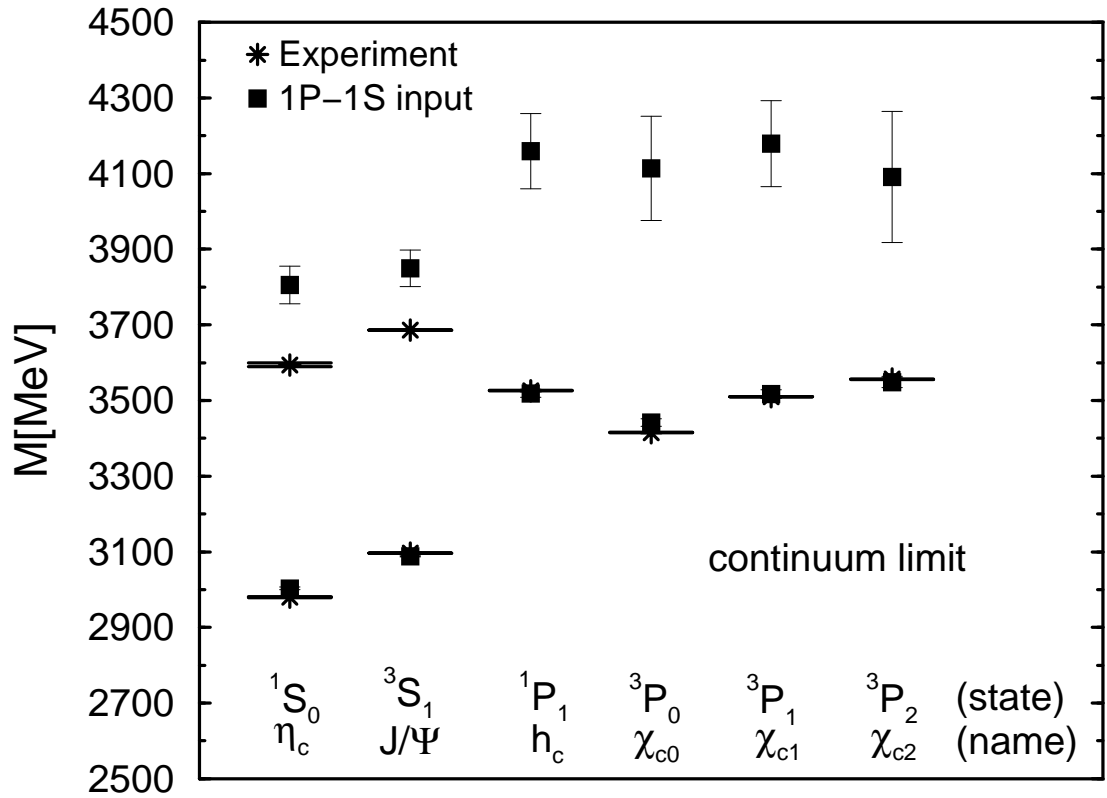}
    \caption{Charmonium spectrum in the continuum limit. 
The scale is set by $1\bar{P}-1\bar{S}$ splitting.}
    \label{fig:spectrumcont}
  \end{center}
\end{figure}

We summarize the continuum results
for the charmonium spectra obtained with the data of 
$(M_{\rm pole},\zeta^{\rm TI})$ and the $a_s^2$-linear fit ansatz
in Fig.~\ref{fig:spectrumcont},
where the scale is set by $1\bar{P}-1\bar{S}$ splitting.
Numerical values for three scales
are listed in Tables \ref{tab:cc_r0}-\ref{tab:cc_2s1s},
where the errors are only statistical. 
Among three different scales, results with the $1P - 1S$ input are the
closest to the experimental value for the ground state masses. 
The spin splittings such as the hyperfine splitting $\dM(1^3S_1-1^1S_0)$
and the fine structure $\dM(1^3P_1-1^3P_0)$ are always smaller than
the experimental values irrespective of the choice of the scale input, 
which is considered to be quenching effects.

\section{Effect of clover coefficient for hyperfine splitting}
\label{sec:hfsrevisit}

We now come back to the issue of the hyperfine splitting.
In Sec.\ref{sec:hfs}, we have shown that our result of 
the HFS (set A in Table \ref{tab:simparam3}) 
agrees with previous results by Klassen (set B) 
and Chen (set C)
in the continuum limit, with the same choice of the clover coefficients 
Eqs.~(\ref{TImasscB}) and (\ref{TIcst}).
However, as mentioned in the 
Introduction, when Klassen made a different choice of the clover
coefficients (set D), he obtained apparently 
different values of the HFS in the continuum limit.
This choice is given by
\footnote{
This choice corresponds to $\tilde{\omega} = 1$ in the mass
form notation Eq.~(\ref{massform}), while the correct choice 
$\tilde{c_s} = 1$ corresponds to $\tilde{\omega} = \nu$.
}
$\tilde{c_s} = 1/\nu$
where the tilde denotes the tadpole improvement, 
$\tilde{c_s}=u_s^3c_s$. 
Since $\nu \rightarrow 1$ as 
$a_sm_q \rightarrow 0$, it agrees with the correct 
choice $\tilde{c_s} = 1$ in the limit
$a_s \rightarrow 0$ with fixed $m_q$, but is incorrect at finite $a_s$.
The quark action then generates an additional 
$O(a_s^2\Lambda_{\rm QCD}m_q)$ error.
Even with such a choice, {\it if} $a_sm_q$ is small enough, the result 
should converge to a universal value after the continuum
extrapolation. 
However, in Refs.\cite{Klassen:1999fh,Klassen}, Klassen obtained 
${\rm HFS}(a_s=0, r_0~{\rm input}) \approx 95$ MeV with 
$\tilde{c_s} = 1/\nu$, which is much larger than the result 
${\rm HFS}(a_s=0, r_0~{\rm input}) \approx 75$ MeV 
with $\tilde{c_s} = 1$ both by Klassen 
and in the present work.

A possible source of this discrepancy is a large 
mass-dependent error of $O(a_s\Lambda_{\rm QCD}\cdot (a_sm_q)^n)$
($n=1,2,\cdots$)
for the results with $\tilde{c_s} = 1/\nu$.
In fact, Klassen adopted rather coarse lattices with 
$a_sm_q \approx 1-$2, for which such errors may not be negligible. 
Because the HFS is sensitive to the spatial clover term,
the choice of $\tilde{c_s} = 1/\nu$ may then result in
a non-linear $a_s$ dependence for the HFS.
In the following, in order to study the effect of the choice of 
the spatial clover coefficient
$c_s$ to the HFS, we make a leading order analysis 
motivated by the potential
model \cite{Lucha:1991vn} and compare it with numerical results,
which will give us a better understanding of the above problem of the HFS.

The potential model predicts that, at the leading order in both $\alpha$ 
and $1/m_q$,
\begin{equation}
{\rm HFS}_{\rm cont} \sim 
\left({\frac{{\bf S}_q}{m_q}}\right) \cdot 
\left({\frac{{\bf S}_{\bar{q}}}{m_{\bar{q}}}}\right) |\Psi (0)|^2_{\rm cont}\ ,
\label{hfspot}
\end{equation}
where $m_q = m_{\bar{q}}$ for the quarkonium, 
${\bf S}_{q,\bar{q}}$ are quark and anti-quark spins, and 
$\Psi (0)$ is the wavefunction at origin.
${\rm HFS}_{\rm cont}$ is the hyperfine splitting in the continuum
quenched ($n_f=0$) theory, which is not necessarily equal to the experimental
value. 
In non-relativistic QCD, the ${\bf S}_q \cdot {\bf S}_{\bar{q}}$
interaction arises from the ${\bf\Sigma}\cdot{\bf B}$ term for
quark and anti-quark.
Giving the non-relativistic interpretation to 
our anisotropic lattice action, 
we expect that the lattice HFS is effectively given by
\begin{equation}
{\rm HFS}_{\rm lat} \sim \left({\frac{\bf\Sigma}{m_B}}\right) \cdot 
\left({\frac{\bf\Sigma}{m_B}}\right) |\Psi (0)|^2_{\rm lat} \ ,
\label{hfsNR}
\end{equation}
where $m_B$ is the magnetic mass Eq.~(\ref{magneticmass}) in
the effective Hamiltonian.
Therefore, in our approach, HFS is dominated by 
the magnitude of $1/m_B^2$, which depends on the spatial clover
coefficient $c_s$.
The ratio,  
\begin{equation}
 \frac{{\rm HFS}_{\rm lat\ \ }}{{\rm HFS}_{\rm cont}} \sim
\left(\frac{m_q}{m_B}\right)^2 \cdot
\frac{|\Psi (0)|_{\rm lat\ \ }^2}{|\Psi (0)|_{\rm cont}^2} \ ,
\end{equation}
generally deviates from 1 at finite $a_s$, and should approach 1 as 
$a_s \rightarrow 0$.
At the leading order in $\alpha$, $|\Psi (0)|_{\rm cont}^2\propto m_q$, while
$|\Psi (0)|_{\rm lat}^2\propto m_2$ with $m_2$ the kinetic mass 
Eq.~(\ref{kineticmass}).
Since $m_2$ does not depend on the spatial clover coefficient 
$c_s$ at the tree level, we neglect the lattice artifact for 
$|\Psi (0)|_{\rm lat}^2$
and set $|\Psi (0)|_{\rm lat}^2/|\Psi (0)|_{\rm cont}^2=1$
in the following, which is sufficient for the present purpose.
Now we define 
\begin{equation}
R_{\rm HFS} \equiv \left(\frac{m_q}{\tilde{m_B}}\right)^2 = 
\left(\frac{a_tm_q}{a_t\tilde{m_B}}\right)^2 \ ,
\label{effhfs}
\end{equation}
as a measure of lattice artifacts 
for the HFS,
where the tilde denotes the tadpole improvement.
In the continuum limit, $R_{\rm HFS}=1$.
Since $m_q$ is constant independent of $a_s$,
we identify $m_q$ with $\tilde{m_1}$ for the pole mass
tuning ({\it i.e.} when setting the 
measured pole mass to the experimental value 
$M_{\rm pole}=M_{\rm exp}$ for the meson), 
and with $\tilde{m_2}$ for the kinetic mass tuning 
($M_{\rm kin}=M_{\rm exp}$).

At the tree level with the tadpole improvement,
the pole mass $\tilde{m_1}$, the kinetic mass $\tilde{m_2}$ and
the magnetic mass $\tilde{m_B}$ for the quark are given by
\begin{eqnarray} 
a_t\tilde{m_1} &=& \log (1+\tilde{m_0}),\label{TIpolemass}\\
\frac{1}{a_t\tilde{m_2}} &=& \frac{2\nu^2}{\tilde{m_0}
(2+\tilde{m_0})}
+ \frac{\xi r_s \nu}{1+\tilde{m_0}}, 
\label{TIkineticmass}\\
\frac{1}{a_t\tilde{m_B}} &=& \frac{2\nu^2}{\tilde{m_0}
(2+\tilde{m_0})}
+ \frac{\xi \tilde{c_s} \nu}{1+\tilde{m_0}},
\label{TImagneticmass}
\end{eqnarray}
where $\nu =\xi_0/\zeta$, $\tilde{c_s}=u_s^3c_s$, and 
$\tilde{m_0}=a_t \tilde{m_{q0}}$ is given by Eq.~(\ref{TIbearemass}).
To obtain Eqs.~(\ref{TIkineticmass}) and (\ref{TImagneticmass}), 
we use the formula $\xi = \tilde{\xi_0} = (u_t/u_s)\xi_0$.
In the following we present the $a_sm_q$ dependence
of $R_{\rm HFS}$ in the case of $\tilde{c_s}=1$ (set A,B,C) and 
$\tilde{c_s}=1/\nu$ (set D), and compare them with
the corresponding numerical data for the S-state HFS.
For the definition of $\zeta$ (or $\nu$), 
there are two choices adopted so far, 
the tree level tadpole improved value $\zeta^{\rm TI}$ and
nonperturbative one $\zeta^{\rm NP}$.
At $\zeta=\zeta^{\rm TI}$, $\tilde{m_1}=\tilde{m_2}$ for the quark, but
$M_{\rm pole} \neq M_{\rm kin}$ for the measured meson.
On the other hand, at $\zeta=\zeta^{\rm NP}$,
$\tilde{m_1} \neq \tilde{m_2}$ though $M_{\rm pole} = M_{\rm kin}$.
Thus in the case of $\zeta=\zeta^{\rm NP}$, {\it i.e.} 
$M_{\rm pole} = M_{\rm kin}$ tuning, 
the identification of $m_q$ (=$\tilde{m_1}$ or $\tilde{m_2}$)
in $R_{\rm HFS}$ Eq.~(\ref{effhfs}) mentioned above is ambiguous. 
Although such an ambiguity should vanish in the continuum limit,
we present $R_{\rm HFS}$ with both 
$m_q=\tilde{m_1}$ and $m_q=\tilde{m_2}$ to check 
consistency. 
For actual numerical data of the HFS, 
we focus on the results with the $r_0$ input 
because Klassen has adopted $r_0$ for the scale setting.

\subsection{The case of $\tilde{c_s} = 1/\nu$
}\label{sec:HFS_TK_lat98}

\begin{figure}[t]
  \begin{center}
    \leavevmode
    \epsfxsize=0.65 \hsize
      \epsffile{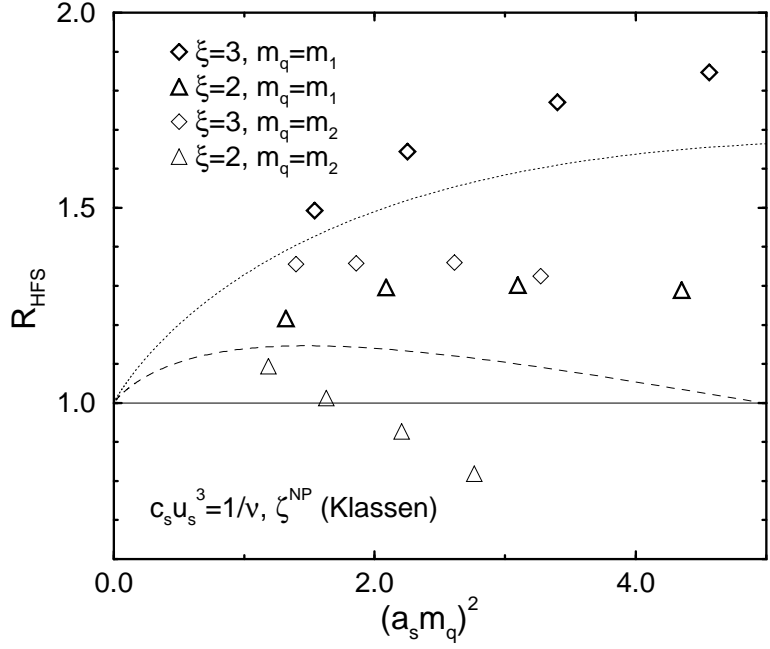}
    \caption{
$R_{\rm HFS}$ with $\tilde{c_s} = 1/\nu$ and 
$\zeta = \zeta^{\rm NP}$ at $\xi=3$ and 2.
The thick symbols are the results with $m_q=\tilde{m_1}$, 
while the thin symbols are those with $m_q=\tilde{m_2}$.
The results with $\tilde{c_s} = 1/\nu$ but $\zeta = \zeta^{\rm TI}$ 
(where $m_q=\tilde{m_1}=\tilde{m_2}$)
are also shown by the dotted line ($\xi =3$) and dashed line ($\xi =2$).}
    \label{fig:effhfs_ZNP_TK}
   \end{center}
\end{figure}

First we consider the case of $\tilde{c_s} = 1/\nu$ (set D),
which is correct only for $a_sm_q=0$ at the tree level.
In Fig.~\ref{fig:effhfs_ZNP_TK} we plot the $(a_sm_q)^2$ dependence of 
$R_{\rm HFS}$ at $\xi =3$ and 2 for  
$\tilde{c_s} = 1/\nu$ with $\nu=\nu^{\rm NP}=\xi_0/\zeta^{\rm NP}$.
Numerical values of $\nu^{\rm NP}$ were taken from Ref.\cite{Klassen}.
Because of the ambiguity for $m_q$ mentioned above, we show 
the results with $m_q=\tilde{m_1}$ and $m_q=\tilde{m_2}$;
the difference between them decreases as $a_s \rightarrow 0$, 
as expected.
We have checked that plotting $R_{\rm HFS}$ as a function of $a_s^2$,
instead of $(a_sm_q)^2$, does not change the figure qualitatively.
We also plot the results with $\tilde{c_s} = 1/\nu$ but 
$\nu=\nu^{\rm TI}=\xi_0/\zeta^{\rm TI}$, where $\tilde{m_1}=\tilde{m_2}$
holds, as a dotted line ($\xi =3$) and a dashed line ($\xi =2$) for 
a guide to the eye.
As shown in this figure, $R_{\rm HFS}$ has a non-linear $a_s^2$
dependence toward the continuum limit (=1),
indicating that the mass dependent error is large for the region
$a_sm_q=1-$2.
$R_{\rm HFS}$ is larger than 1 even at $(a_sm_q)^2 \sim 1$,
which suggests that the
actual HFS should rapidly decrease toward $a_s^2 \rightarrow 0$, 
and data at $(a_sm_q)^2 < 1$ 
are needed for a reliable continuum extrapolation
for the HFS.

\begin{figure}[t]
  \begin{center}
    \leavevmode
    \epsfxsize=0.65 \hsize
     \epsffile{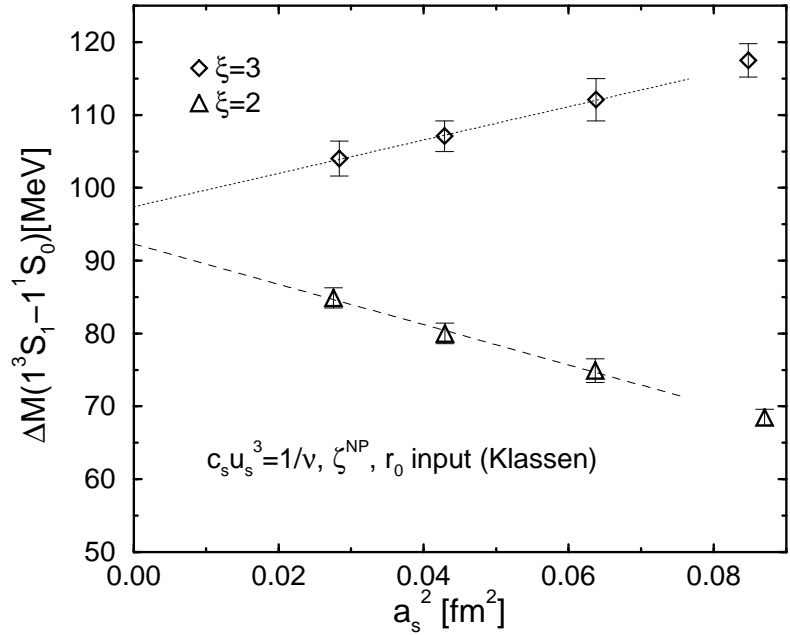}
    \caption{Klassen's results of 
     S-state hyperfine splitting $\dM(1^3S_1-1^1S_0)$ with  
     $\tilde{c_s} = 1/\nu$ and $\zeta = \zeta^{\rm NP}$ (set D). 
     The scale is set by $r_0$. Lines denote $a_s^2$-linear extrapolations.}
    \label{fig:hfsKlassenLat98}
   \end{center}
\end{figure}

Now let us compare $R_{\rm HFS}$ with numerical results of HFS.
In Fig.~\ref{fig:hfsKlassenLat98}, we plot 
corresponding results of HFS 
by Klassen for $\tilde{c_s} = 1/\nu$\cite{Klassen}.
The results at $\xi=3$ for $\tilde{c_s} = 1/\nu$ 
are clearly larger than the results for $\tilde{c_s} = 1$
(see filled circles in Fig.~\ref{fig:hfs}),
and the results at $\xi=3$ and 2 {\it appear} 
to converge to $\approx$ 95 MeV
in the continuum limit with an $a_s^2$-linear scaling.
However, comparing Fig.~\ref{fig:effhfs_ZNP_TK} and
Fig.~\ref{fig:hfsKlassenLat98}, 
we find that the lattice spacing dependence of the
numerical data of HFS qualitatively agrees with that of $R_{\rm HFS}$: 
for both HFS and $R_{\rm HFS}$, 
data at $\xi=3$ are larger than data at $\xi=2$, and the difference 
between $\xi=3$ and 2 decreases as $a_s \rightarrow 0$.
From an $a_s^2$-linear extrapolation of $R_{\rm HFS}$
using the finest three data points, we obtain 
$R_{\rm HFS} \approx 1.2-$1.3 at $a_s=0$.
Because the correct continuum limit of $R_{\rm HFS}$ is 1, 
this suggests a $20-$30\% overestimate from the neglect of non-linear 
dependence of $R_{\rm HFS}$ on $a_s^2$. Hence
the result with $\tilde{c_s} = 1/\nu$, HFS($a_s=0$) $\approx$ 95 MeV,
reported in Refs.\cite{Klassen:1999fh,Klassen} 
is likely an overestimate by $20-$30\%.

These analyses indicate
that the origins of this overestimate
are, first, 
the choice for the spatial clover coefficient
$\tilde{c_s} = 1/\nu \ (=1/\nu^{\rm NP})$, and second, the use of 
coarse lattices with $a_sm_q > 1$.
As shown in Fig.~\ref{fig:tunedzeta}, $\nu$ ($=1/\tilde{c_s}$ in this case)
should eventually start to 
move up to 1 linearly 
around $a_tm_{q0}^{\rm TI}\lsim 0.3$, which corresponds to
$(a_sm_q)^2\lsim 0.6$ in Fig.~\ref{fig:effhfs_ZNP_TK}, 
but Klassen's data of $\nu^{\rm NP}$ (open diamonds)
do not reach such a region. 
We conclude that
the continuum extrapolation for the HFS should not be performed using 
the data on such coarse lattices, and 
results at finer lattice spacing are required.

\subsection{The case of $\tilde{c_s} = 1$}
\label{sec:HFS_Mpole}

Next we consider the case of 
$\tilde{c_s} = 1$ (set A, B and C),
which is correct for any $a_sm_q$ at the tree level.
In this case, there are two choices for $\zeta$, $\zeta^{\rm TI}$ and
$\zeta^{\rm NP}$. 
As mentioned in Sec.\ref{sec:effecttune}, 
$\tilde{m_B}=\tilde{m_2}$ holds for both choices of $\zeta$, 
with $\tilde{c_s} = 1$.

\begin{figure}[t]
  \begin{center}
    \leavevmode
    \epsfxsize=0.65 \hsize
    \epsffile{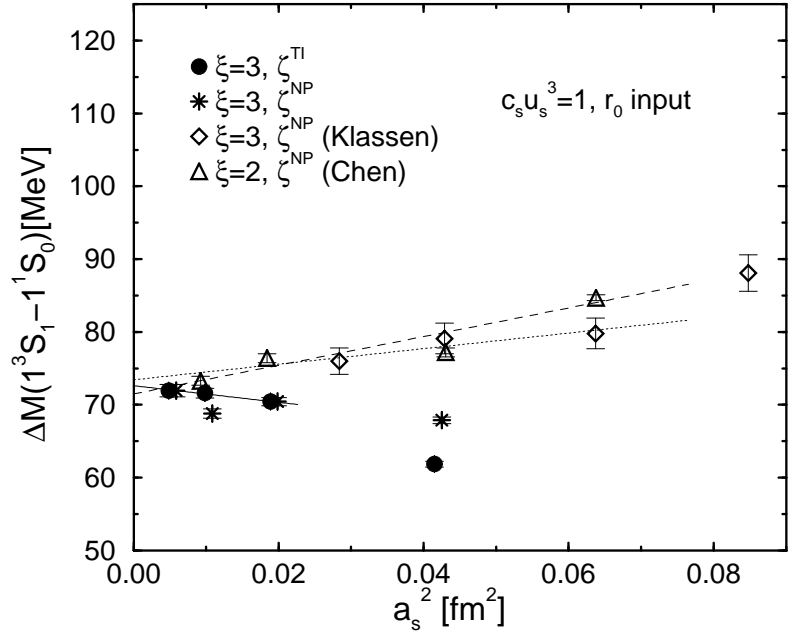}
    \caption{The results of 
     S-state hyperfine splitting $\dM(1^3S_1-1^1S_0)$ with $\tilde{c_s} = 1$.
     The scale is set by $r_0$.}
    \label{fig:comp_hfs}
   \end{center}
\end{figure}

In the case of $\zeta = \zeta^{\rm TI}$, which is adopted only in our
work (set A) so far,
$R_{\rm HFS} = 1$ is always satisfied,
since $\tilde{m_1} = \tilde{m_2} = \tilde{m_B}$ by definition.
This suggests that the scaling violation of HFS 
for $\tilde{c_s} = 1$ should be much smaller than 
that for $\tilde{c_s} = 1/\nu$.
The numerical result for the HFS with the pole mass tuning
has already been shown in Fig.~\ref{fig:hfs}
and re-plotted in Fig.~\ref{fig:comp_hfs} by filled circles, 
which gives our best estimate, HFS($a_s=0$) $= 73$ MeV.

\begin{figure}[t]
  \begin{center}
    \leavevmode
    \epsfxsize=0.65 \hsize
      \epsffile{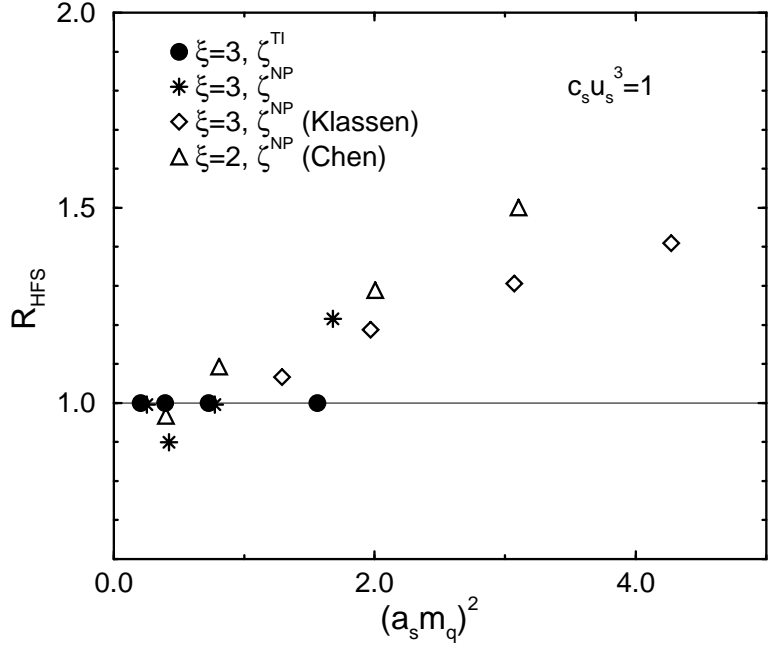}
    \caption{
$R_{\rm HFS}$ with $\tilde{c_s} = 1$.  Here $m_q=\tilde{m_1}$.
The stars are slightly shifted along the
 horizontal axis for the distinguishability.}
  \label{fig:effhfs_ZNP}
   \end{center}
\end{figure}

We next consider 
the case of $\zeta = \zeta^{\rm NP}$, where 
$M_{\rm pole}=M_{\rm kin}$ for the measured meson.
When we identify $m_q=\tilde{m_2}$, $R_{\rm HFS} = 1$ is always 
satisfied again because $\tilde{m_2} = \tilde{m_B}$ even at 
$\zeta = \zeta^{\rm NP}$.
When we identify $m_q=\tilde{m_1}$, $R_{\rm HFS} \not= 1$ in general,
due to the deviation of $\zeta^{\rm NP}$ from $\zeta^{\rm TI}$. 
The results of $R_{\rm HFS}$ with $m_q=\tilde{m_1}$
at $\zeta = \zeta^{\rm NP}$ are shown in Fig.~\ref{fig:effhfs_ZNP},
and corresponding numerical results for the HFS are shown in 
Fig.~\ref{fig:comp_hfs}.
Comparing Fig.~\ref{fig:effhfs_ZNP} with Fig.~\ref{fig:comp_hfs}
we again note that the lattice spacing dependence of the HFS
qualitatively agrees with that of $R_{\rm HFS}$, {\it i.e.},
for both HFS and $R_{\rm HFS}$, 
data at $\xi=3$ by Klassen (open diamonds, set B) and those at $\xi=2$ 
by Chen (open triangles, set C) are close to each other and 
larger than our data at $\zeta =\zeta^{\rm TI}$.
An $a_s^2$-linear extrapolation using the finest three data points gives
HFS $\approx 70-$75 MeV and $R_{\rm HFS} \approx 0.9-$1.0 at $a_s=0$.
The latter confirms that 
a continuum estimate of HFS with $\tilde{c_s} = 1$ is more reliable
than that with $\tilde{c_s} = 1/\nu$.

Concerning our results at $\xi=3$, 
as shown in Fig.~\ref{fig:effhfs_ZNP},
$R_{\rm HFS}$ for $\zeta =\zeta^{\rm NP}$ (stars) does not scale 
smoothly around $(a_sm_q)^2 \lsim 1$, 
while that for $\zeta =\zeta^{\rm TI}$ (filled circles) is always unity.
This behavior is caused by the fact that
the difference, $\zeta^{\rm NP}-\zeta^{\rm TI}$, is not monotonic in $a_sm_q$
(see Fig.~\ref{fig:tunedzeta}).
Correspondingly the numerical value of the HFS,
displayed in Fig.~\ref{fig:comp_hfs},
also shows a slightly non-smooth lattice spacing dependence near 
$a_s^2 \sim 0$, 
which qualitatively agrees with the
$(a_sm_q)^2$ dependence of $R_{\rm HFS}$ in this region.
A possible source of this behavior is
the statistical error of $\zeta^{\rm NP}$ itself, 
because HFS ($R_{\rm HFS}$) is also sensitive to the value of 
$\zeta$ as well as $c_s$.
Due to this reason, we have not used the results with 
$\zeta = \zeta^{\rm NP}$ for our main analysis in Sec.\ref{sec:results}.

\section{Conclusion}\label{sec:conclusion}

In this article, we have investigated the properties of
anisotropic lattice QCD for heavy quarks 
by studying the charmonium spectrum in detail. 
We performed simulations adopting lattices finer than 
those in the previous studies by
Klassen and Chen, and made a more careful analysis for $O((a_sm_q)^n)$
errors. In addition, using derivative operators, 
we obtained the complete P-state fine structure,
which has not been addressed in the previous studies. 

From the tree-level analysis for the effective Hamiltonian,
we found that the mass dependent tuning of parameters
is essentially important.
In particular, with the choice of $r_s=1$ for the spatial Wilson coefficient, 
an explicit $a_sm_{q0}$ dependence remains for the parameters $\zeta$ and 
$c_{t}$ even at the tree level. Moreover
we have shown in the leading order analysis that,
unless the spatial clover coefficient $\tilde{c_s}$
is correctly tuned,
the hyperfine splitting has a large $O((a_sm_q)^n)$ errors,
which can explain  a large value of the hyperfine splitting 
in the continuum limit from rather coarse lattices in
the previous calculation by Klassen.
On the other hand, if $\tilde{c_s}$ is mass-dependently tuned,
the continuum extrapolation is expected to be smooth for the
hyperfine splitting.

Based on these observations, we employed
the anisotropic clover action with $r_s=1$ and tuned the parameters
mass-dependently at the tree level combined with the tadpole improvement.
We then computed the charmonium spectrum in the quenched approximation
on $\xi = 3$ lattices with spatial lattice spacings of $a_sm_q < 1$.
A fine resolution in the temporal direction enabled a precise determination 
of the masses of S- and P- states which is accurate enough to be compared
with the experimental values.
Our results are consistent with previous results at $\xi = 2$
obtained by Chen\cite{Chen:2001ej}, and 
the scaling behavior of the hyperfine splitting is well explained
by the theoretical analysis.  
We then conclude that the anisotropic clover action with
the mass-dependent parameters at the tadpole-improved tree level
is sufficiently accurate for the charm quark 
to avoid large discretization errors due to heavy quark.
We note, however, that $a_sm_q < 1$ is still necessary for a reliable 
continuum extrapolation.

We found in our results that 
the gross features of the spectrum are consistent with the experiment. 
Quantitatively, however, the S-state hyperfine splitting deviates from 
the experimental value by about 30\% (7$\sigma$), 
and the P-state fine structure differs by about 20\% (2.5$\sigma$), 
if the scale is set from the $1\bar{P}-1\bar{S}$ splitting.
We consider that a major source for these deviations is 
the quenched approximation.

Certainly further investigations are necessary to conclude that
the anisotropic QCD can be used for quarks heavier than the charm.
In particular it is important to determine the clover coefficients
as well as other parameters non-perturbatively, since
the spin splittings are very sensitive to the clover coefficients.
It is also interesting to calculate the spectrum with $r_s=1/\xi$ 
and compare the result with the current one in this paper,
since the notorious $a_sm_{q0}$ dependence vanishes from the 
parameters with this choice at the tree level. 
Finally full QCD calculations including dynamical quarks are 
needed to establish the theoretical prediction without systematic errors 
for an ultimate comparison with the experimental spectrum.

\section*{Acknowledgments}

This work is supported in part by Grants-in-Aid of the Ministry of 
Education 
(Nos.~10640246, 
10640248, 
11640250, 
11640294, 
12014202, 
12304011, 
12640253, 
12740133, 
13640260). 
VL is supported by JSPS Research for the Future Program
(No. JSPS-RFTF 97P01102).
KN and M.~Okamoto are JSPS Research Fellows. 
M.~Okamoto would like to 
thank T.R.\ Klassen for giving us his manuscript 
of Ref.\cite{Klassen}. MO also would like to thank A.S.\ Kronfeld
for useful discussions.

\appendix

\section{Derivation of Hamiltonian on the anisotropic lattice}
\label{sec:hamiltonian}

The lattice Hamiltonian $\hat{H}$ is identified with the logarithm of
the transfer matrix $\hat{T}$:
\begin{equation}
\hat{H} = - \log \hat{T}.
\end{equation}
$\hat{T}$ and $\hat{H}$ for the asymmetric clover quark
action on the isotropic lattice have been derived 
in Ref.\cite{El-Khadra:1997mp}. 
An extension to the anisotropic lattice is straightforward.
Using
the fields $\hat{\Psi}$ and $\hat{\bar{\Psi}}=
\hat{\Psi}^{\dagger}\gamma_0$ which satisfy canonical
anti-commutation relations, 
the Hamiltonian in temporal lattice units $\hat{H}$ for the anisotropic 
quark action is given by
\begin{equation}
\hat{H} = \hat{\bar{\Psi}} \left[ a_t m_{1} - \frac{\zeta_{\rm F}
 a_s^2}{2(1+m_0)}(r_s {\bf D}^2 + ic_s{\bf\Sigma}\cdot{\bf B})
 -i\zeta_{\rm F} f_1(m_0)a_s\Theta - \zeta_{\rm F}^2 f_2(m_0)a_s^2\Theta^2
 \right]\hat{\Psi}+O({\bf p}^3a_s^3),
\end{equation}
where 
$(\Sigma_i,\alpha_i)=(-\frac{1}{2}\epsilon_{ijk}\sigma_{jk},-i\sigma_{0i})$,
$(B_i,E_i)=(\frac{1}{2}\epsilon_{ijk} F_{jk},F_{0i})$ and 
\begin{equation}
a_tm_{1} = \log(1+m_0),
\end{equation}
\begin{equation}
\Theta = i({\bf \gamma}\cdot{\bf D} +
\frac{1}{2}(1-c_t)a_t{\bf\alpha}\cdot{\bf E}),
\end{equation}
and
\begin{equation}
f_1(x) = \frac{2(1+x)\log (1+x)}{x(2+x)},\;\;\;\;
f_2(x) = \frac{f_1^2(x)}{2\log (1+x)} - \frac{1}{x(2+x)}.
\end{equation}
Therefore the lattice Hamiltonian in physical units is given by
\begin{eqnarray}
\frac{1}{a_t}\hat{H} &=& \hat{\bar{\Psi}} \left[ m_{1} - \frac{\zeta_{\rm F}
 \xi_0^2a_t}{2(1+m_0)}(r_s {\bf D}^2 + ic_s{\bf\Sigma}\cdot{\bf B})
 -i\zeta_{\rm F} f_1(m_0)\xi_0\Theta - \zeta_{\rm F}^2 f_2(m_0)\xi_0^2a_t\Theta^2
 \right]\hat{\Psi}+O({\bf p}^3a_s^2)\\
  &=& \hat{\bar{\Psi}} \left[ m_{1} - \frac{\zeta_{\rm F}^{\prime}
 a_t}{2(1+m_0)}(r_s^{\prime} {\bf D}^2 + ic_s^{\prime}{\bf\Sigma}\cdot{\bf B})
 -i\zeta_{\rm F}^{\prime} f_1(m_0)\Theta - {\zeta_{\rm F}^{\prime}}^2 f_2(m_0)a_t\Theta^2
 \right]\hat{\Psi}+O({\bf p}^3a_s^2),\label{anisoH}
\end{eqnarray}
where 
\begin{equation}
\zeta_{\rm F}^{\prime} = \xi_0\zeta_{\rm F},\;\;\; r_s^{\prime} = \xi_0
r_s,\;\;\; c_s^{\prime} = \xi_0 c_s.\label{dashedparam}
\end{equation}
Note that Eq.~(\ref{anisoH}) for the anisotropic lattice is the
same as that for the isotropic lattice except for use of
\{$a_t, \zeta_{\rm F}^{\prime}, r_s^{\prime}, c_s^{\prime}$\} instead of 
\{$a, \zeta_{\rm F}, r_s, c_s$\}. Thus one can repeat the derivation of 
the tree level value of bare parameters
($\zeta_{\rm F}$ and $c_{s,t}$)
in Ref.\cite{El-Khadra:1997mp} even for the
anisotropic lattice, after replacing \{$a, \zeta_{\rm F}, r_s, c_s$\}
by \{$a_t, \zeta_{\rm F}^{\prime}, r_s^{\prime}, c_s^{\prime}$\}.

When the lattice Hamiltonian is expressed in more
continuum-like form 
\begin{equation}
\frac{1}{a_t}\hat{H} = \hat{\bar{\Psi}} [ b_0m_q 
+ b_1{\bf \gamma}\cdot{\bf D} + a_t b_2 {\bf D}^2 + ia_t b_B 
{\bf\Sigma}\cdot{\bf B} + a_t b_E {\bf \alpha}\cdot{\bf E} +
a_t^2 b_{so} \gamma_0 [{\bf \gamma}\cdot{\bf D},{\bf \gamma}\cdot{\bf E}]
 + \cdots  ]\hat{\Psi},
\label{contlikeH}
\end{equation}
the coefficients $b$ are given by 
\begin{eqnarray} 
b_0 &=& m_1/m_q, \\
b_1 &=& \zeta_{\rm F}^{\prime} f_1(m_0), \\
b_2&=&-\frac{r_s^{\prime}\zeta_{\rm F}^{\prime}}{2(1+m_0)}+{\zeta_{\rm F}^{\prime}}^2 
f_2(m_0),\\
b_B&=&-\frac{c_s^{\prime}\zeta_{\rm F}^{\prime}}{2(1+m_0)}+{\zeta_{\rm F}^{\prime}}^2 
f_2(m_0),\\
b_E&=& \frac{1}{2}(1-c_t)\zeta_{\rm F}^{\prime}f_1(m_0),\\
b_{so}&=& -\frac{1}{2}(1-c_t){\zeta_{\rm F}^{\prime}}^2f_2(m_0).
\end{eqnarray}
In order to determine tree level parameters, the lattice
Hamiltonian should be matched to the continuum one to the
desired order in $a_s$. 
The continuum Hamiltonian to which the lattice one is matched is 
either the Dirac  Hamiltonian  $\hat{H}_{\rm Dirac} = 
a_t\hat{\bar{\Psi}}(m_q + {\bf \gamma}\cdot{\bf D})\hat{\Psi}$,
or the non-relativistic Hamiltonian 
$\hat{H}_{\rm NR} = 
a_t\hat{\bar{\Psi}}(m_q + \gamma_0 A_0 - \frac{{\bf D}^2}{2m_q} 
+ \cdots )\hat{\Psi}$.
Both choices give the same tree level parameters.

In the Hamiltonian formalism, the
unitary transformation $U$ is possible because the
eigenvalues of $\hat{H}$ are invariant under it.
For example, consider a unitary transformation, 
\begin{equation}
\hat{\Psi} \rightarrow U\hat{\Psi}, \;\;\;
\hat{\Psi}^\dagger \rightarrow \hat{\Psi}^\dagger U^{-1}
\end{equation}
with 
\begin{equation}
U = \exp (-a_t \theta_1 {\bf \gamma}\cdot{\bf D} - a_t^2 \theta_E
{\bf \alpha}\cdot{\bf E}),  
\label{FWT}
\end{equation}
where $\theta_1$ and $\theta_E $ are parameters.
This is called the Foldy-Wouthuysen-Tani (FWT)
transformation, whose element is a spin off-diagonal matrix.
After this transformation the coefficients $b$ become 
\begin{eqnarray} 
b_0^U &=& b_0, \\
b_1^U &=& b_1 - 2m_q a_t b_0 \theta_1, \\
b_2^U &=& b_2 - 2b_1 \theta_1 + 2m_q a_t b_0 \theta_1^2, \\
b_B^U &=& b_B - 2b_1 \theta_1 + 2m_q a_t b_0 \theta_1^2, \\
b_E^U &=& b_E - \theta_1 -  2m_q a_t b_0 \theta_E, \\
b_{so}^U &=& b_{so} - \frac{1}{2}\theta_1^2 + b_E\theta_1 +
b_1\theta_E - 2m_q a_t b_0 \theta_1\theta_E.
\end{eqnarray} 
The transformed Hamiltonian $\hat{H}^{U}$ with $b^U$
is matched to either $\hat{H}_{\rm Dirac}$ or $\hat{H}_{\rm NR}$
so as to obtain tree level parameters.

\begin{table} [h]
\vspace{0mm}
\vspace{5mm}
\begin{center}
  \begin{tabular}{ccccccr@{$\,\times\,$}lc}
$\beta$ &$\xi$ &$\xi_0$ &$c_s$ &$c_t$ & $a_s^{r_0}$[fm]
&$L^3 $&$ T$  & $L a_s $[fm]   \\ \hline
5.70 & 3 & 2.346 & 1.966  & 2.505  &   0.204  &$8^3 $&$ 48$ &     1.63\\
5.90 & 3 & 2.411 & 1.840  & 2.451  &   0.137  &$12^3 $&$ 72$ &    1.65\\
6.10 & 3 & 2.461 & 1.762  & 2.416  &   0.099  &$16^3 $&$ 96$ &    1.59\\
6.35 & 3 & 2.510 & 1.690  & 2.382  &   0.070  &$24^3 $&$ 144$&    1.67\\
\end{tabular}
\end{center}
\caption{Simulation parameters. 
$L a_s$ is calculated using $a_s^{r_0}$, 
the lattice spacing determined from $r_0$.}
\label{tab:simparam}
\end{table}

\begin{table} [h]
\vspace{0mm}
\vspace{5mm}
\begin{center}
 \begin{tabular}{cr@{$\,\times\,$}lrlcrll}
$\beta$ & $L^3 $&$ T$ & $a_tm_{q0}$ &  $\;\;\;\zeta$   & sweep/conf
& \#conf &$c_{\rm PS}$ & $c_{\rm V}$ \\ \hline
5.70 & $8^3 $&$ 48$  & 0.320  & 2.88 (NP)  &100 &1000 & 1.005(10) &  1.008(11)  \\
5.70 & $8^3 $&$ 48$  & 0.253  & 2.85 (NP) &100 &1000  & 1.005(10) &  1.008(11)  \\
5.70 & $8^3 $&$ 48$  & 0.320  & 3.08 (TI) &100 &1000  & 0.962(9) &  0.965(10)   \\
5.70 & $8^3 $&$ 48$  & 0.253  & 3.03 (TI) &100 &1000  & 0.966(9) &  0.969(10)   \\
5.90 & $12^3 $&$ 72$ & 0.144  & 2.99 (NP/TI) &100 &1000  & 0.991(8) &  0.993(9)  \\
5.90 & $12^3 $&$ 72$ & 0.090  & 2.93 (NP/TI) &100 &1000  & 0.991(8) &  0.994(9)  \\
6.10 & $16^3 $&$ 96$ & 0.056  & 3.01 (NP) &200 & 600  & 0.997(9) &  0.997(9)  \\
6.10 & $16^3 $&$ 96$ & 0.024  & 2.96 (NP) &200 & 600  & 0.997(9) &  0.997(9)  \\
6.10 & $16^3 $&$ 96$ & 0.056  & 2.92 (TI) &200 & 600  & 1.017(9) &  1.018(9)   \\
6.10 & $16^3 $&$ 96$ & 0.024  & 2.88 (TI) &200 & 600  & 1.017(9) &  1.016(10)  \\
6.35 & $24^3 $&$ 144$& $-$0.005 & 2.87 (NP/TI) &400 & 400  & 1.006(11) &  1.011(11)  \\
6.35 & $24^3 $&$ 144$& $-$0.035 & 2.81 (NP/TI) &400 & 400  & 1.007(12) &  1.009(11)  \\
\end{tabular}
\end{center}
\caption{
Simulation parameters continued. In fourth column, `NP' and `TI' denote 
the nonperturbative and tree level tadpole improved values for $\zeta$
respectively. $c_{\rm PS,V}$ are the speed of light obtained from the
fit for the pseudoscalar ($^1S_0$) and vector ($^3S_1$) mesons.
}
\label{tab:simparam2}
\end{table}

\begin{table} [h]
\vspace{0mm}
\vspace{5mm}
\begin{center}
  \begin{tabular}{lcccccccc}
set &$\xi$ &$\zeta$ &$c_s$ &$c_t$ & $u_{s,t}$ & $M_{\rm lat}(1\bar{S})$
& scale input & HFS($a_s\!=\!0,r_0$)\\ \hline
A. this work & 3 & 
TI($m\geq 0$), NP 
& TI($m\geq 0$)&TI($m=0$) & $u^P$ &
$M_{\rm pole}, M_{\rm kin}$ 
& $r_0$, $1\bar{P}-1\bar{S}$, $2\bar{S}-1\bar{S}$ & $\approx$ 75 MeV\\
B. Klassen\cite{Klassen} & 2,3 & NP & TI($m\geq 0$)&TI($m=0$) & $u^L$ &
$M_{\rm pole} (\simeq M_{\rm kin})$ &$r_0$ & $\approx$ 75 MeV\\
C. Chen\cite{Chen:2001ej} & 2 & NP & TI($m\geq 0$)&TI($m=0$) & $u^L$ &
$M_{\rm pole} (\simeq M_{\rm kin})$ &$r_0$ & $\approx$ 75 MeV\\
D. Klassen\cite{Klassen:1999fh,Klassen} & 2,3 & NP & TI($m=0$)&TI($m=0$) &$u^L$ &
$M_{\rm pole} (\simeq M_{\rm kin})$ &$r_0$ & $\approx$ 95 MeV\\
\end{tabular}
\end{center}
\caption{Comparison of simulation parameters in various anisotropic
 lattice studies of the $c\bar{c}$ spectrum. 
In the third to fifth columns, TI($m\geq 0$), TI($m=0$) and NP
respectively denote the tree level tadpole improved value for massive
quarks, that are correct only 
in the massless limit and the non-perturbative value.
The sixth column shows which method is used for the estimation of the
tadpole factors $u_{s,t}$ (the plaquette prescription $u^P$
or the Landau mean link prescription $u^L$).
The seventh column shows which $1\bar{S}$ mass is 
tuned to the experimental value. The eighth column denotes
quantities used for the scale setting.
The final column is the continuum estimate of the hyperfine splitting
from the $a_s^2$-linear fit with the scale set by $r_0$.}
\label{tab:simparam3}
\end{table}

\begin{table} [h]
\vspace{0mm}
\begin{center}
  \begin{tabular}{ccccl}
$^{2S+1}L_J$ & $J^{PC}$ 
& name &$\Gamma$ operator &$\Gamma \Delta$ operator \\ \hline
$^1S_0$ & $0^{-+}$ & $\eta_c$ & $\bar{\psi} \gamma_5 \psi$ &  \\
$^3S_1$ & $1^{--}$ & $J/\psi$ & $\bar{\psi} \gamma_i \psi$ &\\
$^1P_1$ & $1^{+-}$ & $h_c$ & $\bar{\psi} \sigma_{ij} \psi$ &
 $\bar{\psi} \gamma_5 \Delta_i \psi$   \\
$^3P_0$ & $0^{++}$ & $\chi_{c0}$ & $\bar{\psi}  \psi$ &  
$\bar{\psi} \sum_i \gamma_i \Delta_i \psi$ \\
$^3P_1$ & $1^{++}$ & $\chi_{c1}$ &$\bar{\psi}
\gamma_i\gamma_5 \psi$   & $\bar{\psi} \{ \gamma_i \Delta_j - \gamma_j \Delta_i
\} \psi$  \\
$^3P_2$ & $2^{++}$ & $\chi_{c2}$ &  &   $\bar{\psi} \{ \gamma_i
\Delta_i - \gamma_j \Delta_j \} \psi$ (E rep)\\
 & & & & $\bar{\psi} \{ \gamma_i \Delta_j + \gamma_j
\Delta_i \} \psi$ (T rep)\\
\end{tabular}
\end{center}
\caption{S- and P-state operators. In the first and second columns, the
 state is labeled by $^{2S+1}L_J$ and $J^{PC}$ respectively. The third
 column shows the particle name for the charmonium family. In the fourth
 and fifth columns, we give the corresponding $\Gamma$ operator and 
 $\Gamma \Delta$ operator.
}
\label{tab:SPope}
\end{table}

\begin{table} [h]
\vspace{0mm}
\begin{center}
  \begin{tabular}{ccccccc}
state & fit form & source & \multicolumn{4}{c}{fit range
($t_{\rm min}/t_{\rm max}$)} \\ \hline
 &  & & $\beta = 5.70$ & $\beta = 5.90$ 
& $\beta = 6.10$ & $\beta = 6.35$ \\ \cline{4-7}
$1S, 2S$               & 2-cosh & 00+01+11 & 11/24 &17/36 &22/48 & 32/72 \\
$1P, 2P$               & 2-cosh & 00+11+02+12 & 7/18 &11/25 &15/35 &21/50 \\
$1\bar{S}, \Delta S$   & 1-cosh & 01 & 13/24 &19/36 &26/48 & 38/72 \\
$1S({\bf p \ne 0})$   & 1-cosh & 01 & 13/22 & 20/32 & 26/45 & 40/66  \\
$\Delta P$  & 1-cosh & 12 & 11/18 &17/25 &23/35 & 33/50 \\
\end{tabular}
\end{center}
\caption{Fit ranges we adopted. In the first column, $\Delta S$ and $\Delta P$ 
denote the S- and P-state spin mass splitting respectively.}
\label{tab:frange}
\end{table}

\begin{table} [h]
\vspace{0mm}
\vspace{5mm}
\begin{center}
  \begin{tabular}{ccccccc}
$\beta$ &$r_0/a_s$ & $L^3 \times T$ &$La_s$[fm] & smear\# &conf
  & sweep/conf   \\ \hline
5.70 &  2.449(35)  & $12^3 \times 72$ &  2.45 &  4 &  150  &  100\\
5.90 &  3.644(36)  & $12^3 \times 36$ &  1.65 &  5 &  220  &  100\\
6.00 &  4.359(51)  & $12^3 \times 48$ &  1.38 &  6 &  150  &  100\\
6.10 &  5.028(35)  & $16^3 \times 48$ &  1.59 &  6 &  150  &  100\\
6.20 &  5.822(33)  & $16^3 \times 64$ &  1.37 & 10 &  220  &  100\\
6.35 &  7.198(52)  & $24^3 \times 72$ &  1.67 & 12 &  150  &  200\\
\end{tabular}
\end{center}
\caption{Simulation parameters and results for the Sommer scale $r_0$. 
The fifth column shows the number of smearing steps we adopted.
}
\label{tab:R0}
\end{table}

\begin{table} [h]
\vspace{0mm}
\vspace{5mm}
\begin{center}
\begin{tabular}{cr@{.}lr@{.}lr@{.}lr@{.}lr@{.}lr@{.}l}
$\beta$ & \multicolumn{4}{c}{$r_0$} & \multicolumn{4}{c}{$1\bar{P}-1\bar{S}$} & 
\multicolumn{4}{c}{$2\bar{S}-1\bar{S}$}    \\ \hline
&\multicolumn{2}{c}{$m_0^{\rm charm}$} &
\multicolumn{2}{c}{$a_s^{r_0}$[fm]} & \multicolumn{2}{c}{$m_0^{\rm charm}$} & 
\multicolumn{2}{c}{$a_s^{1\bar{P}-1\bar{S}}$[fm]} & 
\multicolumn{2}{c}{$m_0^{\rm charm}$} & 
\multicolumn{2}{c}{$a_s^{2\bar{S}-1\bar{S}}$[fm]} \\ \cline{2-13}
5.70 &  0&2843(3) &  0&2037(0) &  0&2994(115) &  0&2077(30) &  0&3782(190) &  0&2272(45)  \\
5.90 &  0&1106(2) &  0&1374(0) &  0&0972(58) &  0&1333(18) &  0&1664(150) &  0&1544(44)  \\ 
6.10 &  0&0319(1) &  0&0991(0) &  0&0155(60) &  0&0934(21) &  0&0632(110) &  0&1099(37)  \\ 
6.35 & $-$0&0179(1) &  0&0697(0) & $-$0&0301(43) &  0&0650(18) &  0&0115(84) &  0&0808(30)  \\  
\end{tabular}
\end{center}
\caption{Bare charm quark mass $m_0^{\rm charm}$ and lattice spacing $a_s^Q$ for 
$Q=r_0,\ 1\bar{P}-1\bar{S}$ and $2\bar{S}-1\bar{S}$.}
\label{tab:m0ch-as}
\end{table}

\begin{table} [h]
\vspace{0mm}
\vspace{5mm}
\begin{center}
  \begin{tabular}{ccccccc}
$(M_{\rm lat},\zeta)$ & $\dM(1\bar{P}-1\bar{S})$ & $\dM(2\bar{S}-1\bar{S})$
 & $\dM(1^3S_1-1^1S_0)$  & $\dM(1^3P_1-1^3P_0)$      \\ \hline
$(M_{\rm pole},\zeta^{\rm TI})$  & 426.7(104)& 676(30) & 71.6(07) & 57.3(37)\\
$(M_{\rm pole},\zeta^{\rm NP})$  & 423.1(096) & 671(29) & 68.8(06) & 55.3(34)\\
$(M_{\rm kin},\zeta^{\rm TI})$   & 424.1(097) & 671(31) & 69.2(14)& 55.2(38)\\
$(M_{\rm kin},\zeta^{\rm NP})$   & 423.6(097) & 672(30) & 69.2(13)& 55.7(37)\\
\end{tabular}
\end{center}
\caption{Comparison of mass splittings for different choices 
of $(M_{\rm lat},\zeta)$ at $\beta =6.10$. 
The results are presented in units of MeV, and the scale is set by $r_0$.}
\label{tab:comptuning}
\end{table}

\begin{table} [h]
\vspace{0mm}
\vspace{5mm}
\begin{center}
  \begin{tabular}{lr@{(}lr@{(}lr@{(}lr@{(}lr@{(}lr}
state  
& \multicolumn{2}{c}{$\beta=5.70$} & \multicolumn{2}{c}{$\beta=5.90$}  
& \multicolumn{2}{c}{$\beta=6.10$} & \multicolumn{2}{c}{$\beta=6.35$} 
& \multicolumn{2}{c}{$a_s \rightarrow 0$}  & Exp.~ \\ \hline
$1^1S_0$             &   3020.9&7) &   3013.8&8) &   3014.0&10) &   3012.7&9) &   3012.7&11) &   2979.8 \\ 
$1^3S_1$             &   3082.0&7) &   3083.1&8) &   3085.1&8) &   3083.7&8) &   3084.6&10) &   3096.9 \\ 
$1^1P_1$             &   3526.6&79) &   3506.7&57) &   3489.7&66) &   3483.8&83) &   3474.2&94) &   3526.1 \\ 
$1^3P_0$             &   3496.0&94) &   3462.4&65) &   3438.7&58) &   3420.2&86) &   3408.5&95) &   3415.0 \\ 
$1^3P_1$             &   3526.7&84) &   3506.6&61) &   3490.5&62) &   3480.8&80) &   3472.3&91) &   3510.5 \\ 
$1^3P_{2E}$          &   3555.2&106) &   3515.6&116) &   3509.8&199) &   3506.7&219) &   3503.6&250) &   3556.2 \\ 
$1^3P_{2T}$          &   3555.0&100) &   3512.4&115) &   3508.9&179) &   3502.5&213) &   3501.2&238) &   3556.2 \\ 
$1\bar{S}$           &   3067.6&0) &   3067.6&0) &   3067.6&0) &   3067.6&0) &   3067.6&0) &   3067.6 \\ 
$1\bar{P}$           &   3536.0&85) &   3506.7&73) &   3494.0&104) &   3487.3&120) &   3480.4&137) &   3525.5 \\ 
$1^1P_1-1\bar{S}$    &    459.9&79) &    440.9&59) &    422.4&67) &    417.8&84) &    407.2&95) &    458.5 \\ 
$1^3P_0-1\bar{S}$    &    429.2&93) &    396.7&66) &    371.3&61) &    354.2&87) &    341.2&97) &    347.4 \\ 
$1^3P_1-1\bar{S}$    &    459.9&84) &    440.9&62) &    423.2&64) &    414.9&81) &    405.2&93) &    442.9 \\ 
$1^3P_2-1\bar{S}$    &    488.5&106) &    449.9&117) &    442.5&198) &    440.7&218) &    436.6&249) &    488.6 \\ 
$1\bar{P}-1\bar{S}$  &    469.3&85) &    441.0&74) &    426.7&104) &    421.3&121) &    413.4&138) &    457.9 \\ 
$1^3S_1-1^1S_0$      &     61.9&4) &     70.4&6) &     71.6&7) &     72.0&8) &     72.6&9) &    117.1 \\ 
$1^3P_1-1^3P_0$      &     32.3&34) &     46.7&34) &     57.3&37) &     62.7&42) &     68.4&50) &     95.5 \\ 
$1^3P_2-1^3P_1$      &     18.1&43) &     18.2&41) &     20.4&68) &     30.4&72) &     31.1&84) &     45.7 \\ 
$1^3P_{2T}-1^3P_{2E}$ &     $-$0.8&23) &     $-$2.3&28) &     $-$2.6&33) &     $-$2.0&41) &     $-$2.2&47) &      0.0 \\ 
$1^1P_1-1^3P$        &     $-$6.0&18) &     $-$3.5&21) &     $-$0.7&29) &     $-$3.5&36) &     $-$1.4&40) &      0.9 \\ 
$\frac{1^3P_2-1^3P_1}{1^3P_1-1^3P_0}$    &     0.56&13) &     0.39&9) &     0.36&12) &     0.49&11) &     0.47&14) &     0.48 \\ \hline 
$2^1S_0$             &     3719&22) &     3700&28) &     3699&32) &     3746&40) &     3739&46) &     3594 \\ 
$2^3S_1$             &     3767&20) &     3773&27) &     3758&31) &     3786&34) &     3777&40) &     3686 \\ 
$2^1P_1$             &     4248&68) &     4411&70) &     4214&70) &     4161&79) &     4053&95) &      -   \\ 
$2^3P_0$             &     4175&93) &     4226&89) &     4148&94) &     4049&100) &     4008&122) &      -   \\ 
$2^3P_1$             &     4228&75) &     4388&77) &     4256&90) &     4140&84) &     4067&105) &      -   \\ 
$2^3P_{2E}$          &     4238&109) &     4254&99) &     4190&144) &     4023&148) &     3992&175) &      -   \\ 
$2^3P_{2T}$          &     4230&111) &     4281&100) &     4223&157) &     4082&146) &     4047&177) &      -   \\ 
$2\bar{S}$           &     3755&20) &     3755&27) &     3744&30) &     3776&34) &     3768&40) &     3663 \\ 
$2\bar{P}$           &     4233&74) &     4324&68) &     4209&86) &     4089&86) &     4027&105) &      -   \\ 
$2\bar{P}-2\bar{S}$  &      478&73) &      569&70) &      466&90) &      313&88) &      256&107) &      -   \\ 
$2^3S_1-2^1S_0$      &       48&9) &       74&16) &       60&17) &       40&22) &       34&25) &       92 \\ 
$2^1S_0-1^1S_0$      &      698&22) &      686&28) &      685&32) &      733&40) &      726&46) &      614 \\ 
$2^3S_1-1^3S_1$      &      685&20) &      690&27) &      673&31) &      702&34) &      692&40) &      589 \\ 
$2^1P_1-1^1P_1$      &      721&68) &      904&69) &      724&69) &      678&79) &      579&94) &      -   \\ 
$2^3P_0-1^3P_0$      &      679&95) &      763&90) &      709&95) &      629&103) &      601&124) &      -   \\ 
$2^3P_1-1^3P_1$      &      701&76) &      881&77) &      766&90) &      659&84) &      595&105) &      -   \\ 
$2^3P_2-1^3P_2$      &      683&109) &      738&93) &      681&129) &      516&136) &      490&160) &      -   \\ 
$2\bar{S}-1\bar{S}$  &      688&20) &      689&27) &      676&30) &      710&34) &      701&40) &      595 \\ 
$2\bar{P}-1\bar{P}$  &      697&75) &      817&66) &      715&81) &      602&83) &      547&100) &      -   \\ 
\end{tabular}
\end{center}
\caption{Results of charmonium masses $M$ and mass splittings $\dM$
in units of MeV at $\zeta = \zeta^{\rm TI}$ 
using the pole mass tuning. The scale is set by $r_0$. 
}
\label{tab:cc_r0}
\end{table}

\begin{table} [h]
\vspace{0mm}
\vspace{5mm}
\begin{center}
  \begin{tabular}{lr@{(}lr@{(}lr@{(}lr@{(}lr@{(}lr}
state  
& \multicolumn{2}{c}{$\beta=5.70$} & \multicolumn{2}{c}{$\beta=5.90$}  
& \multicolumn{2}{c}{$\beta=6.10$} & \multicolumn{2}{c}{$\beta=6.35$} 
& \multicolumn{2}{c}{$a_s \rightarrow 0$}  & Exp.~ \\ \hline
$1^1S_0$             &   3023.0&16) &   3010.3&16) &   3007.1&27) &   3004.3&33) &   3003.0&35) &   2979.8 \\ 
$1^3S_1$             &   3081.4&8) &   3084.0&10) &   3087.1&12) &   3086.0&12) &   3087.5&14) &   3096.9 \\ 
$1^1P_1$             &   3515.6&29) &   3523.3&46) &   3520.7&88) &   3519.9&98) &   3518.6&106) &   3526.1 \\ 
$1^3P_0$             &   3486.6&49) &   3476.2&51) &   3464.0&91) &   3446.4&92) &   3441.6&104) &   3415.0 \\ 
$1^3P_1$             &   3515.8&35) &   3523.5&44) &   3522.3&96) &   3516.8&102) &   3516.8&112) &   3510.5 \\ 
$1^3P_{2E}$          &   3543.2&40) &   3532.9&60) &   3541.3&128) &   3544.9&139) &   3548.9&151) &   3556.2 \\ 
$1^3P_{2T}$          &   3543.0&38) &   3529.3&69) &   3539.8&122) &   3540.0&155) &   3546.0&160) &   3556.2 \\ 
$1\bar{S}$           &   3067.6&0) &   3067.6&0) &   3067.6&0) &   3067.6&0) &   3067.6&0) &   3067.6 \\ 
$1\bar{P}$           &   3524.7&7) &   3523.4&7) &   3525.0&9) &   3523.4&8) &   3524.1&9) &   3525.5 \\ 
$1^1P_1-1\bar{S}$    &    448.8&29) &    457.8&46) &    453.6&89) &    454.3&100) &    452.0&108) &    458.5 \\ 
$1^3P_0-1\bar{S}$    &    419.8&47) &    410.6&51) &    396.9&93) &    380.9&95) &    375.2&106) &    347.4 \\ 
$1^3P_1-1\bar{S}$    &    448.9&34) &    457.9&44) &    455.3&98) &    451.3&104) &    450.3&114) &    442.9 \\ 
$1^3P_2-1\bar{S}$    &    476.4&40) &    467.4&58) &    474.2&126) &    479.4&136) &    482.4&148) &    488.6 \\ 
$1\bar{P}-1\bar{S}$  &    457.9&0) &    457.9&0) &    457.9&0) &    457.9&0) &    457.9&0) &    457.9 \\ 
$1^3S_1-1^1S_0$      &     59.2&18) &     74.9&21) &     80.4&34) &     82.7&42) &     85.3&44) &    117.1 \\ 
$1^3P_1-1^3P_0$      &     30.6&37) &     49.9&39) &     64.6&45) &     72.6&65) &     79.2&66) &     95.5 \\ 
$1^3P_2-1^3P_1$      &     17.4&41) &     19.2&43) &     22.3&75) &     34.7&81) &     35.0&90) &     45.7 \\ 
$1^3P_{2T}-1^3P_{2E}$ &     $-$0.8&22) &     $-$2.5&30) &     $-$3.2&39) &     $-$2.1&51) &     $-$2.7&53) &      0.0 \\ 
$1^1P_1-1^3P$        &     $-$5.9&17) &     $-$3.7&22) &     $-$0.8&35) &     $-$3.7&44) &     $-$1.5&46) &      0.9 \\ 
$\frac{1^3P_2-1^3P_1}{1^3P_1-1^3P_0}$    &     0.57&12) &     0.39&9) &     0.35&13) &     0.48&12) &     0.45&14) &     0.48 \\ \hline 
$2^1S_0$             &     3704&22) &     3722&30) &     3746&39) &     3801&45) &     3806&50) &     3594 \\ 
$2^3S_1$             &     3749&21) &     3800&29) &     3811&41) &     3847&43) &     3849&49) &     3686 \\ 
$2^1P_1$             &     4217&70) &     4458&75) &     4294&79) &     4238&87) &     4159&100) &      -   \\ 
$2^3P_0$             &     4146&95) &     4260&95) &     4222&105) &     4121&124) &     4114&138) &      -   \\ 
$2^3P_1$             &     4196&78) &     4434&83) &     4339&100) &     4222&96) &     4179&114) &      -   \\ 
$2^3P_{2E}$          &     4203&107) &     4303&96) &     4263&145) &     4096&155) &     4091&173) &      -   \\ 
$2^3P_{2T}$          &     4194&111) &     4329&98) &     4287&163) &     4147&153) &     4131&177) &      -   \\ 
$2\bar{S}$           &     3738&21) &     3781&29) &     3794&39) &     3836&42) &     3839&47) &     3663 \\ 
$2\bar{P}$           &     4200&76) &     4371&68) &     4286&81) &     4165&88) &     4132&100) &      -   \\ 
$2\bar{P}-2\bar{S}$  &      462&72) &      590&72) &      492&95) &      329&97) &      290&112) &      -   \\ 
$2^3S_1-2^1S_0$      &       45&9) &       78&18) &       65&20) &       47&27) &       43&29) &       92 \\ 
$2^1S_0-1^1S_0$      &      681&23) &      712&30) &      738&40) &      797&46) &      803&51) &      614 \\ 
$2^3S_1-1^3S_1$      &      668&21) &      716&29) &      723&40) &      762&43) &      762&48) &      589 \\ 
$2^1P_1-1^1P_1$      &      701&69) &      935&73) &      773&76) &      718&84) &      641&97) &      -   \\ 
$2^3P_0-1^3P_0$      &      659&96) &      783&96) &      758&106) &      674&122) &      671&137) &      -   \\ 
$2^3P_1-1^3P_1$      &      681&77) &      910&82) &      817&99) &      705&94) &      662&111) &      -   \\ 
$2^3P_2-1^3P_2$      &      660&107) &      770&93) &      722&135) &      551&147) &      543&164) &      -   \\ 
$2\bar{S}-1\bar{S}$  &      671&21) &      715&28) &      727&39) &      770&42) &      772&47) &      595 \\ 
$2\bar{P}-1\bar{P}$  &      675&76) &      847&68) &      761&81) &      641&87) &      608&100) &      -   \\ 
\end{tabular}
\end{center}
\caption{
The same as Table \ref{tab:cc_r0}, but the scale is set by $1\bar{P}-1\bar{S}$
splitting.
}
\label{tab:cc_1p1s}
\end{table}

\begin{table} [h]
\vspace{0mm}
\vspace{5mm}
\begin{center}
  \begin{tabular}{lr@{(}lr@{(}lr@{(}lr@{(}lr@{(}lr}
state  
& \multicolumn{2}{c}{$\beta=5.70$} & \multicolumn{2}{c}{$\beta=5.90$}  
& \multicolumn{2}{c}{$\beta=6.10$} & \multicolumn{2}{c}{$\beta=6.35$} 
& \multicolumn{2}{c}{$a_s \rightarrow 0$}  & Exp.~ \\ \hline
$1^1S_0$             &   3032.3&21) &   3026.4&30) &   3024.9&33) &   3028.6&38) &   3027.4&45) &   2979.8 \\ 
$1^3S_1$             &   3079.1&8) &   3079.8&10) &   3082.0&13) &   3079.5&12) &   3080.5&15) &   3096.9 \\ 
$1^1P_1$             &   3467.1&113) &   3446.7&139) &   3440.5&158) &   3415.3&170) &   3412.6&208) &   3526.1 \\ 
$1^3P_0$             &   3445.3&112) &   3412.8&124) &   3398.6&130) &   3370.2&128) &   3361.5&165) &   3415.0 \\ 
$1^3P_1$             &   3467.8&117) &   3446.1&142) &   3440.1&158) &   3412.4&168) &   3409.7&207) &   3510.5 \\ 
$1^3P_{2E}$          &   3490.4&124) &   3453.4&153) &   3460.0&198) &   3433.8&200) &   3437.7&244) &   3556.2 \\ 
$1^3P_{2T}$          &   3490.1&120) &   3451.6&155) &   3460.0&185) &   3431.2&180) &   3435.3&226) &   3556.2 \\ 
$1\bar{S}$           &   3067.6&0) &   3067.6&0) &   3067.6&0) &   3067.6&0) &   3067.6&0) &   3067.6 \\ 
$1\bar{P}$           &   3475.2&114) &   3446.5&140) &   3445.0&164) &   3418.5&170) &   3418.2&209) &   3525.5 \\ 
$1^1P_1-1\bar{S}$    &    399.7&114) &    380.2&141) &    372.8&159) &    348.5&172) &    345.1&210) &    458.5 \\ 
$1^3P_0-1\bar{S}$    &    377.9&113) &    346.4&126) &    330.8&131) &    303.4&131) &    294.2&168) &    347.4 \\ 
$1^3P_1-1\bar{S}$    &    400.4&118) &    379.7&144) &    372.3&159) &    345.6&171) &    342.2&210) &    442.9 \\ 
$1^3P_2-1\bar{S}$    &    423.0&126) &    386.9&155) &    392.2&199) &    367.0&202) &    370.4&246) &    488.6 \\ 
$1\bar{P}-1\bar{S}$  &    407.8&116) &    380.1&142) &    377.3&164) &    351.7&173) &    350.8&212) &    457.9 \\ 
$1^3S_1-1^1S_0$      &     47.4&25) &     54.4&38) &     57.7&43) &     51.5&48) &     53.9&58) &    117.1 \\ 
$1^3P_1-1^3P_0$      &     23.2&29) &     35.2&35) &     45.8&46) &     43.9&54) &     50.5&62) &     95.5 \\ 
$1^3P_2-1^3P_1$      &     14.1&32) &     14.4&30) &     17.3&51) &     22.2&52) &     23.7&61) &     45.7 \\ 
$1^3P_{2T}-1^3P_{2E}$ &     $-$1.0&15) &     $-$1.7&17) &     $-$1.6&23) &     $-$1.9&24) &     $-$1.8&29) &      0.0 \\ 
$1^1P_1-1^3P$        &     $-$5.4&12) &     $-$2.7&14) &     $-$0.6&21) &     $-$3.0&23) &     $-$1.5&26) &      0.9 \\ 
$\frac{1^3P_2-1^3P_1}{1^3P_1-1^3P_0}$    &     0.60&12) &     0.41&8) &     0.38&11) &     0.50&10) &     0.49&13) &     0.48 \\ \hline 
$2^1S_0$             &     3637&6) &     3618&8) &     3624&10) &     3641&11) &     3644&13) &     3594 \\ 
$2^3S_1$             &     3671&2) &     3676&3) &     3676&3) &     3669&4) &     3669&4) &     3686 \\ 
$2^1P_1$             &     4078&59) &     4241&69) &     4087&70) &     4015&76) &     3930&95) &      -   \\ 
$2^3P_0$             &     4020&77) &     4103&76) &     4031&80) &     3914&88) &     3877&108) &      -   \\ 
$2^3P_1$             &     4057&66) &     4222&73) &     4125&82) &     3985&81) &     3929&103) &      -   \\ 
$2^3P_{2E}$          &     4049&85) &     4078&85) &     4076&120) &     3884&106) &     3872&134) &      -   \\ 
$2^3P_{2T}$          &     4037&87) &     4109&84) &     4120&128) &     3958&104) &     3948&133) &      -   \\ 
$2\bar{S}$           &     3663&1) &     3662&1) &     3663&1) &     3662&1) &     3663&1) &     3663 \\ 
$2\bar{P}$           &     4056&61) &     4157&65) &     4087&79) &     3945&73) &     3900&93) &      -   \\ 
$2\bar{P}-2\bar{S}$  &      393&61) &      495&65) &      424&79) &      283&73) &      237&93) &      -   \\ 
$2^3S_1-2^1S_0$      &       34&7) &       59&11) &       52&13) &       29&14) &       26&17) &       92 \\ 
$2^1S_0-1^1S_0$      &      605&5) &      592&8) &      600&10) &      612&10) &      616&13) &      614 \\ 
$2^3S_1-1^3S_1$      &      592&2) &      597&3) &      594&3) &      590&3) &      588&4) &      589 \\ 
$2^1P_1-1^1P_1$      &      611&57) &      794&63) &      647&64) &      600&73) &      517&88) &      -   \\ 
$2^3P_0-1^3P_0$      &      575&77) &      690&74) &      633&79) &      543&86) &      514&105) &      -   \\ 
$2^3P_1-1^3P_1$      &      589&64) &      776&67) &      685&78) &      573&76) &      520&96) &      -   \\ 
$2^3P_2-1^3P_2$      &      559&85) &      624&77) &      616&109) &      450&104) &      443&128) &      -   \\ 
$2\bar{S}-1\bar{S}$  &      595&0) &      595&0) &      595&0) &      595&0) &      595&0) &      595 \\ 
$2\bar{P}-1\bar{P}$  &      581&60) &      710&58) &      642&72) &      526&70) &      487&87) &      -   \\ 
\end{tabular}
\end{center}
\caption{
The same as Table \ref{tab:cc_r0}, but the scale is set by $2\bar{S}-1\bar{S}$
splitting.
}
\label{tab:cc_2s1s}
\end{table}

\vspace{1cm}


\vspace{1cm}

\vspace{1cm}





\end{document}